\documentclass[lettersize,journal]{IEEEtran}
\usepackage{amsmath,amsfonts}
\usepackage{algorithmic}
\usepackage{algorithm}
\usepackage{array}
\usepackage[caption=false,font=normalsize,labelfont=sf,textfont=sf]{subfig}
\usepackage{textcomp}
\usepackage{stfloats}
\usepackage{url}
\usepackage{verbatim}
\usepackage{graphicx}
\usepackage{cite}
\usepackage{comment}
\usepackage{multirow}
\usepackage{booktabs}
\usepackage{url}
\usepackage{balance} % For balanced columns on the last page
\usepackage{xcolor}
\usepackage{color}

\begin{document}

\title{Tradeoffs in Processing Queries and Supporting Updates over an ML-Enhanced R-tree}

\author{Abdullah-Al-Mamun, Ch. Md. Rakin Haider, 
%Raymond A. Yeh, 
Jianguo Wang, Walid G. Aref,~\IEEEmembership{Fellow,~IEEE,}
        % <-this % stops a space
% \thanks{This paper was produced by the IEEE Publication Technology Group. They are in Piscataway, NJ.}% <-this % stops a space
% \thanks{Manuscript received April 19, 2021; revised August 16, 2021.}
}

% The paper headers
% \markboth{Journal of \LaTeX\ Class Files,~Vol.~14, No.~8, August~2021}%
% {Shell \MakeLowercase{\textit{et al.}}: A Sample Article Using IEEEtran.cls for IEEE Journals}

%\IEEEpubid{0000--0000/00\$00.00~\copyright~2021 IEEE}
% Remember, if you use this you must call \IEEEpubidadjcol in the second
% column for its text to clear the IEEEpubid mark.

\maketitle

\begin{abstract}
Machine Learning (ML) techniques have been successfully applied to design various learned database index structures for both the one- and multi-dimensional spaces. Particularly, a class of traditional multi-dimensional indexes 
%have 
has
been augmented with ML models to design ML-enhanced variants of their traditional counterparts. This paper focuses on the R-tree multi-dimensional index structure as it is widely used for indexing multi-dimensional 
%structure. 
data.
The R-tree has been augmented with machine learning models to enhance the R-tree performance. The AI+R-tree is an ML-enhanced R-tree index structure that augments a traditional disk-based R-tree with an ML model to enhance the R-tree's query processing performance, mainly, to avoid navigating the overlapping branches of the R-tree that do not yield query results,   
%disk-based traditional 
%traditional disk-based
%R-tree 
e.g., 
in the presence of high-overlap among the rectangles of the R-tree nodes.
We investigate the empirical tradeoffs in processing dynamic query workloads and 
%design tradeoffs for supporting 
and in supporting
updates over the AI+R-tree. Particularly, we investigate the impact of the choice of ML models over the AI+R-tree query processing performance. Moreover, we present a case study of designing a custom loss function for a neural network model tailored to the query processing requirements of the AI+R-tree. Furthermore, we present the design tradeoffs for adopting 
%different 
various
strategies for supporting dynamic inserts, updates, and deletes with the vision of realizing a mutable AI+R-tree. Experiments on real datasets demonstrate that the AI+R-tree can enhance the query processing performance of a traditional R-tree for high-overlap range queries by up to 5.4X while achieving up to 99\% average query recall. 
\end{abstract}


\section{Introduction}
%In the past few years
Recently, machine learning techniques (ML techniques, for short) have been successfully applied to build various learned database system components~\cite{cong2024machine}. Particularly, ML techniques have been applied to database indexes to build learned index structures, e.g.,~\cite{kraska2018case, Ferragina_PGM, ding2019alex, LIPP_index_vldb} for the one-dimensional space. The concept of learned one-dimensional learned indexes has been extended to 
%build 
realize 
learned multi-dimensional indexes~\cite{kraska2019sagedb, al2020tutorial, liu2024good, li2024survey, al2024survey}. Moreover, learned indexes can be broadly categorized into two types: {\em Pure} and {\em Hybrid} learned indexes~\cite{al2024survey}. The core idea behind the class of hybrid learned indexes is that they incorporate ML models to enhance the performance of a traditional index structure~\cite{llavesh2019accelerating, dai2020wisckey}. These hybrid learned indexes can also be viewed as ML-enhanced variants of their traditional counterparts.

On the other hand, traditional multi-dimensional (e.g., spatial) index structures have been used successfully over the years as efficient access methods for multi-dimensional data (e.g., location data). In the area of spatial databases, the R-tree~\cite{guttman1984r} is a widely-used index structure. In the multi-dimensional space, the R-tree is analogous to the one-dimensional index structure B$^{+}$-tree~\cite{comer1979ubiquitous}. In the R-tree, objects are stored using Minimum Bounding Rectangles (MBRs). Notice that in the B$^{+}$-tree, nodes do not overlap in space. However, the MBRs  of non-leaf and leaf nodes of an R-tree can overlap in space.

\begin{figure}[t!]
  \centering
  \includegraphics[width=\linewidth]{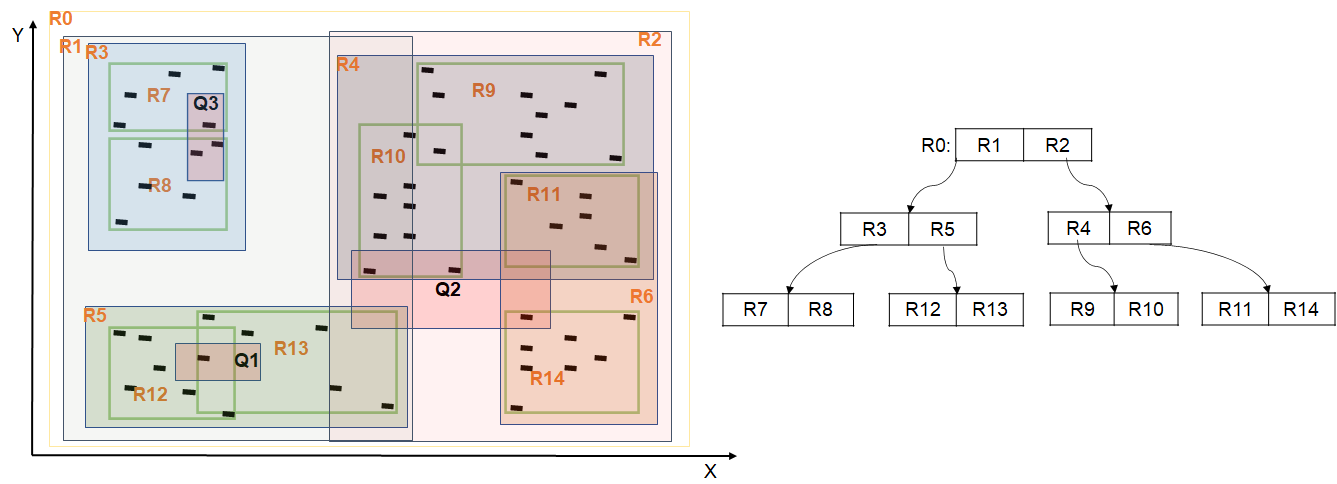}
  \caption{An example of an R-tree with overlapping nodes}
  \label{fig:node_overlapping}
\end{figure}

In this paper, we focus on processing range queries over an R-tree due to their wide applicability in spatial databases~\cite{manolopoulos2010r}. Figure~\ref{fig:node_overlapping} illustrates the impact of node overlap in an R-tree to answer a range query. Notice that the number of accessed leaf nodes directly impacts the query response time of an R-tree~\cite{manolopoulos2010r}. For a disk-based R-tree, descending multiple paths in the R-tree incurs high I/O cost~\cite{LISA}. In Figure~\ref{fig:node_overlapping}, the leaf nodes of the R-tree are labeled R7-R14. Consider Range Queries Q1, Q2, and Q3. For Q1, we search the R-tree down the 2 paths (from root to leaf): $R1\rightarrow R5 \rightarrow R12$ and $R1 \rightarrow R5 \rightarrow R13$ while only the latter path contains actual query results. Hence, accessing the leaf node R12 is wasted.
For Q2, we search the R-tree along 4 paths: $R1\rightarrow R5\rightarrow R13$, $R2\rightarrow R4\rightarrow R10$, $R2\rightarrow R6\rightarrow R11$, and $R2\rightarrow R6\rightarrow R14$ while only the 2nd and 4th paths contain output data objects. Hence, accessing the leaf nodes R11 and R13  negatively impacts  query processing performance.
For Query Q3, the R-tree searches two paths: $R1\rightarrow R3\rightarrow R7$ and $R1\rightarrow R3\rightarrow R8$, where both paths contain output data objects.
In both Q1 and Q2, the R-tree accesses $50\%$ more leaf nodes than the true number of leaf nodes containing the output data objects. In contrast, for Query Q3, the R-tree searches both R7 and R8, and data objects are exactly found in both nodes. In this case, the number of visited leaf nodes by the R-tree matches the true number of leaf nodes required to answer Q3. Thus, in terms of the number of leaf node accesses, we can identify Q1 and Q2 as high-overlap queries and Q3 as a low-overlap query. Observe that the R-tree searches extraneous leaf nodes to answer Q1 or Q2 but performs optimally for Q3.
We define an overlap ratio $\alpha$ to quantify the degree of extraneous leaf node accesses required by a query.
\begin{figure}[htbp]
  \centering
  \includegraphics[width=0.7\linewidth]{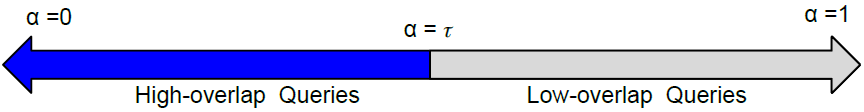}
  \caption{Spectrum of the overlap ratio $\alpha$ with Threshold $\tau$ to identify high- and low-overlap queries}
  \label{figure_overlap_ratio}
\end{figure}
Specifically, for a range query, we divide the number of true leaf nodes by the total number of visited leaf nodes to estimate $\alpha$, e.g., in Figure~\ref{fig:node_overlapping}, to answer Q2, the number of visited leaf nodes is $4$ while the number of true leaf nodes is $2$ making $\alpha = 0.50$. Similarly, for Q1 and Q3, $\alpha$ is $0.50$ and $1$, respectively. Notice that the number of true leaf nodes cannot exceed the number of visited leaf nodes. Hence, $\alpha$ ranges from $[0 - 1]$. 

For the purposes of this paper, the high- and low-overlap queries are determined as follows: 
Based on a pre-defined threshold $\tau$, queries with overlap ratio $\alpha \leq \tau$ (i.e., closer to $0$) are high-overlap while queries with $\alpha > \tau$ (i.e., closer to $1$) are low-overlap. 
The spectrum of the 
%of the 
overlap ratio $\alpha$ with Threshold $\tau$ is given in Figure~\ref{figure_overlap_ratio}.

\begin{figure}[htbp]
  \centering
  \includegraphics[width=\linewidth]{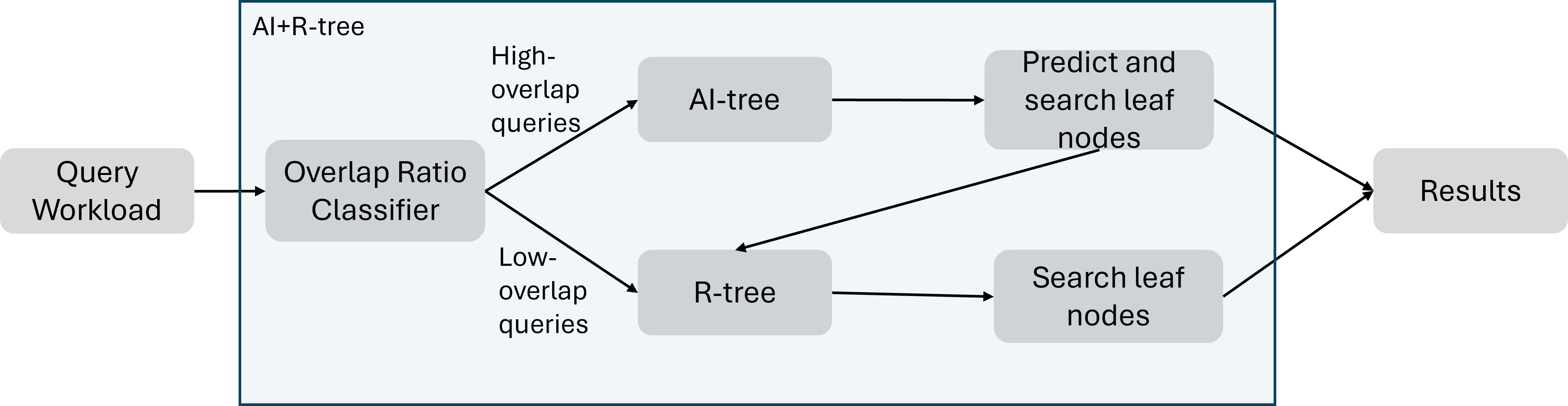}
  \caption{The AI+R-tree}
  \label{figure_the_AI+R_hybrid_approach}
\end{figure}

Motivated by the benefits of hybrid learned indexes and considering the issue of node overlap in the R-tree, the following important questions arise: 
\textit{Which workloads degrade the performance of range query processing in a traditional R-tree? Can we leverage ML techniques to make R-tree range query processing faster? }
Hence, we formulate the problem as follows: 
\textit{\bf Given a range query $Q(X_{min}, Y_{min}, X_{max}, Y_{max})$, we need to predict the true leaf nodes of the R-tree that contain output data objects, and only access these  nodes without accessing  extraneous ones.} 
In this paper, the proposed AI-tree transforms this problem into a multi-label classification~\cite{herrera2016multilabel} task by treating the leaf node IDs as the class labels. For example, classifying a research paper into a Systems, Theory, or ML paper is a multi-label classification task as a paper can be both a Systems and ML paper. Analogously, we can cast answering a range query over the R-tree as a multi-label classification task, where the classes are the R-tree leaf nodes, and we need to find these nodes that overlap the range query and that contain the output objects to the query.
At query time, the trained multi-label classifier predicts the true leaf node IDs that contain data entries that fall inside the query region. Hence, with correct prediction, the AI-tree only needs to access the true leaf nodes (i.e., without accessing any extraneous nodes) to process the query. As a result, \emph{the choice of ML models plays an important role in the query processing performance of the AI-tree~\cite{AI_R_tree_mdm_2022}.}
Moreover, for leveraging the benefit of both a traditional R-tree and an ML-enhanced AI-tree, the hybrid AI+R-tree~\cite{AI_R_tree_mdm_2022}
has been proposed (refer to Figure~\ref{figure_the_AI+R_hybrid_approach}). 
The AI+R-tree uses the overlap ratio $\alpha$ to identify the high-overlap queries for which an R-tree accesses many extraneous leaf nodes. Here, the Overlap Ratio Classifier (a binary classifier~\cite{aggarwal2015data_book}) routes the high-overlap queries to the AI-tree component and the low-overlap queries to the R-tree component. After that, the AI-tree and the R-tree process the high- and low-overlap queries, respectively. For an input query, if the AI-tree returns an empty result set (i.e., predicting all false positives/negatives), we call the regular R-tree to verify and resolve this case.
%\abdullah{
Moreover, the AI+R-tree indexes the learned models~\cite{handwritten_trie} as it uses a grid to decide which ML model to invoke from the multiple models that the AI-tree component has. For fixed query workloads, experiments on real spatial data demonstrate that the AI+R-tree significantly enhances the performance over a traditional R-tree.
%}

This paper extends over the previous conference paper on the AI+R-tree~\cite{AI_R_tree_mdm_2022}, and makes the following new contributions:

\begin{itemize}
    \item We extend the AI+R-tree to process dynamic query workloads, and present  empirical tradeoffs in query processing by varying the choice of ML models. 
    We adopt three performance metrics: Query Recall, Query Processing Time, and ML Model Size. Moreover, we include a case study of designing a custom loss function for a neural-network-based model tailored to the query processing requirements of the AI+R-tree.
    \item We present  design tradeoffs to extend the immutable AI+R-tree to support dynamic data insert/update/delete with the vision of realizing a mutable AI+R-tree.
    \item For dynamic query workloads, experiments on real datasets demonstrate that the AI+R-tree can enhance  query performance %of a traditional R-tree 
    for high-overlap range queries by up to 5.4X while maintaining up to 99\% recall.
\end{itemize}

The rest of this paper proceeds as follows: 
%\abdullah{change this after finishing the paper.}\jianguo{This papagraph can be deleted if you don't have enough space}
Section~\ref{section:AI-tree} provides an overview of the AI-tree. Section~\ref{section:AI+R-tree} presents the hybrid AI+R-tree. Section~\ref{section:Choice of ML Models} investigates the choice of ML models in the AI+R-tree. Section~\ref{section: design choices for updates} 
%includes the 
presents 
design tradeoffs for supporting dynamic data inserts and updates. Section~\ref{section:Evaluation} presents 
experimental results. Section~\ref{section:related_work} 
%gives an overview of 
discusses 
the related work. Section~\ref{section: future directions} provides several directions for future research. Finally, Section~\ref{section:conclusion} concludes the paper.

\section{The AI-tree }~\label{section:AI-tree}
%In this section, we provide an overview of 
%We overview the AI-tree. Moreover, we present the extensions made to the training steps of the AI-tree to process dynamic query workloads.
We overview the AI-tree with the extensions made to the training steps of the AI-tree to process dynamic query workloads.

\subsection{The Preprocessing Phase}

\subsubsection{Assigning Unique Identifiers to the R-tree Leaf Nodes}
In the preprocessing step, each R-tree node is assigned a unique  identifier (ID) based on Depth First Search (DFS) order. 

\subsubsection{Definition (The Overlap Ratio $\alpha$)}
Given a range query, say Q, to calculate the value of $\alpha$, we use two metrics: TN(Q), the {\underline t}rue {\underline n}umber of leaf-node accesses required to process Q, and VN(Q), the {\underline n}umber of leaf nodes {\underline v}isited by the R-tree index to answer Q. 
For the range query Q, the definition of $\alpha$ is as follows (the value of $\alpha$ is in the interval 
 $[0,1]$):
\[ \alpha = \frac{TN(Q)}{VN(Q)} \]

\subsubsection{Query Workload Categorization} \label{Query Workload Generation}
Given a query workload, we categorize each query based on its selectivity.
After identifying the selectivity of a query, the overlap ratio $\alpha$ of the query is calculated to further categorize the queries based on their value of $\alpha$. This is achieved by  executing the query during the preprocessing phase, computing the query's selectivity, the leaf nodes being touched, and the true leaf nodes.

\subsubsection{Training Data and Features} \label{training_data_generation}
%\abdullah{
We prepare the training data following a two-step process. In the first step, all queries in the query workload are executed one at a time on the constructed R-tree over the given dataset. For each executed query, we collect the following information: The IDs of the leaf nodes that the R-tree visits to answer the query, and the true leaf node IDs that contain the output data objects that are actually inside the query region.
%}
Moreover, for an input range query Q, we use the values $(X_{min},Y_{min},X_{max},Y_{max})$ of the query rectangle as input features to the ML model without any additional transformation. Thus, the same input can be processed seamlessly by both the AI-tree and the R-tree. For multi-label classification, the output labels are encoded using one-hot encoding~\cite{bishop2023deep}, where we represent the class labels using binary values.

\subsection{Learning the R-tree Index: ML Model Training and Testing}
\begin{figure}[h!]
  \centering
  \includegraphics[width=0.8\linewidth]{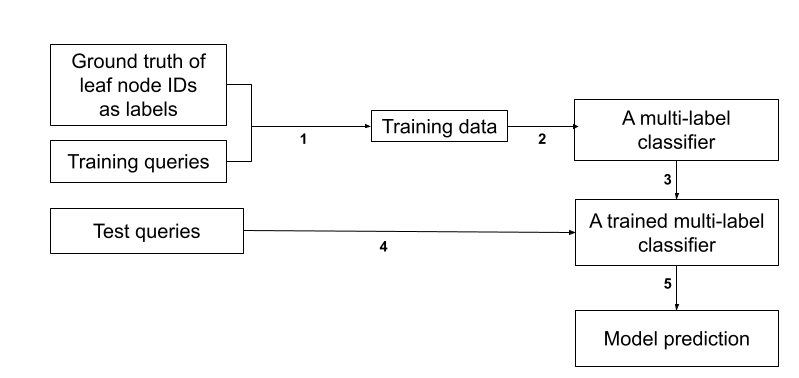}
  \caption{Workflow of ML model training and testing}
  \label{figure_workflow_model_training}
\end{figure}

Refer to Figure~\ref{figure_workflow_model_training}. The workflow for training and testing the multi-label classifier is as follows.
(1)~While the given queries are executed by the R-tree, for each query, the IDs of the visited and true leaf nodes are captured.
(2)~Then, the training queries are used to train the multi-label classifier. \emph{In this paper, the goal is to enhance the performance of the AI-tree for dynamic query workloads. As a result, the training queries are further divided into two sets: training and validation. ML models are trained on the training set and the hyperparameters are tuned based on the performance on the validation set.}
(3)~A trained multi-label classifier is created after the training phase.
(4)~The test queries are given as input to the trained multi-label classifier.
(5)~At query time, the pre-trained multi-label classifier is invoked to directly predict the true leaf node IDs that contain the query result.

\subsection{Indexing the Learned Models: A Multi-model Approach} \label{indexing the learned model}

The idea of indexing multiple learned models using a traditional index structure has been used in the context of handwritten and time series data~\cite{handwritten_trie}. In the AI-tree, we use a simple coarse grid structure to group the training queries. We train a separate ML model over queries inside each grid partition. The grid serves as an index to the localized learned ML models. 
%At query time, we only invoke the ML models whose grid cell overlap the query rectangle~\cite{AI_R_tree_mdm_2022}. 
%\abdullah{Updated with details without referring to the MDM paper}

\begin{figure}[htbp]
  \centering
  \includegraphics[width=0.6\linewidth]{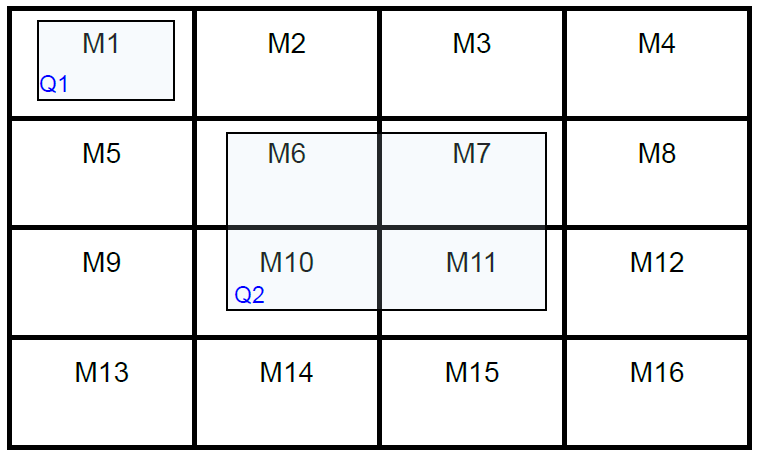}
  \caption{Indexing the learned models} 
  \label{fig:multi-model_approach}
\end{figure}
%\abdullah{
For example, in Figure~\ref{fig:multi-model_approach}, the space is partitioned using a $4X4$ grid. In the training phase, we incrementally search for the grid size that produces the best fit over the validation set~\cite{bergstra2011algorithms}. In Figure~\ref{fig:multi-model_approach}, as Query Q1 falls completely inside the top-left grid cell, only model M1 is trained for  Q1. On the other hand, as Query Q2 overlaps four grid cells, the four models M6, M7, M10, and M11 are trained for Q2. If multiple models are trained for a particular query, during query processing, we aggregate their prediction results from all the overlapping cells.
%}

\section{The AI+R-tree }~\label{section:AI+R-tree} 
%\abdullah{
To achieve the best of both the AI-tree and the R-tree, the hybrid AI+R-tree (see Figure~\ref{figure_the_AI+R_hybrid_approach}) processes the high-overlap queries using the AI-tree and the low-overlap queries using the traditional R-tree.
%}
However, this is non-trivial because the overlap ratio $\alpha$ of a query is unknown until we process the query. Hence, we  leverage ML techniques to learn how to distinguish between high- and low-overlap queries.
Specifically, the problem of classifying the range queries based on the value of $\alpha$ and the threshold $\tau$ can be formulated as a binary classification task~\cite{aggarwal2015data_book}. 
To prepare the training data for a particular dataset, we combine the queries for each of the $\alpha$ values. Then, we assign Label $0$ for the queries whose $\alpha$ value is less than or equal to the threshold $\tau$, and assign Label $1$ for the queries whose $\alpha$ value is greater than the threshold $\tau$. Next, a binary classifier is trained on the training queries. At query time, the pre-trained binary classifier (i.e., overlap ratio classifier) is invoked to classify an incoming range query into either a high- or a low-overlap query.

\subsection{Query Processing over the AI+R-tree}
Given a range query, say Q, the binary classifier is invoked first (see Figure~\ref{figure_the_AI+R_hybrid_approach}) to predict whether Q is high- or low-overlap. If Q is classified as a high-overlap query, the AI-tree processes the query. Otherwise, the R-tree processes the query. Notice that query processing using the AI+R-tree incurs a prediction cost before accessing the leaf nodes. Hence, the cost of query processing of the AI+R-tree is: ML model prediction cost + I/O cost. Thus, we expect to get the benefit of the AI-tree for processing the high-overlap queries whose $\alpha$ value is closer to zero. On the other side of the spectrum of $\alpha$ (Figure~\ref{figure_overlap_ratio}), for the queries with $\alpha$ closer to one, the R-tree is expected to perform better than the AI-tree.
For example, consider the three queries in Figure~\ref{fig:node_overlapping}. For Queries Q1 and Q2, the overlap ratio $\alpha = 0.50$. If the AI+R-tree can accurately predict the leaf nodes, $50\%$ less number of leaf nodes will be accessed to answer the query. Notice that 
%we have 
there is
room for improvement to process Q1 and Q2 using the AI-tree component of the AI+R-tree. In contrast, for Q3, $\alpha = 1$. Thus, both the visited leaf nodes contain data entries that fall inside the query rectangle. Thus, 
it is not possible for the AI-tree to process the query using less leaf node accesses than the R-tree. Thus, 
we use the R-tree component in this case.

\subsubsection{Query Processing using the AI-tree Component}
Given a range query, the ML models are identified whose grid cells overlap the range query. Then, only these models are executed to process the query. The resulting leaf node IDs from the ML model prediction are aggregated. Then, only the corresponding leaf nodes of these IDs are accessed. Finally, the data entries in these leaf nodes are checked against the input query rectangle, and only the  ones   actually contained are reported.  This ensures that the AI-tree never returns a false-positive query result. However, for a given query, the AI-tree can produce query results with false negatives, which 
%negatively 
impacts 
%the 
query recall. 

Notice that the AI-tree is expected to only access the predicted leaf nodes without accessing the extraneous leaf nodes. Thus, if the ML models inside the AI-tree predict the leaf nodes accurately, the minimum number of leaf nodes is accessed to answer a query. This reduces the number of disk I/Os for processing a range query. 
Notice further that for a given query, if the trained multi-label classifier 
%predicts all false-negative
mispredicts all
leaf node IDs, the AI-tree returns an empty result set. This can happen either due to the misprediction of the ML models or the range query indeed returns an empty set if processed by the R-tree. As a result, if the AI-tree returns an empty result set for a particular query, a regular R-tree search is invoked to verify and resolve the case (see Figure~\ref{figure_the_AI+R_hybrid_approach}).

\section{Choice of ML Models}~\label{section:Choice of ML Models}
\begin{figure*}[tbph] %[h!]
     \centering
     \captionsetup[subfloat]{labelfont=scriptsize,textfont=scriptsize}
     \subfloat[Workflow of the custom loss function\label{fig:custom_loss_workflow}]{
      \includegraphics[width=0.45\linewidth] {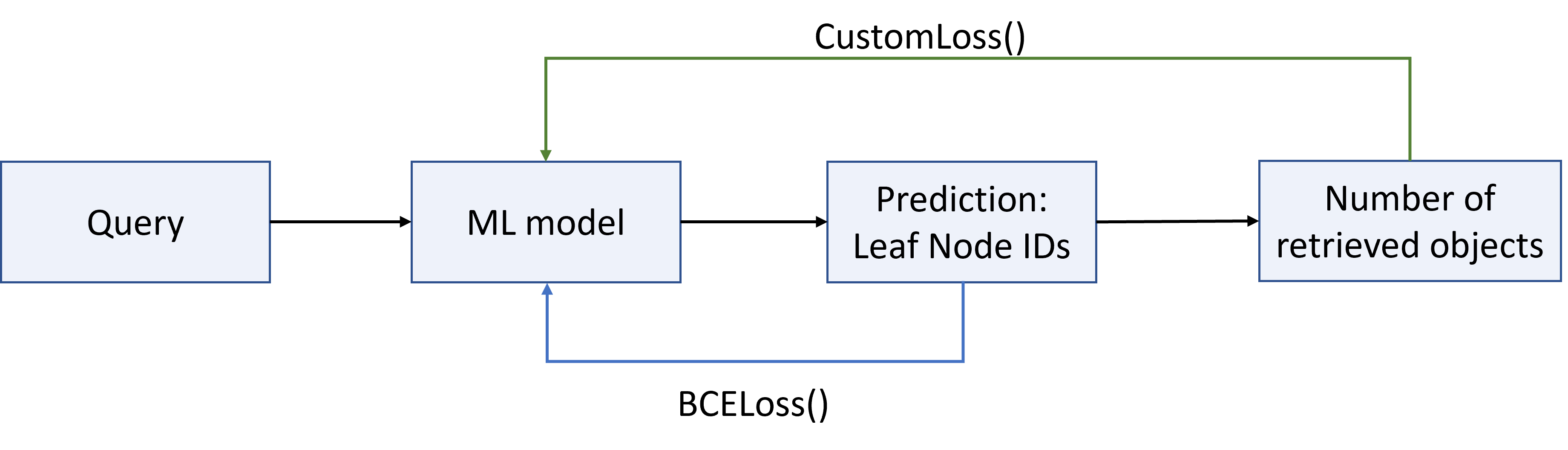}}
     %\hfill
     \hspace{1em}
     \subfloat[An example of the custom loss function\label{fig:custom_loss_example}]{
      \includegraphics[width=0.50\linewidth]{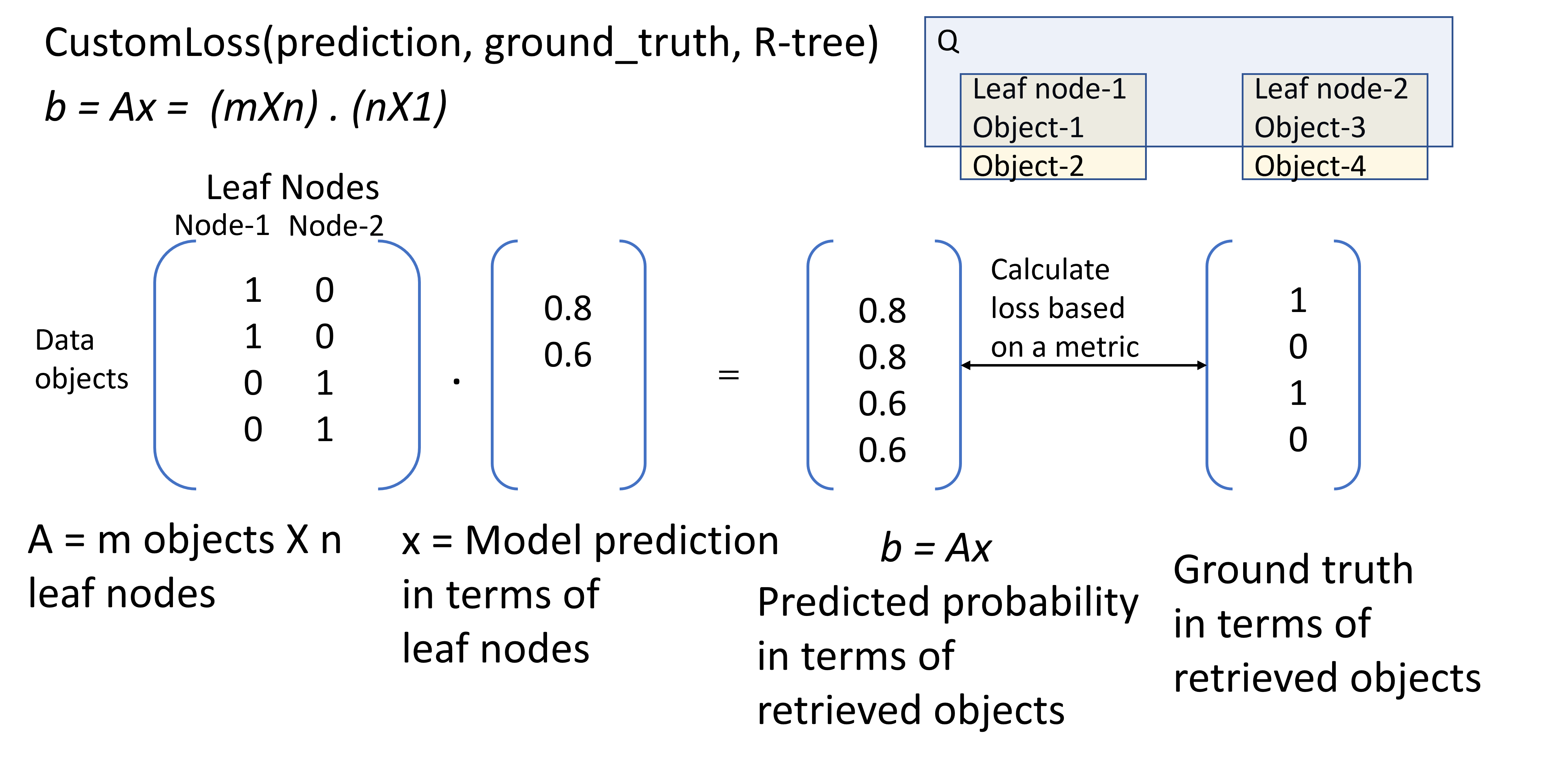}}
      
        \caption{Designing a custom loss function for the NN model} 
        \label{fig:custom_loss_design}
\end{figure*}

For the binary classifier (i.e., overlap ratio classifier), the goal is to train an ML model to classify an incoming query as high- or low-overlap. We use a Random Forest classifier (RF, for short)~\cite{breiman2001random} as the binary classifier. The prediction accuracy of the binary RF classifier is around $80\%$ over different values of $\alpha$. Although, with proper training any binary classifier can be used in the AI+R-tree, the chosen RF classifier generalizes well across different datasets. As a result, in this paper, we do not vary the choice of ML models for this binary classification task.
On the other hand, to investigate the impact of the chosen ML model on the average query recall, average query processing time, and average ML model size, we use three variants of Decision-Tree-based (DT, for short)~\cite{gibaja2015tutorial,quinlan1986induction} classifiers, and two variants of a neural-network-based classifier. However, depending on the dataset at hand, any multi-label classifier~\cite{szymanski2019scikit_multilearn} can be used to build the AI+R-tree.

\subsection{The Binary Classifier}
The steps to train the binary classifier are as follows: Assign Label $0$ for queries where $\alpha \leq \tau$ (e.g.,  $\alpha \leq 0.75$), and Label $1$ for queries where $\alpha > \tau$ (e.g., $\alpha$ $>$ $0.75$). The training and test sets are split using an 80-20 ratio. The scikit-learn python library~\cite{scikit-learn} is used for the RF classifier. 

\subsection{The Multi-label Classifier}

\subsubsection{DT-based Classifiers}
We  use three DT-based classifiers for  the multi-label classification task: the basic Decision Tree (DCT, for short), RF, and XGBoost (XG, for short)~\cite{chen2016xgboost}.

\subsubsection{NN-based Classifier}
In the context of learned indexes, it is generally recommended to adopt a simple ML model (e.g., linear regression, decision tree) whenever possible~\cite{al2024survey}. However, it has  been observed that a customized ML model tailored to the problem definition has the potential to achieve high performance compared to its basic counterpart. For example, in~\cite{eppert2021tailored}, a tailored regression model achieves high performance. Although neural-network-based models (NN-based models, for short) require 
%significantly 
longer training time, 
%(without GPU/other modern hardware support), 
there is  growing interest in designing 
%and using 
NN-based ML models~\cite{amato2022suitability, amato2023neural, ferragina2023nonlinear} for use in learned indexes. We investigate NN-based models as a multi-label classifier, and study their flexibility and the benefit of tailoring NN-based models for the AI+R-tree. 
%Particularly, we perform experiments 
We experiment with two NN-based model variants based on the underlying loss function. We leverage a standard Binary Cross Entropy Loss function (BCELoss, for short)~\cite{BCELoss_pytorch}. The BCELoss is used as a criterion to calculate the binary cross entropy between the input and the target probabilities. Then, we design a custom loss function for the AI+R-tree's classification task. 

%\paragraph{Designing a Custom Loss Function for NN-based Classifer} 
% \jianguo{Why a)? No need to have a paragraph title here.}
% \abdullah{Omitted the paragraph title}

%positioning the insert figure
\begin{figure*}[tbph] %[h!]
     \centering
     \captionsetup[subfloat]{labelfont=scriptsize,textfont=scriptsize}
     \subfloat[Case-1: No overlap and no split.\label{fig:case-1}]{
      \includegraphics[width=0.21\linewidth] {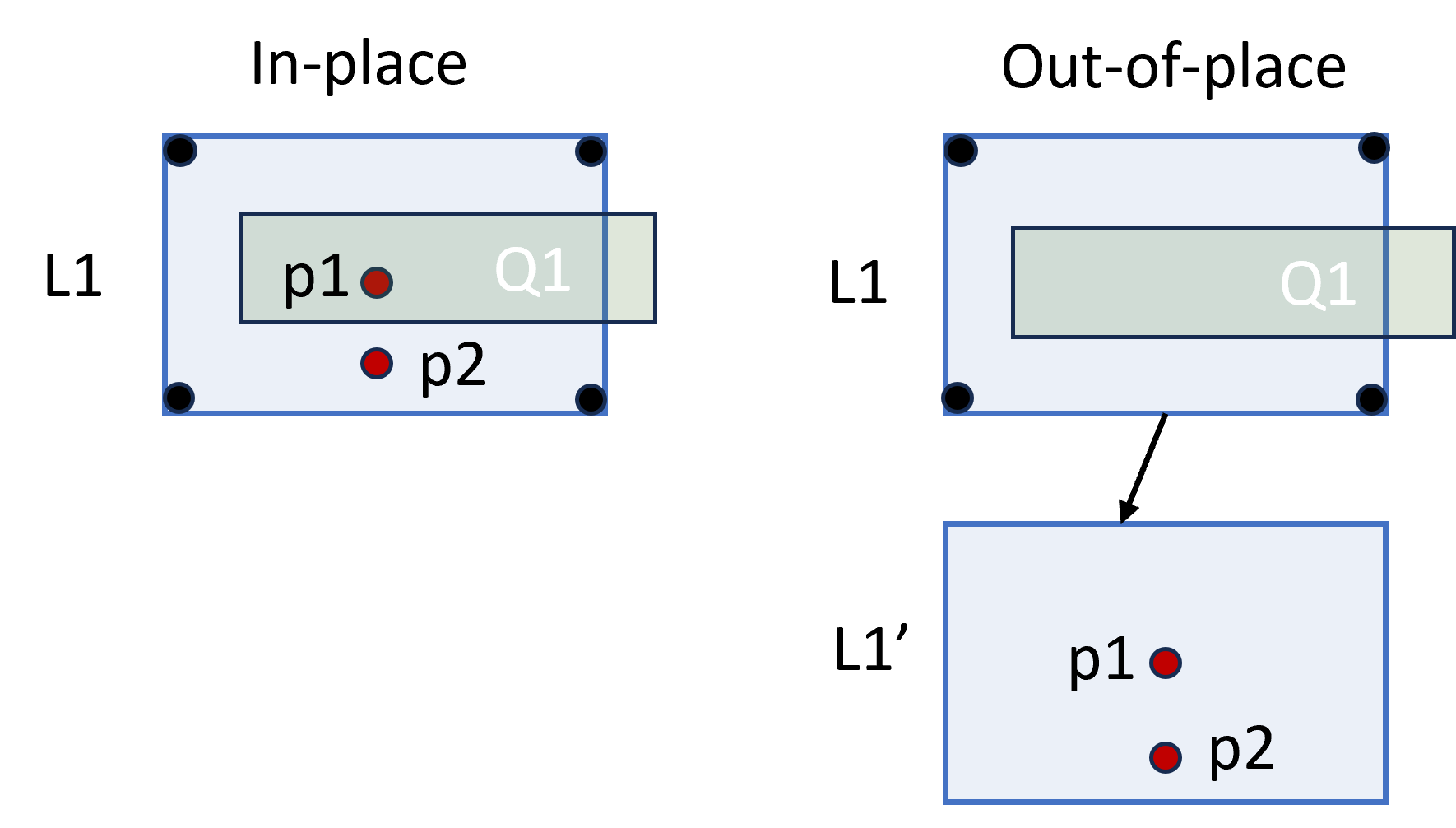}}
     %\hfill
     %\hspace{1em}
     \subfloat[Case-2: No overlap and split.\label{fig:case-2}]{
      \includegraphics[width=0.21\linewidth] {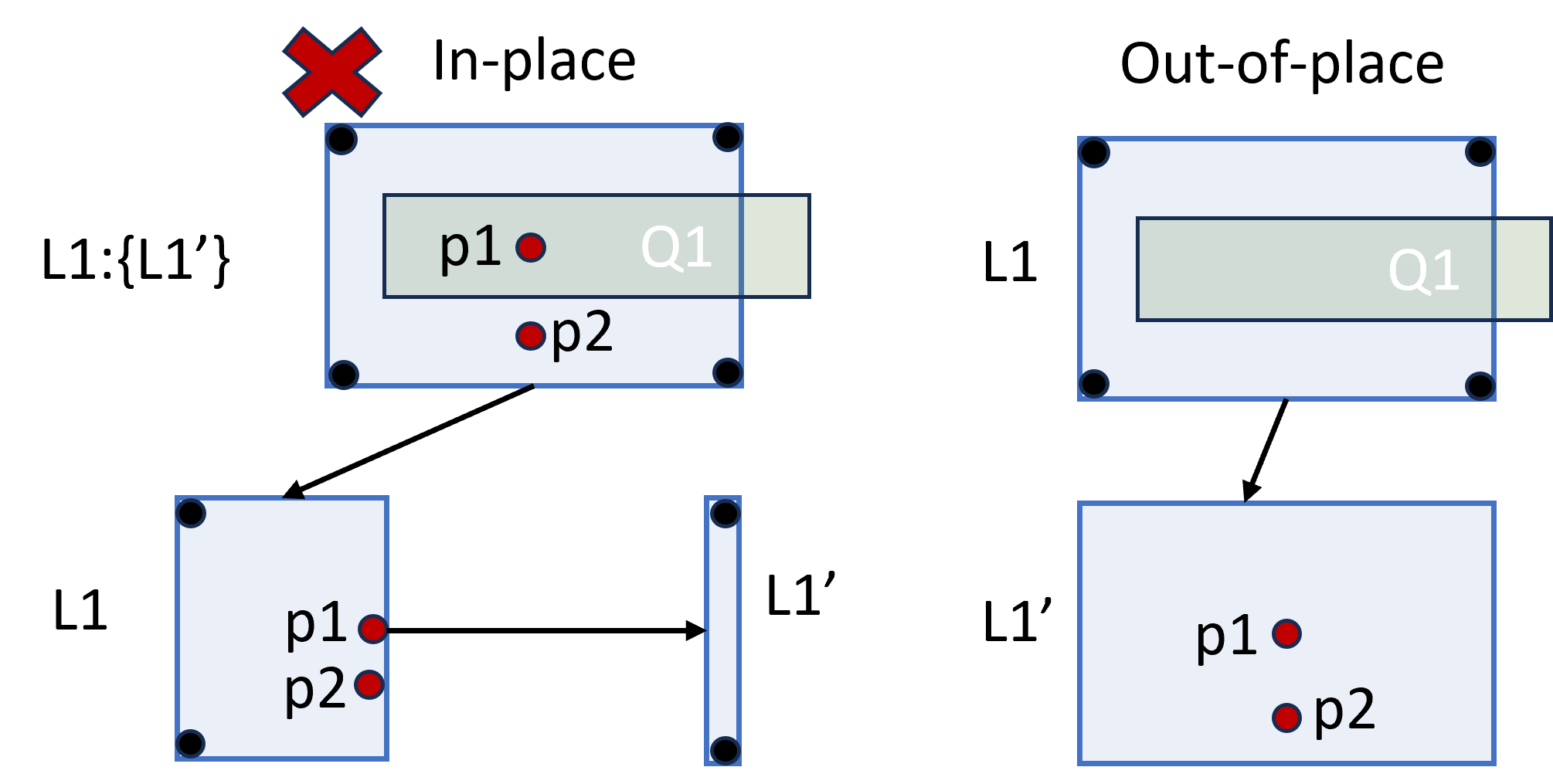}}
       %\hfill
       %\hspace{1em}
     \subfloat[Case-3: Overlap and no split.\label{fig:case-3}]{
      \includegraphics[width=0.29\linewidth] {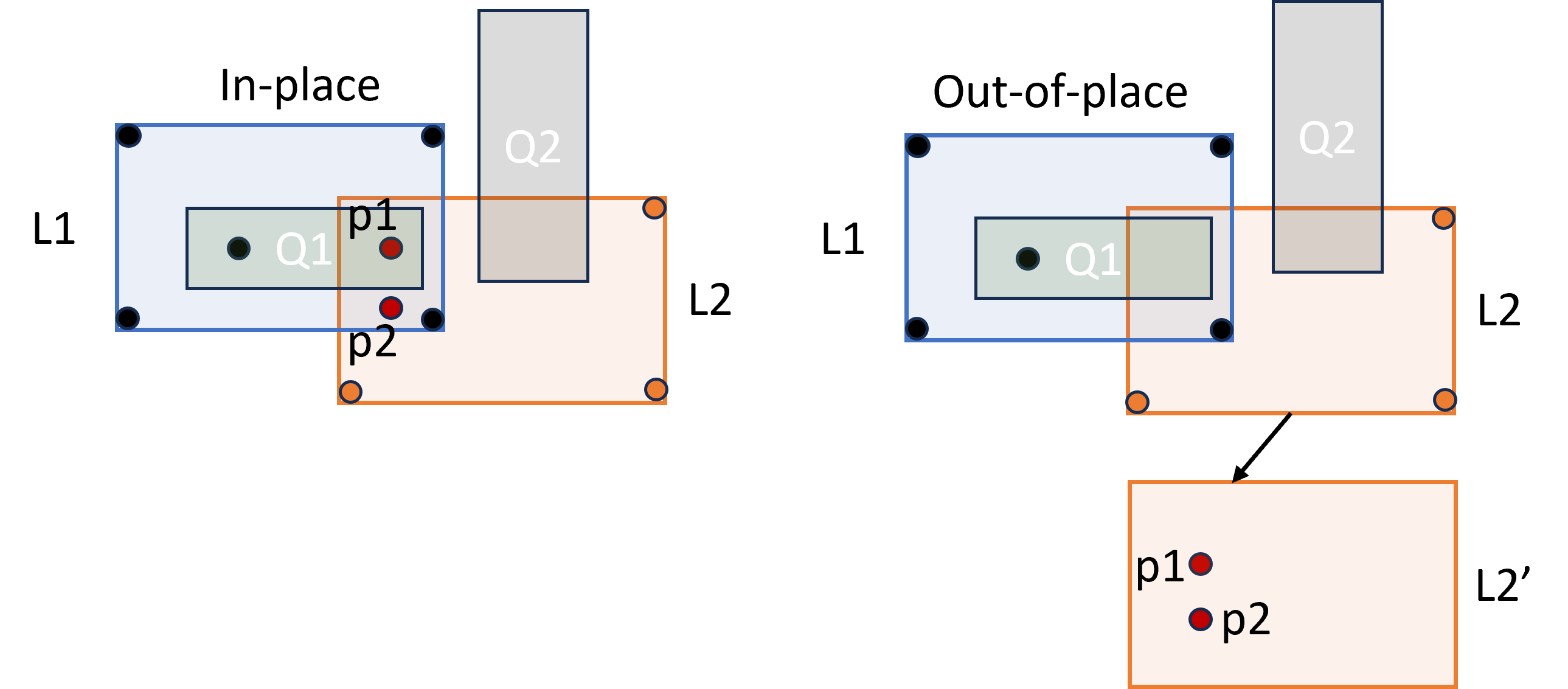}}
       %\hfill
       %\hspace{1em}
     \subfloat[Case-4: Overlap and split.\label{fig:case-4}]{
      \includegraphics[width=0.29\linewidth] {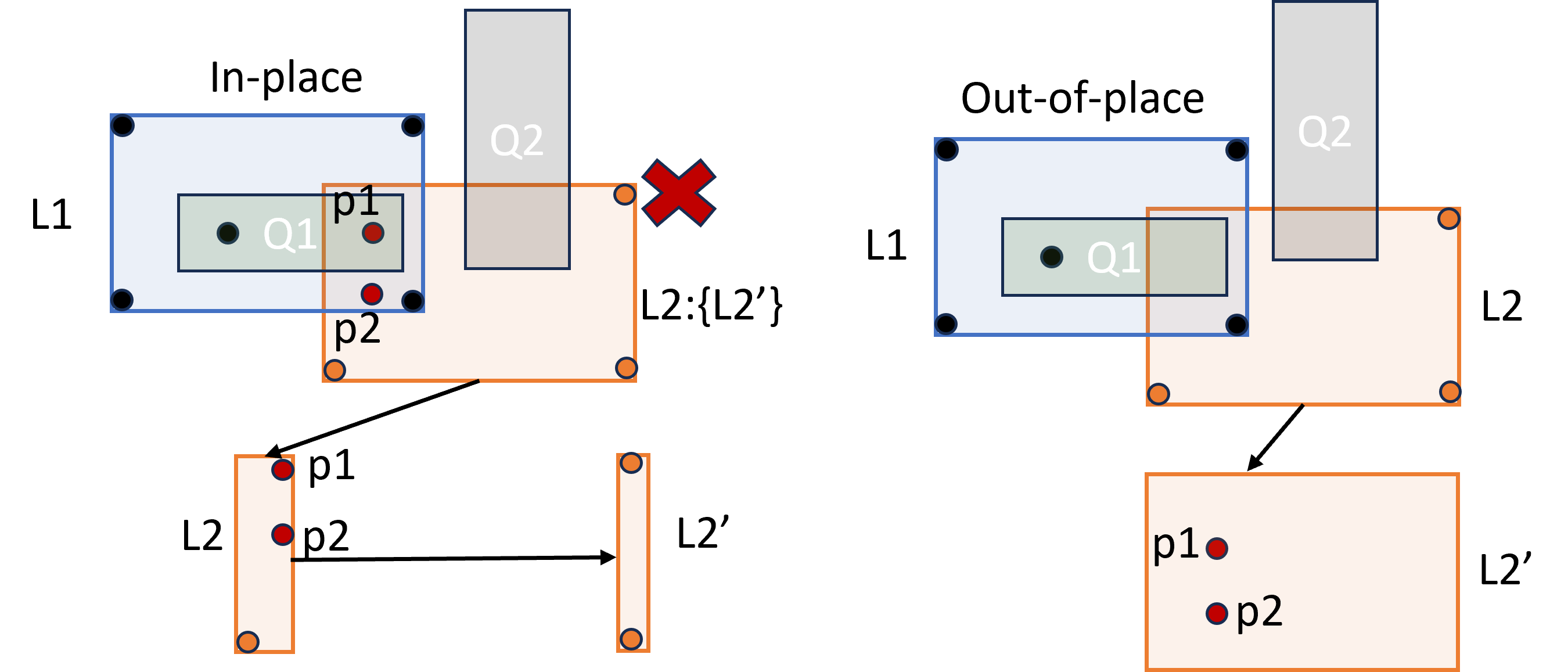}}
        %\abdullah{Updated the figure format to fit all the sub-figures in a single row.}
        \caption{In-place vs. Out-of-place insertion strategies} 
        \label{fig:Cases for new data insertions.}
\end{figure*}

Based on the problem formulation of the AI+R-tree, during the model training phase, given an input query, the multi-label classifier predicts a set of leaf node IDs. In the context of the multi-label NN-based classifier, the training and validation loss is calculated based on a pre-defined criterion (e.g., BCELoss). The standard BCELoss function takes two parameters as input: Model prediction and the ground truth. Notice that our 
%eventual 
goal is to maximize the average query recall in terms of number of retrieved data objects while reducing the number of false positive/negative predictions. However, the goal of maximizing the average query recall might not be fully captured by calculating the loss in terms of leaf node predictions only. For example, for an input range query, say $Q$, 
%let us 
assume that the ground truth labels (i.e., leaf node ids) are: ${L1, L2, L3}$, and the number of qualifying data objects within $L1, L2, L3$ (e.g., inside $Q$) are $5, 10, 15$, respectively. 
%Now, i
If an ML model predicts only $L1$ correctly, the query recall is $\frac{5}{30} = 0.16$. On the other hand, if the same ML model predicts only $L3$, the query recall becomes $\frac{15}{30} = 0.50$. However, this information is not captured if we calculate the NN model loss in terms of leaf node predictions only.  
As a result, we design a custom loss function by including the information about the number of retrieved objects. The workflow of the custom loss function is given in Figure~\ref{fig:custom_loss_workflow}.

In Figure~\ref{fig:custom_loss_example}, we present an example to illustrate the process of the custom loss calculation. Consider an R-tree with two leaf nodes and four data objects. In Figure~\ref{fig:custom_loss_example}, the given range query $Q$ overlaps 
%with 
both 
%the 
leaf nodes. However, only Data Object-1 and Object-2 are contained inside the query rectangle. Notice that the custom loss function takes three parameters: model prediction, ground truth, and the R-tree. As we provide the R-tree as input to the custom loss function, we can build an $mXn$ matrix $A$, where $m$ and $n$ refer to the number of data objects and the number of leaf nodes, respectively. Although $A$ can be a large matrix, 
%notice that 
the number of $1's$ in a column of Matrix $A$ is bounded by the maximum capacity of a leaf node. In other words, the total number of 1's in $A$ is exactly equal to $m$. Hence, $A$ can be constructed and maintained efficiently as a sparse matrix. During training, we get the prediction of the model in terms of the predicted probabilities of the leaf node ids. We can treat the predictions as a vector, say $x$, and generate the predicted probabilities in terms of retrieved data objects by performing the operation $b=Ax$. Notice that we can also create a similar ground truth vector in terms of data object (instead of leaf node ids). Finally, we can use a standard metric (e.g., binary cross entropy) to calculate the loss between the newly formulated prediction and the ground truth. In this paper, we refer to the NN-based models by BCE, and Custom loss functions by nnBCE and nnCustom, respectively. 
\emph{Notice that the reason behind choosing the NN-based models is not to optimize for outperforming the other DT-based models but rather to show the potential benefit of designing a custom loss function tailored to the query processing requirements of the AI+R-tree.}

\section{Design Choices for Supporting Inserts, Updates, and Deletes}~\label{section: design choices for updates}
In the context of mutable (i.e., updateable) learned indexes, the two main strategies to support new data insertion are: In-place and Out-of-place (e.g., delta buffer)~\cite{al2024survey}. However, supporting inserts/updates comes at the cost of periodic re-organization (i.e., re-training) of the learned index structures or incorporating mechanisms that require additional space~\cite{al2024survey}. In this section, we present the design tradeoffs in adopting both In-place and Out-of-place strategies with the vision of realizing a mutable AI+R-tree.

\subsection{Supporting New Data Inserts}
In the context of a mutable AI+R-tree, to avoid frequent ML model re-training, it is desirable to keep the R-tree component intact by deferring all structural modifications (whenever possible) for preserving the leaf node ID assignment as long as possible. Moreover, the correctness of the R-tree query processing should always be preserved to maintain the correctness of the AI+R-tree query processing. As a result, during the model re-training phase, the AI+R-tree can process all the incoming queries correctly using only the R-tree component. Hence, we advocate for using the R-tree component for new data inserts instead of 
using
the AI-tree-based (i.e., ML-model-based) insert. We identify all the cases for supporting new data inserts using the R-tree component of the AI+R-tree. 
Moreover, for each of the cases, we present the design tradeoffs from the perspectives of both the AI-tree and the R-tree components, respectively. 

We can identify two scenarios in the context of new data inserts into the R-tree component. In Scenario-1, the new data object is to be inserted into a leaf node and the leaf node does not overlap 
%with 
any other leaf node. Moreover, in Scenario-1, the leaf node might (or might not) have space to accommodate the newly inserted object without (or with) invoking a node split operation. As a result, 
%the 
Scenario-1 leads to 
%form 
forming 
two cases. Case-1: No overlap and no split (Figure~\ref{fig:case-1}), and Case-2: No overlap and split (Figure~\ref{fig:case-2}).
In Scenario-2, the new data object is to be inserted into a leaf node and the leaf node overlaps 
%with 
other leaf nodes. Moreover, in Scenario-2, the leaf node might (or might not) have space to accommodate the newly inserted object without (or with) invoking a node split operation. As a result, 
%the 
Scenario-2 leads to form two cases. Case-3: Overlap and no split (Figure~\ref{fig:case-3}), and Case-4: Overlap and split (Figure~\ref{fig:case-4}).
% \jianguo{Each case is not well abstracted. The current writing is based on a concrete example. What if there's another example? It's not clear which case will a new example fall into. It should be abstracted in a higher-level first and then use examples to illustrate.}
% \abdullah{Discussed the cases with proper abstraction at the start of this subsection}

\subsubsection{Case-1}
In Figure~\ref{fig:case-1}, $L1$ and $Q1$ represent a leaf node and a range query, respectively. New data points $p1$ and $p2$ are being inserted 
%in 
into
$L1$. In 
%case-1, 
Case-1,
we assume that $p1$ and $p2$ can be inserted into $L1$ without overflowing the leaf node, and there is no other leaf nodes overlapping 
%with 
$L1$. As a result, we refer to Case-1 as no-overlap-no-split.
For the AI-tree component, if 
%the 
$Q1$ is trained to predict the leaf node $L1$, it continues to do so for similar queries irrespective of the adopted insert strategy. Moreover, with an in-place strategy, only $L1$ will be accessed. For an out-of-place strategy, both $L1$ and $L1'$ will be accessed. 
For the R-tree component, with an in-place strategy, only $L1$ will be accessed. For an out-of-place strategy, both $L1$ and $L1'$ will be accessed.
In Case-1, the in-place strategy potentially saves both space and time. However, the overlap ratio of $Q1$ is impacted due to the in-place insert of point $p1$. As a result, the changes 
%of 
in
overlap ratio of queries need to be considered in the re-training phase of the overlap ratio (i.e., the binary) classifier. 

%\subsubsection{Case-2: No overlap and split}
\subsubsection{Case-2}
In Figure~\ref{fig:case-2}, we assume that $L1$ overflows upon inserting $p1$ and $p2$, and no other leaf nodes overlaps $L1$. As a result, $L1$ will be split into $L1$ and $L1'$. Hence, we refer to Case-2 as no-overlap-and-split. Moreover, with an in-place strategy, after the split, we assign the initial leaf node id $L1$ to one of the newly created leaf nodes. Moreover, the other newly created leaf node is given the id $L1'$, and we maintain a link from $L1$ to $L1'$ as: $L1:\{L1’\}$. 
In contrast, for out-of-place inserts, we  defer the split to keep the leaf node ids intact as long as possible. Once 
%the 
$L1$ reaches its maximum node capacity, an empty copy of $L1$, called $L1'$, will be created and will be linked
%with 
to
$L1$. Moreover, the new data points will be inserted into $L1'$.
For the AI-tree component, if 
%the 
$Q1$ is trained to predict the leaf node $L1$, it is maintained for similar queries irrespective of the adopted insert strategy. For both the in-place and out-of-place policies, the AI-tree needs to access both $L1$ and $L1’$ for processing $Q1$. Similarly, for the R-tree component, both in-place and out-of-place strategies require accessing both $L1$ and $L1'$.
In Case-2, there is no clear winner between the in-place and out-of-place insert strategies.

%\subsubsection{Case-3: Overlap and no split}
\subsubsection{Case-3}
In Figure~\ref{fig:case-3}, $L1, L2$, and $Q1, Q2$ represent leaf nodes and range queries, respectively. New data points $p1$ and $p2$ are being inserted 
%in 
into
$L2$. In 
%case-3, 
Case-3,
we assume that there is 
%overlapping 
an overlap
between $L1$ and $L2$, and $p1, p2$ can be inserted into $L2$ without overflowing $L2$. As a result, we refer to Case-3 as overlap-and-no-split.
For the AI-tree component, if $Q1$ is trained to predict the leaf node $L1$, it continues for $Q1$ and similar queries irrespective of the adopted insert strategy. However, after the insert of $p1, p2$ with an in-place strategy, for processing $Q1$, the AI-tree does not predict $L2$ until model re-training is performed. Hence, the AI-tree will not return $p1$ until model re-training. On the other hand, if $Q2$ is trained to predict none of  $L1$ and $L2$, the trained model is not impacted due to the new data inserts.
In Case-3, adopting an out-of-place insert policy, the AI-tree processes $Q1$ and $Q2$ similar to the in-place strategy.
For the R-tree component, after the inserts of $p1, p2$, with an in-place strategy, the R-tree accesses both $L1, L2$ for processing $Q1$, and only $L2$ for processing $Q2$. However, following an out-of-place strategy, the R-tree component accesses additional $L2'$ for processing both $Q1$ and $Q2$.
In Case-3, the in-place strategy potentially saves both space and time for most of the queries. However, the overlap ratio of $Q1$ is impacted due to the in-place insert of Point $p1$ into $L2$. As a result, the changes 
%of 
in
overlap ratio of queries need to be considered in the re-training phase of the overlap ratio classifier.

%\subsubsection{Case-4: Overlap and split}
\subsubsection{Case-4}
In Figure~\ref{fig:case-4}, we assume an
overlap exists
between $L1$ and $L2$, and $L2$ overflows upon inserts of $p1$ and $p2$. Hence, we refer to Case-4 as overlap-and-split. Moreover, with an in-place strategy, after a split, we assign the initial leaf node id $L2$ to one of the newly created leaf nodes. Moreover, the other newly created leaf node is given the id $L2'$, and we maintain a link from $L2$ to $L2'$ as: $L2:\{L2’\}$. 
In contrast, for out-of-place inserts, we 
%will 
defer the split to keep the leaf node ids intact as long as possible. Once 
%the 
$L2$ reaches its maximum node capacity, an empty copy of $L2$ called $L2'$ 
%will be 
is
created, and
is
linked 
%with 
to
$L2$. Moreover, the new data points will be inserted into $L2'$.
Similar to 
%the 
Case-3, in Case-4, the AI-tree component processes both $Q1$ and $Q2$ similarly irrespective of the adopted insert strategy.
For the R-tree component, after inserting $p1, p2$, with an in-place strategy, the R-tree accesses $L2$, and skips $L2'$ (no overlap between $Q1$ and the newly created $L2'$) for processing $Q1$. However, for processing $Q2$, the R-tree does not access either $L2$ or $L2'$ due to the no overlap with $Q2$. In contrast, following an out-of-place strategy, the R-tree will have to access additionally $L2'$ for processing both $Q1$ and $Q2$.
In Case-4, the in-place strategy potentially saves both space and time for certain queries (e.g., $Q2$) processed by the R-tree component.

%\subsection{Supporting Deletion}
\subsection{Supporting Deletes and Updates}
%\abdullah{Merged deletion and updates in a single subsection as the latter was too short.}
In the AI+R-tree, a delete  is supported by adopting any of the in-place or out-of-place strategies. However, we can avoid the cost of additional space by adopting an in-place delete strategy. Thus, a delete  is treated as a logical operation by marking the data items %as 
deleted so that the marked items can be physically deleted at a later time. Hence, a logical delete can be performed in-place. Notice that the motivation behind performing a logical delete  is to defer structural modification of leaf nodes until model re-training happens.

Updates  are supported by performing a delete  followed by an insert.

%\subsection{Supporting Updates}
%Updates  are supported by performing a delete  followed by an insert.

\subsection{Discussion}~\label{Discussion in supporting updates}
%\abdullah{
Here, we advocate for using the R-tree component for new data inserts instead of using the AI-tree-based (i.e., ML-model-based) inserts. However, using the traditional R-tree component for inserts will bypass the AI-tree adaptation. Hence, the AI-tree component will not be aware of the newly inserted items until the ML models are re-trained. 
Notice that the performance of a mutable AI+R-tree adopting either the in-place or the out-of-place insert strategy is expected to be less impacted in both Cases-1 and-2. However, as illustrated in Cases-3 and-4, the performance of a mutable AI+R-tree can be negatively impacted until the ML models are re-trained. 
ML model re-training can be triggered periodically (e.g., based on a pre-defined performance deterioration threshold) to maintain high query processing performance. Otherwise, a separate data/workload distribution change detection mechanism can be employed to trigger the ML-model re-training phase~\cite{nathan2019learning}. Moreover, the AI+R-tree can fall back to the traditional R-tree component during the re-training process or in case of a significant distribution shift similar to the hybrid structure  in~\cite{davitkova2024learning}. 
As a future direction, 
we plan to investigate the query processing performance over a mutable AI+R-tree by adopting different insert strategies 
%in future 
(see Section~\ref{section: future directions}). 
%}

\section{Evaluation}\label{section:Evaluation}
We  run all  experiments on an Ubuntu $18.04$  with Intel Xeon Platinum $8168$ ($2.70$GHz) and $3$TB of total available memory.

\subsection{Datasets}
We use three datasets from the UCR Spatio-Temporal Active Repository, namely UCR-STAR \cite{GVE+19}. Specifically, we use three real-world datasets with two-dimensional location data (in the form of longitude and latitude). The Tweets location dataset contains the locations of real tweets, the Gowalla dataset contains the locations of users from a social networking website, and the Chicago crimes dataset contains the locations of Chicago crimes. Moreover, we  preprocess the datasets to eliminate duplicate and missing values. For the Tweets locations dataset, we create a processed dataset containing the first $2$ Million tweet locations. For the processed Gowalla and Chicago crimes datasets, there are $1.2$ Million and $872,127$ records, respectively.

\subsection{The AI+R-tree Parameter Settings}

\subsubsection{R-tree Parameters} 
We construct the R-tree using a one-at-a-time tuple insert method to replicate the scenario of a dynamic environment, Moreover, we use a linear node-splitting algorithm for  R-tree construction. On the other hand, due to the observed advantage of larger leaf node capacity in~\cite{AI_R_tree_mdm_2022}, we have set the maximum leaf capacity $M = 1000$ for the R-tree. We  fix the leaf node capacity for all  experiments to observe the AI+R-tree performance by varying the ML model type. 

\subsubsection{Query Selectivity and Values of $\alpha$}
For a particular dataset, to demonstrate the query performance for a particular value of $\alpha$, $1000$ synthetic range queries are used in the experiments with a fixed selectivity. 
For example, in the case of the Tweets locations dataset, a range query with Selectivity $0.00005$ returns approximately $100$ objects, a query with Selectivity $0.0001$ returns approximately $200$ objects, and a query with Selectivity $0.0002$ returns approximately $400$ objects. In the experiments, 
the selectivity varies among $[0.00005, 0.0001, 0.0002]$. Moreover, we categorize the queries into five different values of $\alpha [0.1, 0.25, 0.5, 0.75, 1.0]$. Thus, for each dataset, for all variations of selectivities and $\alpha$ values, we use $15000$ queries in total.
Notice that all evaluations are performed by splitting each query set into 60\%/20\%/20\% (train/validation/test), unless a specific split is specified.

\subsubsection{ The AI-tree Parameters}
The AI-tree has two parameters: The size of the grid (see Section~\ref{indexing the learned model}) and the choice of the threshold $\tau$ (see Figure~\ref{figure_overlap_ratio}). 
Similar to the process of hyperparameter tuning~\cite{bergstra2011algorithms} for ML models, we start from a grid size $2X2$, and increase the size (e.g., $4X4$) to get the best fit for the validation data. As we vary the choices of  ML models, the choices of the grid size also vary based on the type of the ML model. For example, the basic DT-based Model (referred to as DCT in the figures) performs well on the validation set 
%with 
for
a grid with finer granularity: $20X20$. On the other hand, the ensemble methods, e.g., RF and XG, perform best 
%with 
for
a grid with coarser granularity: $4X4$.
Moreover, considering the training overhead, for the NN models, we do not vary the grid size but rather train a single NN model to demonstrate the benefit of the custom loss function.
%On the other hand, similar to the setup in~\cite{AI_R_tree_mdm_2022}, 
%\abdullah{
On the other hand, for a query Q with $\alpha = 0.75$, the $\frac{TN(Q)}{VN(Q)}$ can be e.g., $\frac{15}{20}$. Thus, there is room for improvement unless $\alpha =1$. As a result, we set Threshold $\tau = 0.75$. In other words, for an incoming range query, if $\alpha \leq 0.75$, it is identified as a high-overlap query. If $\alpha > 0.75$, it is considered low-overlap.
%}

\subsection{Implementation and Measurements}
We realize the AI+R-tree using an open-source python library for the R-tree available on Github~\footnote{\url{https://github.com/sergkr/rtreelib}}.
We integrate the AI+R-tree inside the library and run the experiments using Python version 3.6.9. 
%\abdullah{
On the other hand, for a disk-based R-tree index realized inside a practical system, the performance of a query depends on the number of leaf node accesses. In the experiments, we assume that the required number of disk I/Os is equivalent to the number of leaf node accesses~\cite{mahmood2014indexing}. For a query, we measure the CPU time, and count the number of leaf node accesses. Then, we multiply the number of leaf node accesses by a standard disk I/O access time. Finally, we sum the CPU and disk I/O times to report the average query processing time (in milliseconds).
%}
Notice that for a query workload with a particular selectivity, to demonstrate the performance for each value of $\alpha$, we run each experiment individually for each value of $\alpha$. This enables us to report the average query processing time, and the average query recall for each value of $\alpha$.

\subsubsection{ML Model Parameters}
For the DT-based ML models, we use the standard scikit-learn python library~\cite{scikit-learn}. We use the default parameters for the DCT classifier. For the RF classifier, we have varied the $n\_estimators$ parameter on the validation set, and picked the best value for testing. For the XG classifer, we have used the xgboost~\footnote{\url{https://xgboost.readthedocs.io/en/stable/tutorials/multioutput.html}} library with default parameters for multi-label classification. 
For the NN-based classifiers, we implement a Multi Layer Perception (MLP) with three hidden layers and ReLU activation function using the standard pytorch~\cite{paszke2019pytorch} library. Moreover, we use the Adam~\cite{kingma2014adam} optimizer with learning rate $10^{-3}$, and train for $30$ epochs to show the performance comparison between the nnBCE and nnCustom models.

\subsection{Experimental Results}
In each of the figures for each of the datasets, we show the value of overlap ratio $\alpha$ in the X-axis and the average query processing time (in milliseconds) in the Y-axis. Moreover, we report the average query recall of the AI+R-tree for each type of ML model. Notice that the query recall for the traditional R-tree is always $1$. Hence, the R-tree recall is omitted from the figures used for reporting query recall. On the other hand, the precision of the AI+R-tree is always $1$ because the data objects of the predicted leaf nodes are checked against the query range.
\subsubsection{Effect of DT-based Models for the Tweet Location Dataset}

\paragraph{Average Query Recall}
\begin{figure*}[ht] %[h!]
     \centering
     \captionsetup[subfloat]{labelfont=scriptsize,textfont=scriptsize}
     \subfloat[Query Selectivity=0.00005\label{fig:tweet_0.00005_percent_found}]{
      \includegraphics[width=0.25\textwidth] {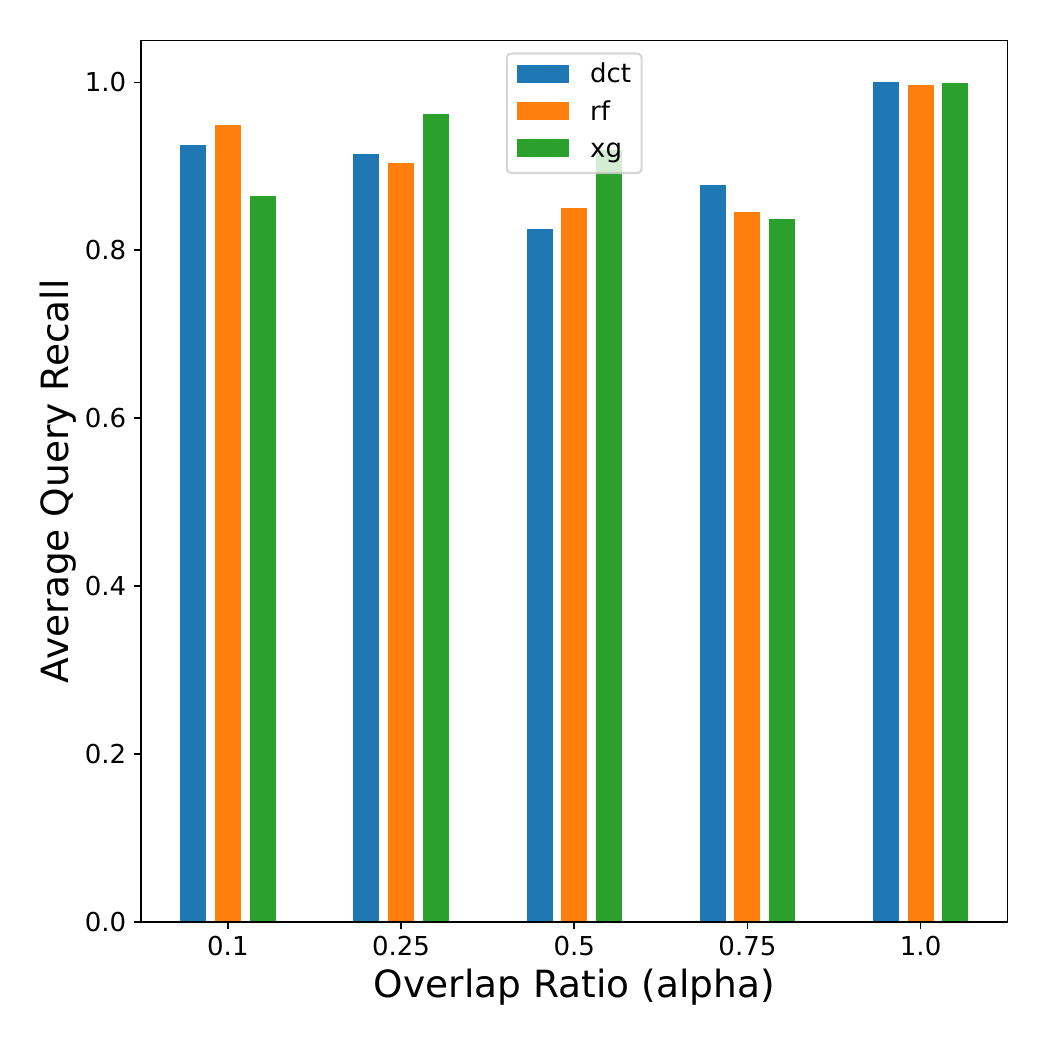}}
    \hspace{1em}
     \subfloat[Query Selectivity=0.0001\label{fig:tweet_0.0001_percent_found}]{
      \includegraphics[width=0.25\textwidth] {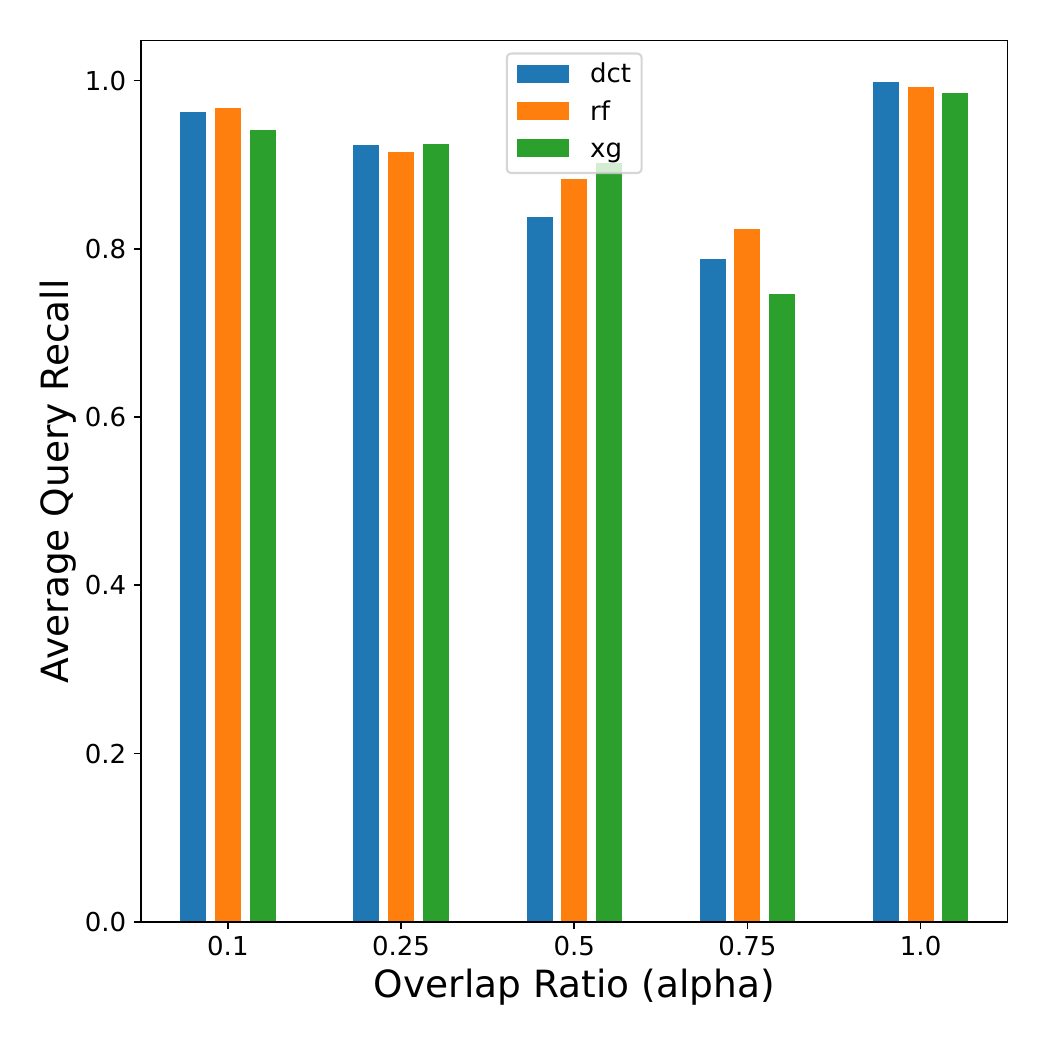}}
        \hspace{1em}
     \subfloat[Query Selectivity=0.0002\label{fig:tweet_0.0002_percent_found}]{
      \includegraphics[width=0.25\textwidth] {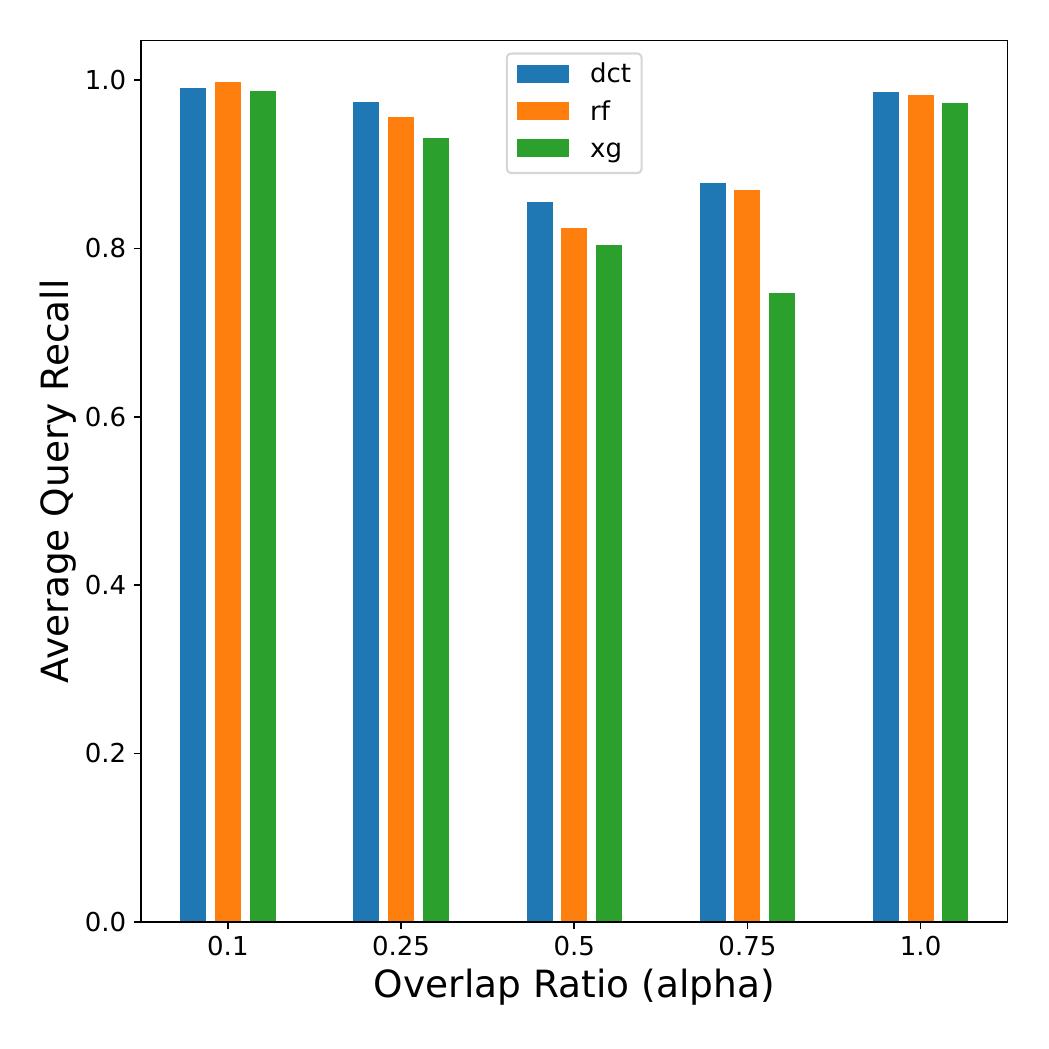}}    
        \caption{Average query recall of DT based classifiers for the Tweets location dataset. The R-tree recall is always 1, hence is omitted in the figures.} 
        \label{fig:tweet_Average query recall}
\end{figure*}

In Figure~\ref{fig:tweet_Average query recall}, we observe that the DT-based models maintain average query recall over 85\% for high-overlap queries with overlap ratio in $0.10$ and $0.25$. For  queries with selectivity $0.0002$, the recall reaches up to 99\%. However, for queries with overlap ratio closer to the threshold, the recall deteriorates for all variants of the DT-based model. Notice that the DT and RF classifiers perform better than the XG models for queries with overlap ratios between $0.50$ and $0.75$. For queries with $\alpha=1$, the AI+R-tree processes most of the queries using its R-tree component. As a result, the recall almost reaches $1$.  

\paragraph{Average Query Processing Time}

\begin{figure*}[ht] %[h!]
     \centering
     \captionsetup[subfloat]{labelfont=scriptsize,textfont=scriptsize}
     \subfloat[Query Selectivity=0.00005\label{fig:tweet_0.00005_time}]{
      \includegraphics[width=0.25\textwidth] {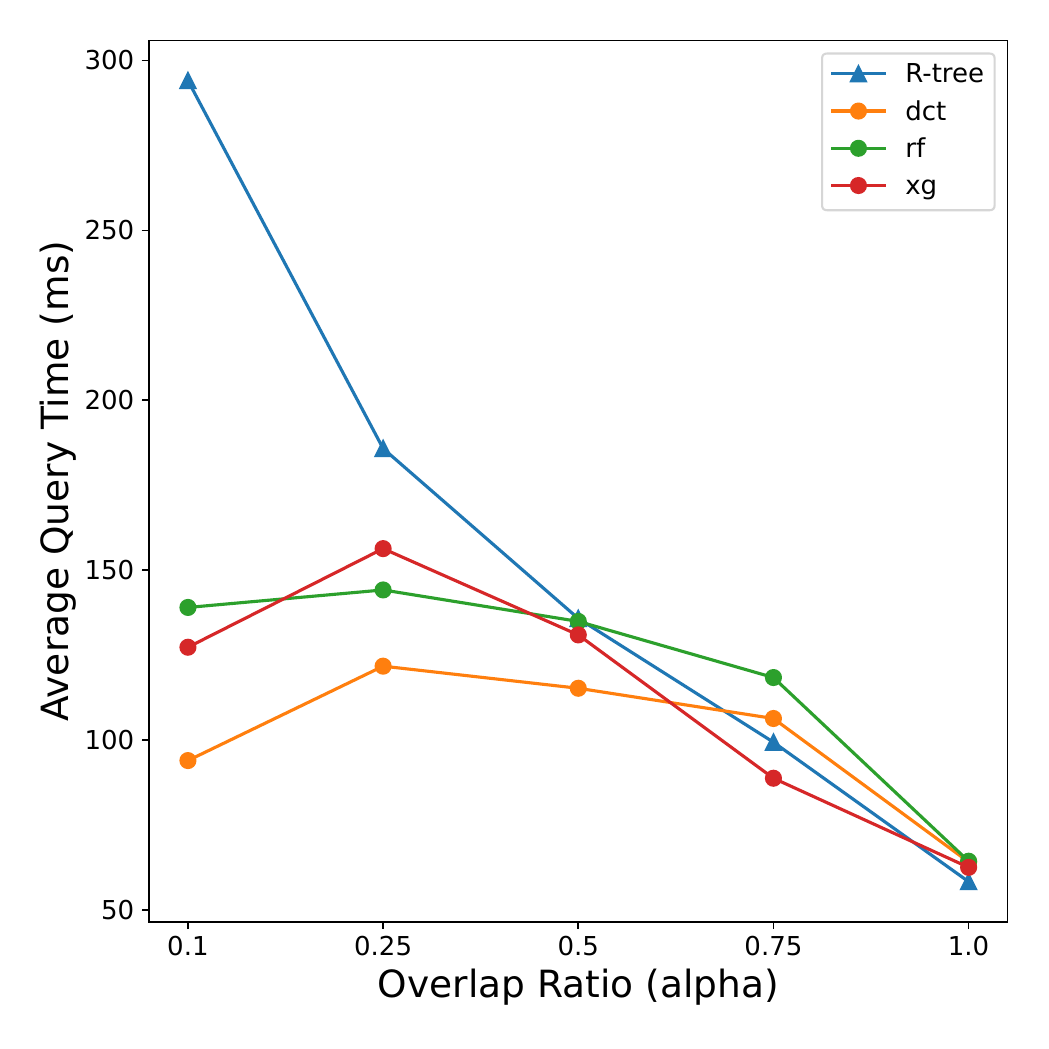}}
     %\hfill
     \hspace{1em}
     \subfloat[Query Selectivity=0.0001\label{fig:tweet_0.0001_time}]{
      \includegraphics[width=0.25\textwidth] {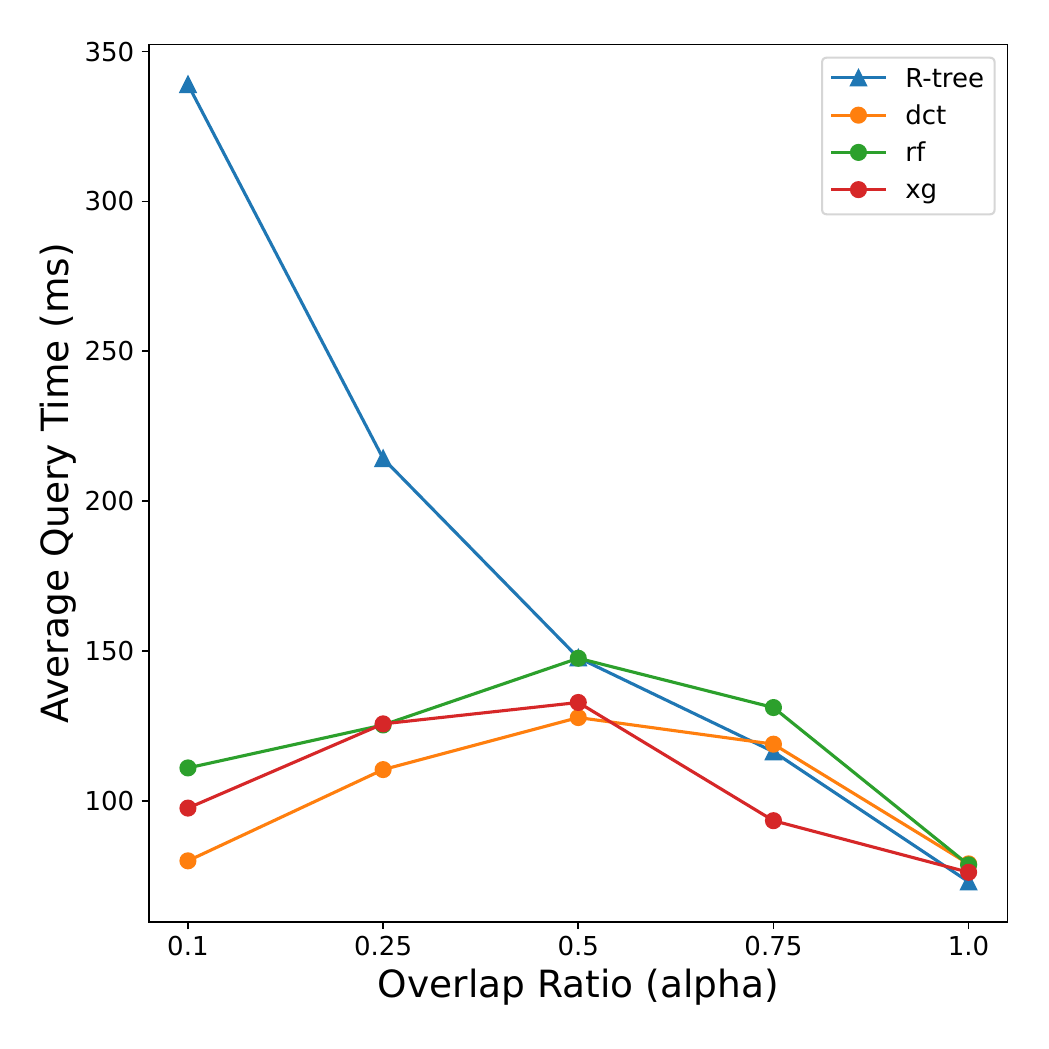}}
       %\hfill
       \hspace{1em}
     \subfloat[Query Selectivity=0.0002\label{fig:tweet_0.0002_time}]{
      \includegraphics[width=0.25\textwidth] {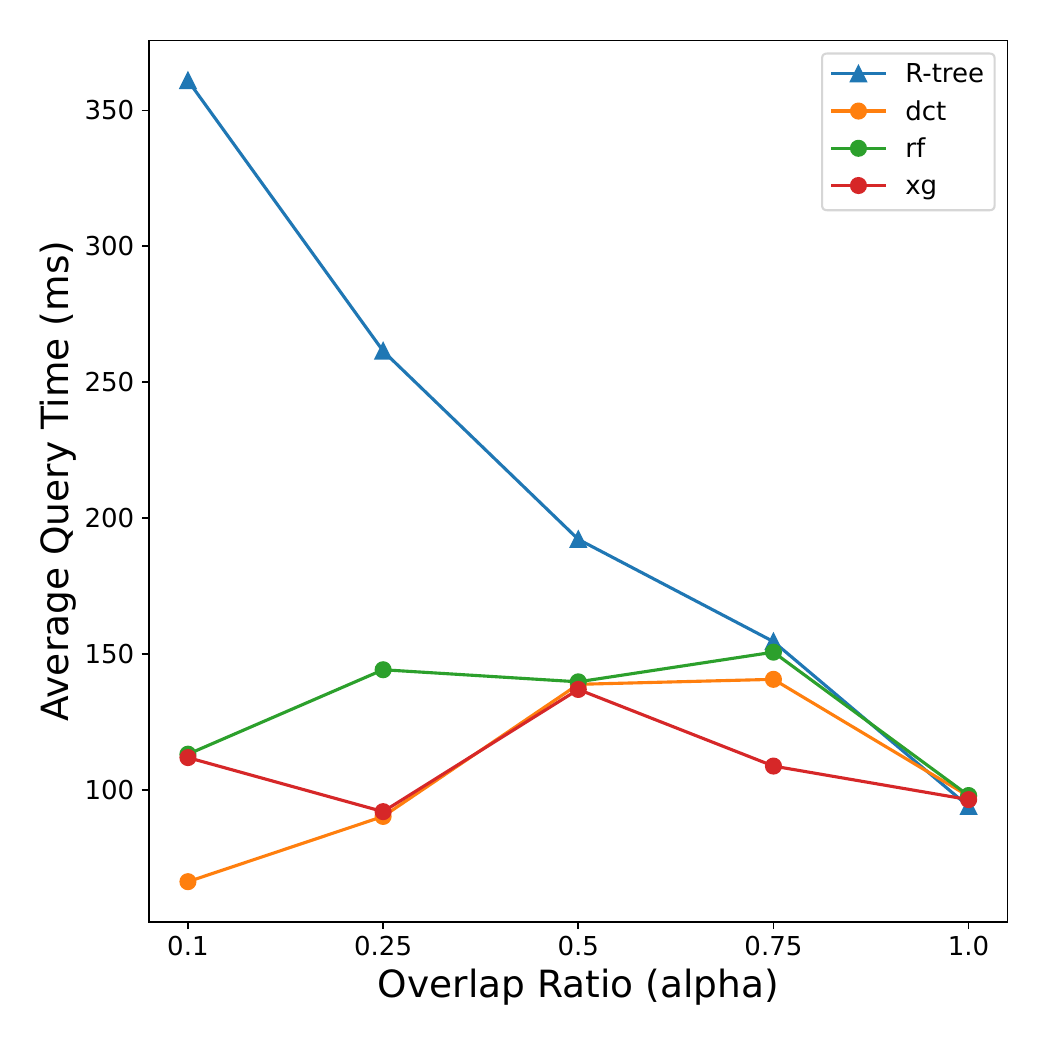}}    
        \caption{Average query time of DT based classifiers for the Tweets dataset} 
        \label{fig:tweet_Average query time}
\end{figure*}

In Figure~\ref{fig:tweet_Average query time}, we observe that the DT-based models maintains low query latency for processing high-overlap queries across all selectivities. Particularly, the DCT classifier enhances the query processing performance by up to 3.1X to 5.4X for queries with overlap ratio $0.10$. Although the query processing time deteriorates for queries with overlap ratio closer to the threshold, it closely follows the performance of a regular R-tree. Notice that the DCT classifier performs slightly better than the RF and XG models accors all selectivities. For queries with $\alpha=1$, the AI+R-tree processes most of the queries using its R-tree component. As a result, the query processing time almost matches the R-tree query processing time.

\subsubsection{Effect of Loss Functions of the NN Model for the Tweets Location Dataset}

\paragraph{Average Query Recall}
\begin{figure*}[ht] %[h!]
     \centering
     \captionsetup[subfloat]{labelfont=scriptsize,textfont=scriptsize}
     \subfloat[Query Selectivity=0.00005\label{fig:tweets_NN_0.00005_percent_found}]{
      \includegraphics[width=0.25\textwidth] {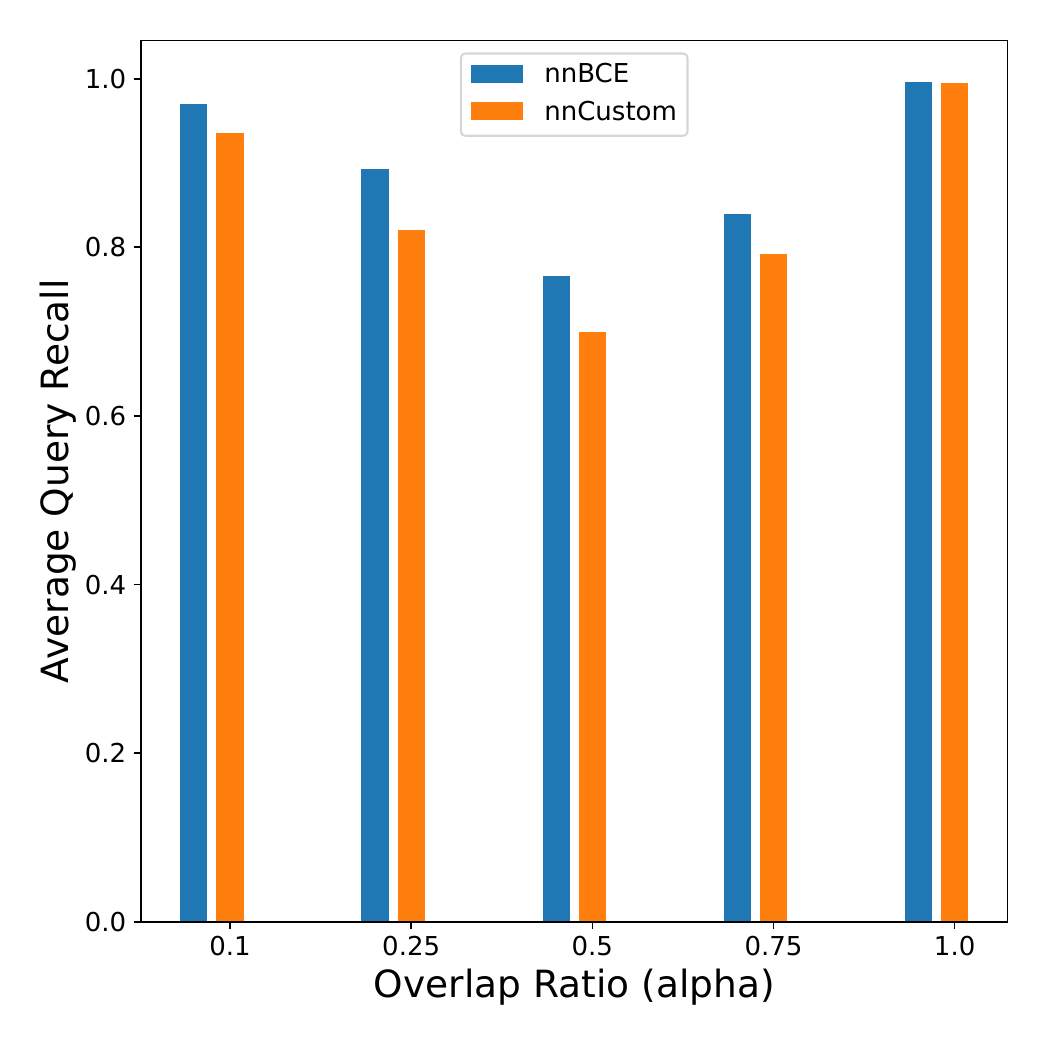}}
      %\hfill
      \hspace{1em}
      \subfloat[Query Selectivity=0.0001\label{fig:tweets_ucr_star_NN_0.0001_percent_found}]{
      \includegraphics[width=0.25\textwidth] {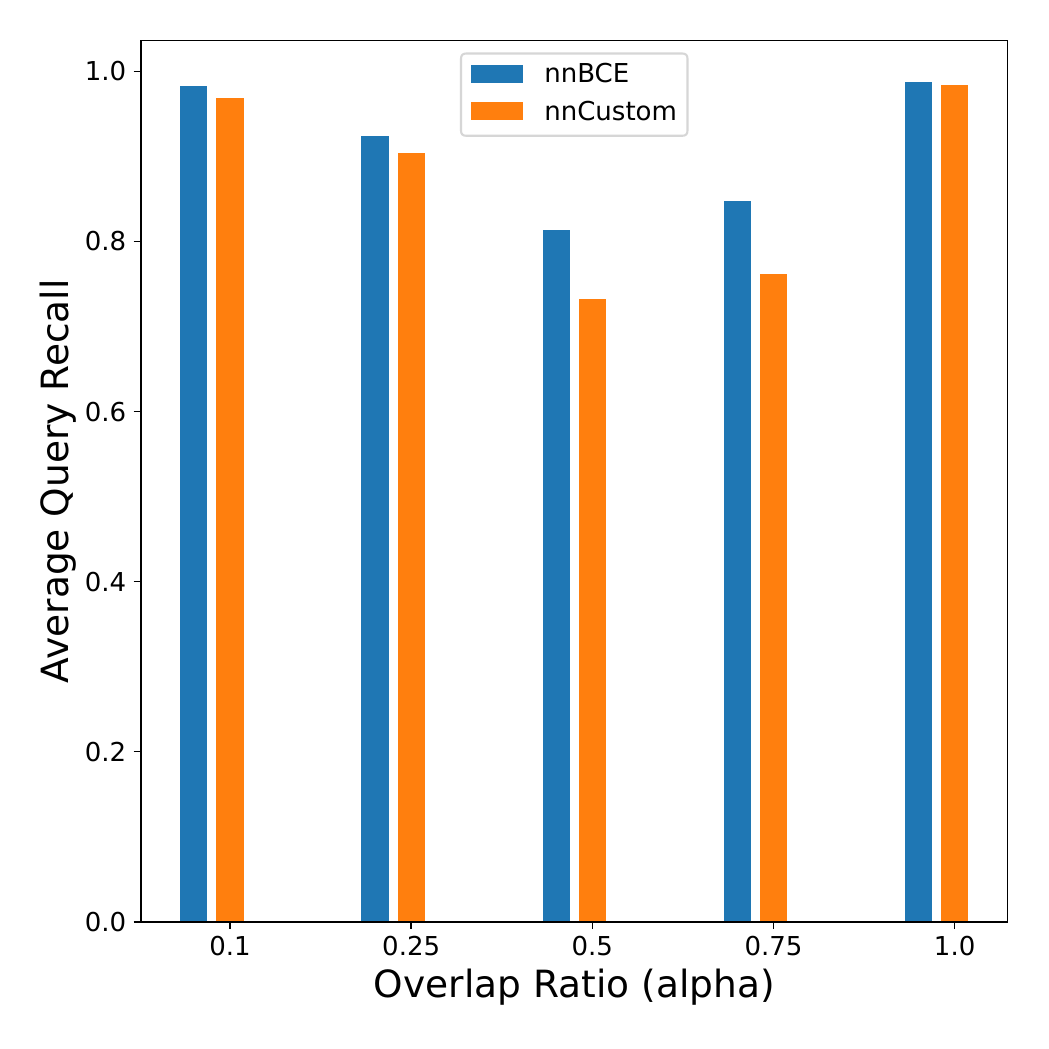}}
      %\hfill
      \hspace{1em}
      \subfloat[Query Selectivity=0.0002\label{fig:tweets_ucr_star_NN_0.0002_percent_found}]{
      \includegraphics[width=0.25\textwidth] {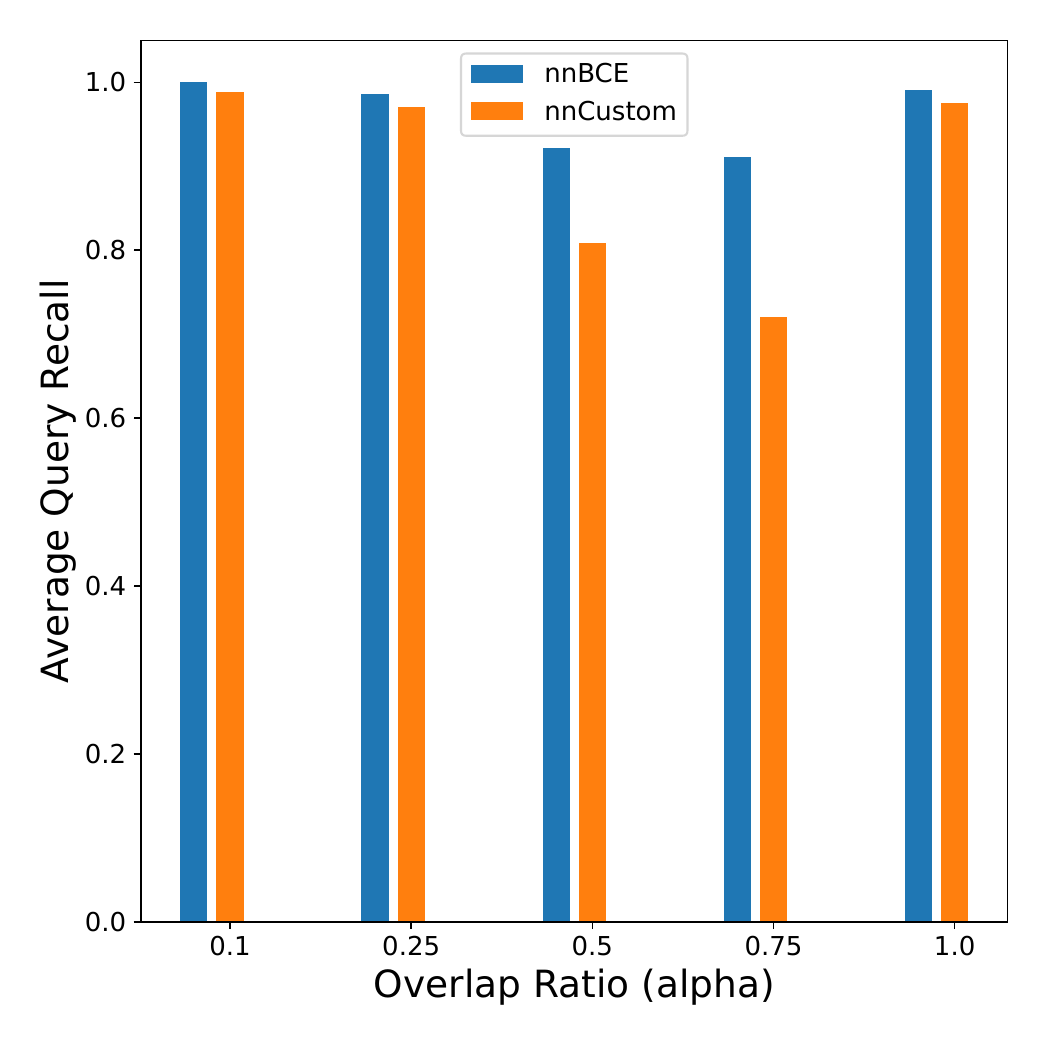}}
      \caption{Average query recall of NN based classifiers for the tweet locations dataset} 
        \label{fig:tweets_ucr_star_NN_Average query recall}
\end{figure*}

In Figure~\ref{fig:tweets_ucr_star_NN_Average query recall}, the average query recall is shown for each NN-based ML models for queries with various selectivities. The average query recall of both nnBCE and the nnCustom model are similar for the queries with very high alpha values (e.g., 0.10 and 0.25). However, the recall of the nnBCE model is higher in case of the queries with $\alpha$ values in $[0.50, 0.75]$.

\paragraph{Average Query Processing Time}
\begin{figure*}[ht] %[h!]
     \centering
     \captionsetup[subfloat]{labelfont=scriptsize,textfont=scriptsize}
     \subfloat[Query Selectivity=0.00005\label{fig:tweets_NN_0.00005_time}]{
      \includegraphics[width=0.25\textwidth] {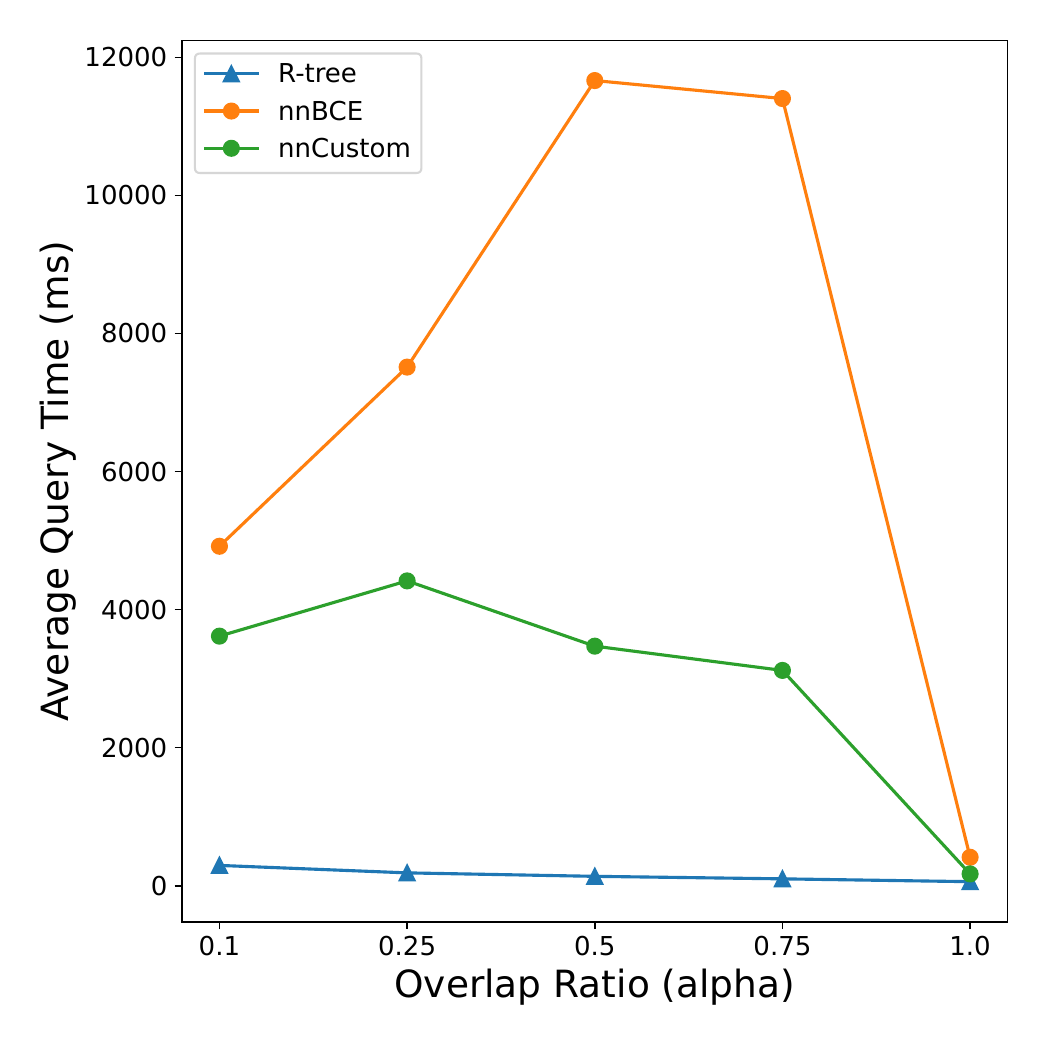}}
      %\hfill
      \hspace{1em}
      \subfloat[Query Selectivity=0.0001\label{fig:tweets_ucr_star_NN_0.0001_time}]{
      \includegraphics[width=0.25\textwidth] {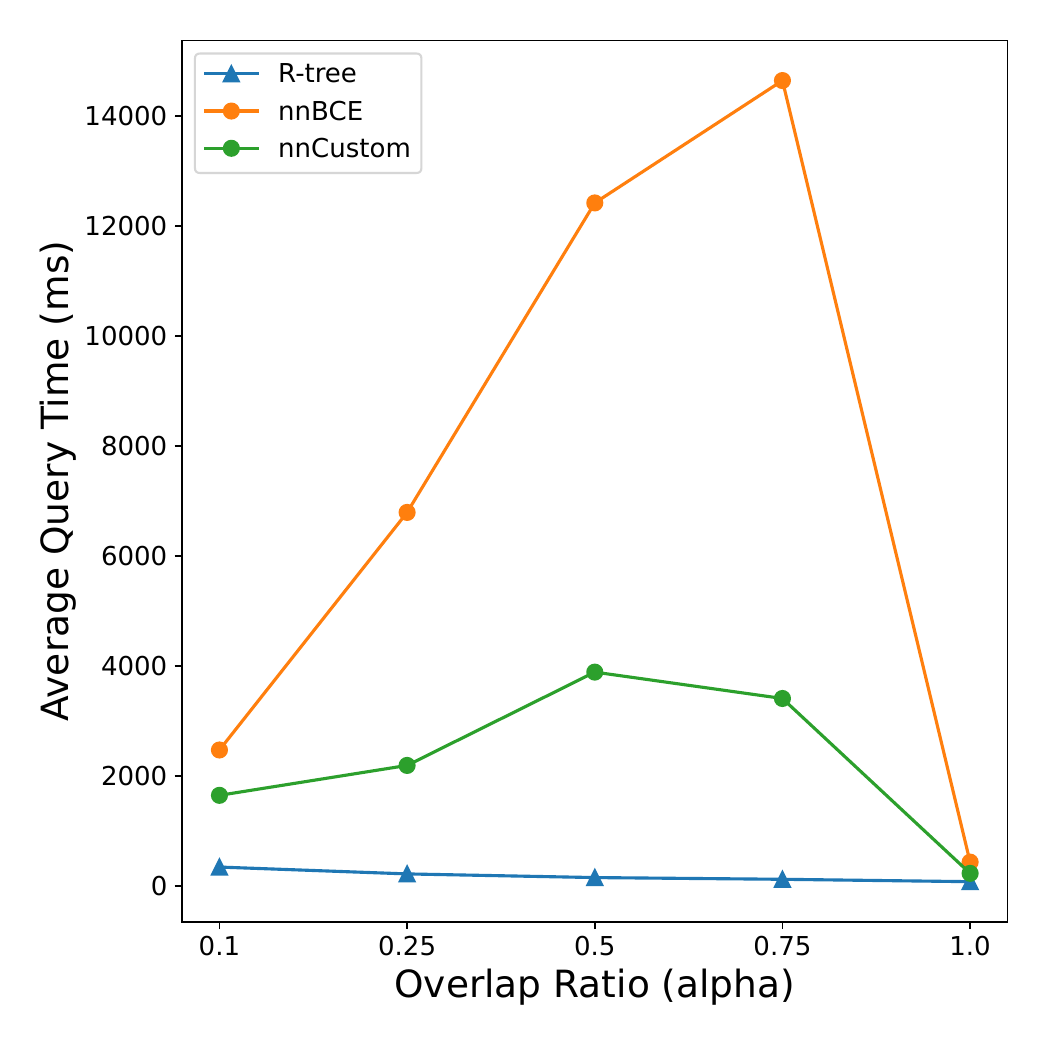}}
      %\hfill
      \hspace{1em}
      \subfloat[Query Selectivity=0.0002\label{fig:tweets_ucr_star_NN_0.0002_time}]{
      \includegraphics[width=0.25\textwidth] {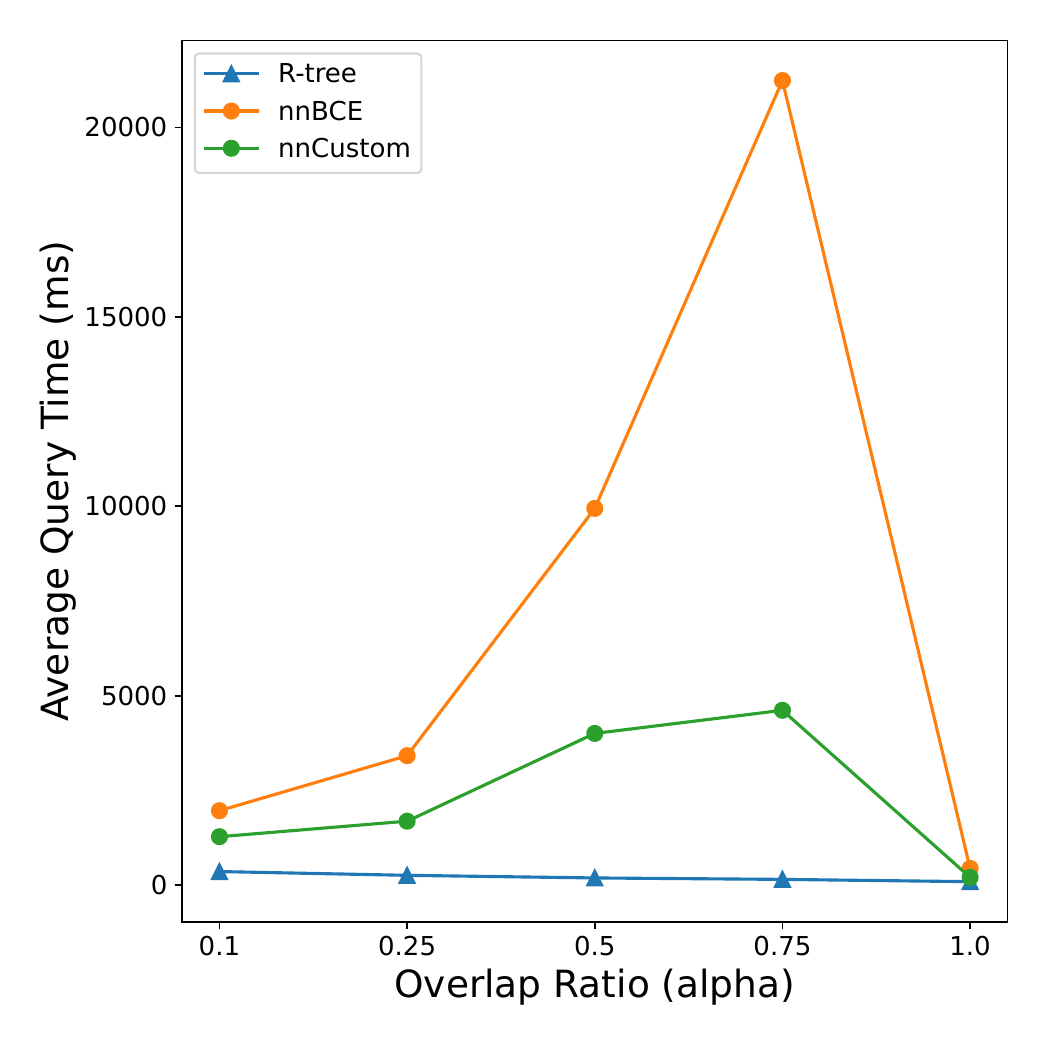}}
      \caption{Average query time of NN based classifiers for the tweet locations dataset} 
        \label{fig:tweets_ucr_star_NN_Average query time}
\end{figure*}

In Figure~\ref{fig:tweets_ucr_star_NN_Average query time}, the average query processing time is shown for each NN-based ML models for queries with various selectivities. Although the average query recall of nnBCE is higher than 
that 
of nnCustom, the nnCustom model achieves this recall with significantly less query latency. The reason behind this performance gain is as follows. The nnBCE model predicts many false positives compared to the nnCustom model. As a result, the nnBCE model requires to search many extraneous leaf nodes compared to the nnCustom model. Moreover, the nnCustom model takes into account the benefit of predicting a leaf node with its contribution to the final result of the range queries. Notice that the goal of the experiments related to the NN-based models is to demonstrate the potential benefit of 
%the 
designing a loss function tailored to the requirements of the AI+R-tree query processing.

\subsubsection{Space Consumption of the ML Models for the Tweets Location Dataset}

\begin{table}[h!]
    \caption{Average ML model size of the AI+R-tree for the Tweets location dataset across all $\alpha$ values (in MBs)}
    \label{tab:modelsize_tweets_location}
        \begin{tabular}{p{3em} p{3em} p{1.5em} p{1.5em} p{1.5em} c c}
    \toprule
        &  &\multicolumn{3}{c}{DT-based models}& \multicolumn{2}{c}{NN-based models}\\
        %&  & \multicolumn{3}{c}{NN based models}\\
        \cmidrule(lr){3-5}
        \cmidrule(lr){6-7}
        Selectivity & R-tree & DCT & RF & XG & nnBCE & nnCustom\\
    \midrule
        %dct
        %0.00005 & 1106.54 & 2.74 & 2.81 & 2.84 & 2.83 & 3.34\\
        %0.0001 & 1106.54 & 2.20 & 2.80 & 2.84 & 2.84 & 3.35\\
        %0.0002 & 1106.54 & 1.72 & 2.79 & 2.84 & 2.85 & 2.82\\

        %rf
        %0.00005 & 1106.54 & 3.30 & 3.41 & 3.57 & 3.51 & 3.41\\
        %0.0001 & 1106.54 & 3.26 & 3.37 & 3.60 & 3.57 & 3.44\\
        %0.0002 & 1106.54 & 3.23 & 3.30 & 3.59 & 3.64 & 3.47\\

        %xg
        %0.00005 & 1106.54 & 0.12 & 0.12 & 0.12 & 0.12 & 0.12\\
        %0.0001 & 1106.54 & 0.12 & 0.12 & 0.12 & 0.12 & 0.12\\
        %0.0002 & 1106.54 & 0.12 & 0.12 & 0.12 & 0.12 & 0.12\\

        %NN: the model size is expected to be independent of the underlying loss function
        %0.00005 & 1106.54 & 0.13 & 0.13 & 0.13 & 0.13 & 0.13\\
        %0.00005 & 1106.54 & 0.13 & 0.13 & 0.13 & 0.13 & 0.13\\
        %0.00005 & 1106.54 & 0.13 & 0.13 & 0.13 & 0.13 & 0.13\\
        
        %Average accros all alpha values
        0.00005 & 1106.54 & 2.91 & 3.44 & 0.12 & 0.13 & 0.13\\
        0.0001 & 1106.54 & 2.80 & 3.44 & 0.12 & 0.13 & 0.13\\
        0.0002 & 1106.54 & 2.61 & 3.44 & 0.12 & 0.13 & 0.13\\
    \bottomrule
\end{tabular}
\end{table}

In Table~\ref{tab:modelsize_tweets_location}, we present the R-tree size and the average ML model overhead across all $\alpha$ values. Notice that the model size includes both the binary and the multi-label classifiers. For the Tweets location dataset, the model overhead varies between $0.01\%$ to $0.26\%$. Notice that the space consumption of the NN-based model does not depend on the underlying loss function.

\subsubsection{Effect of DT-based Models for the Gowalla Dataset}

\paragraph{Average Query Recall}
\begin{figure*}[ht] %[h!]
     \centering
     \captionsetup[subfloat]{labelfont=scriptsize,textfont=scriptsize}
     \subfloat[Query Selectivity=0.00005\label{fig:gowalla_0.00005_percent_found}]{
      \includegraphics[width=0.25\textwidth] {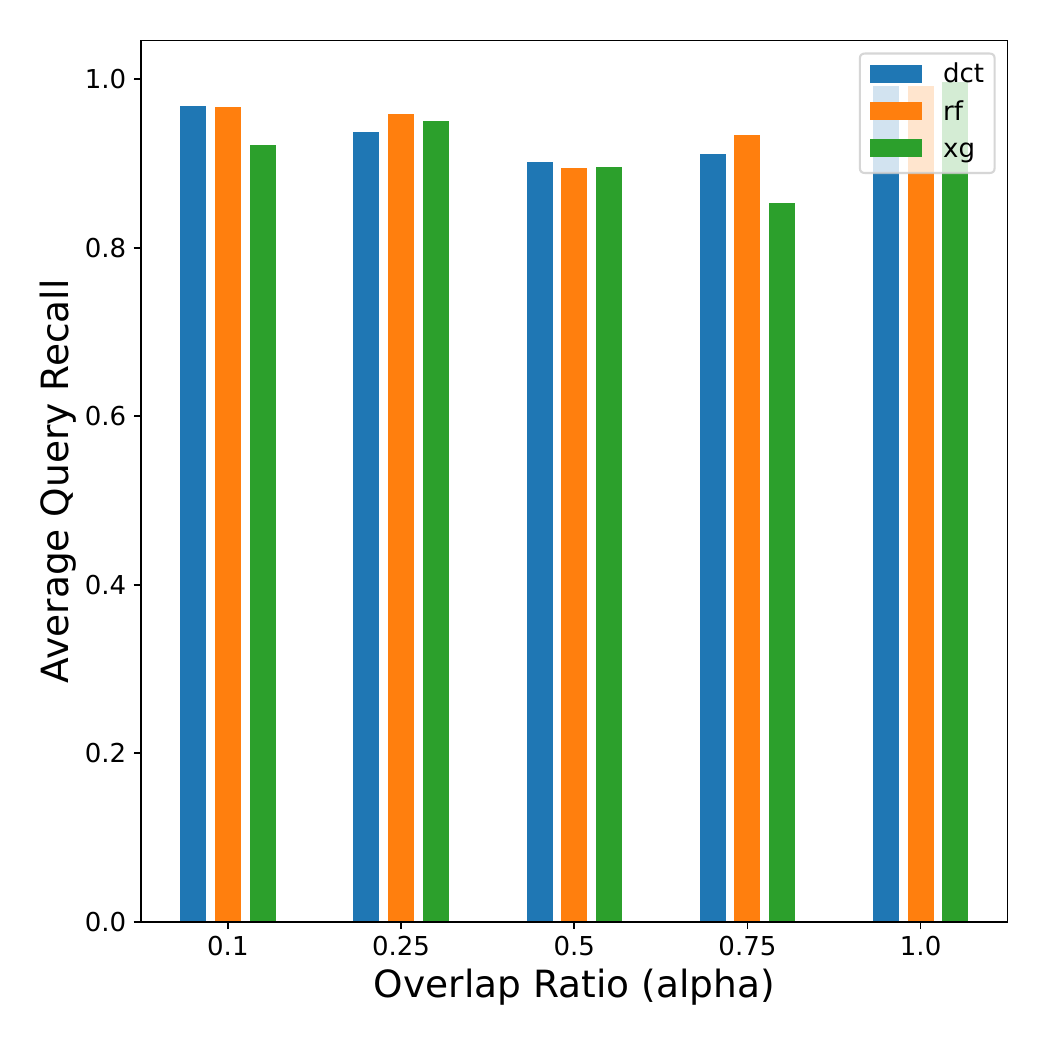}}
     %\hfill
     \hspace{1em}
     \subfloat[Query Selectivity=0.0001\label{fig:gowalla_0.0001_percent_found}]{
      \includegraphics[width=0.25\textwidth] {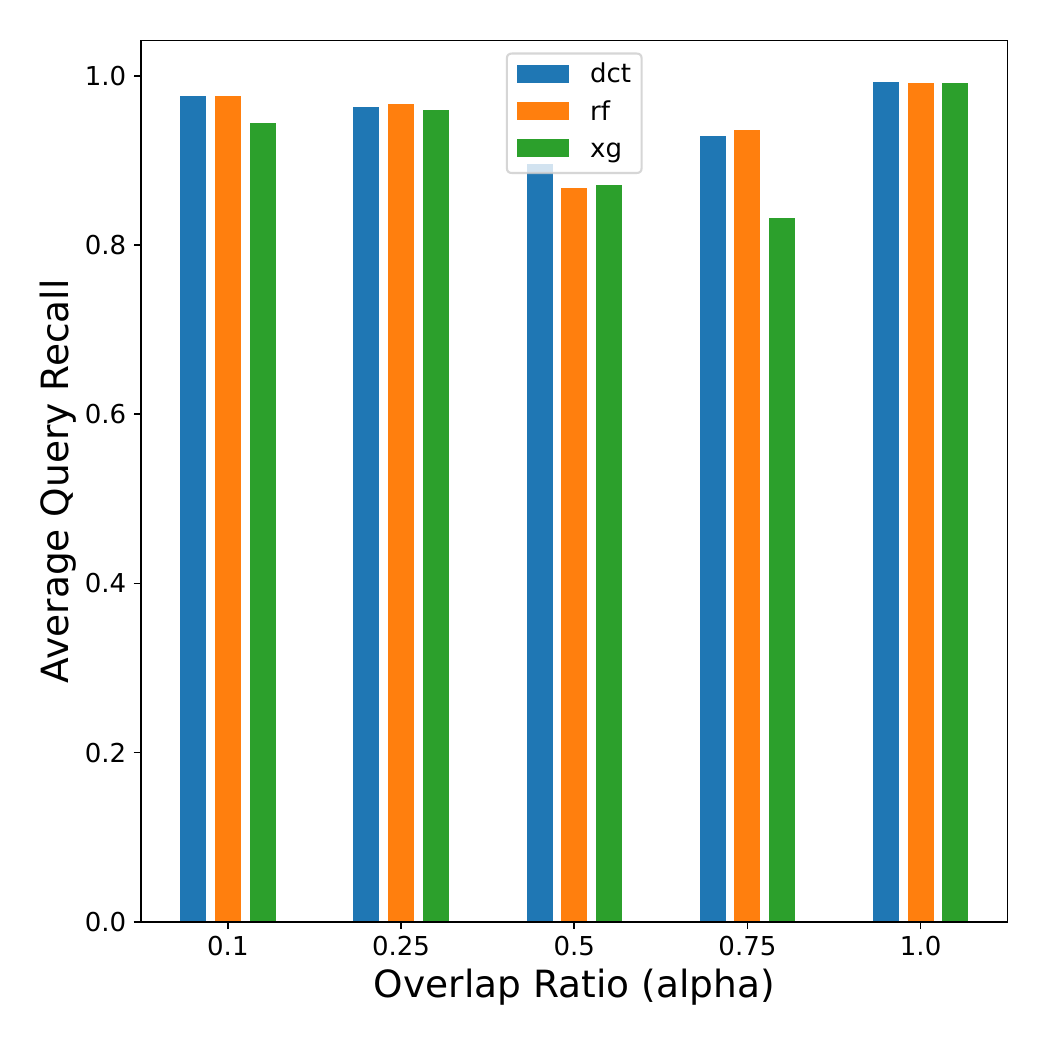}}
       %\hfill
       \hspace{1em}
     \subfloat[Query Selectivity=0.0002\label{fig:gowalla_0.0002_percent_found}]{
      \includegraphics[width=0.25\textwidth] {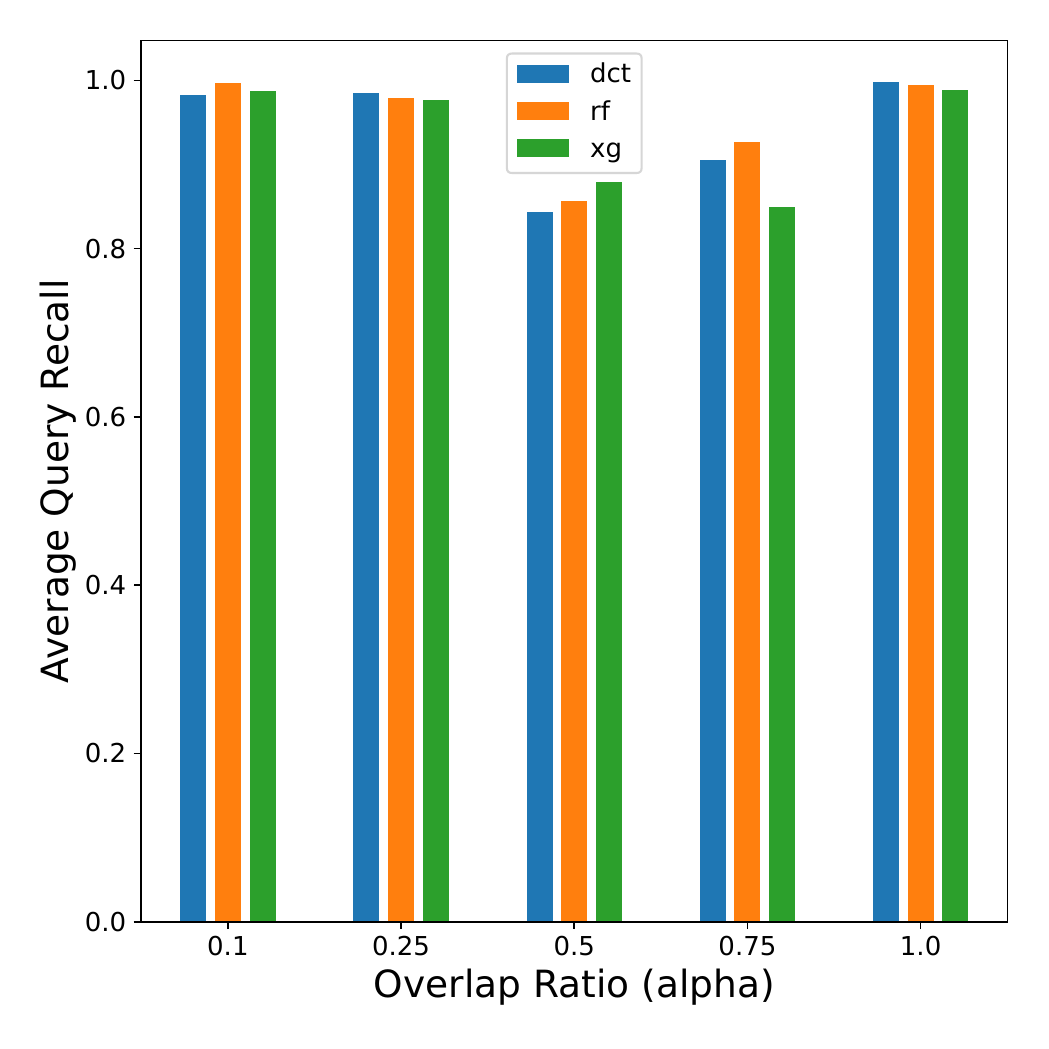}}    
        \caption{Average query recall of DT based classifiers for the gowalla dataset. The R-tree recall is always 1 and is not shown in the figures.} 
        \label{fig:gowalla_Average query recall}
\end{figure*}

In Figure~\ref{fig:gowalla_Average query recall}, we observe that the DT-based models maintain average query recall over 93\% for high-overlap queries with overlap ratio in $0.10$ and $0.25$ across all selectivities. For the queries with selectivity $0.0002$, the recall reaches up to 98\%. However, similar to the Tweets location dataset, for queries with overlap ratio closer to the threshold, the recall deteriorates for all variants of the DT-based model. 

\paragraph{Average Query Processing Time}
\begin{figure*}[ht] %[h!]
     \centering
     \captionsetup[subfloat]{labelfont=scriptsize,textfont=scriptsize}
     \subfloat[Query Selectivity=0.00005\label{fig:gowalla_0.00005_time}]{
      \includegraphics[width=0.25\textwidth] {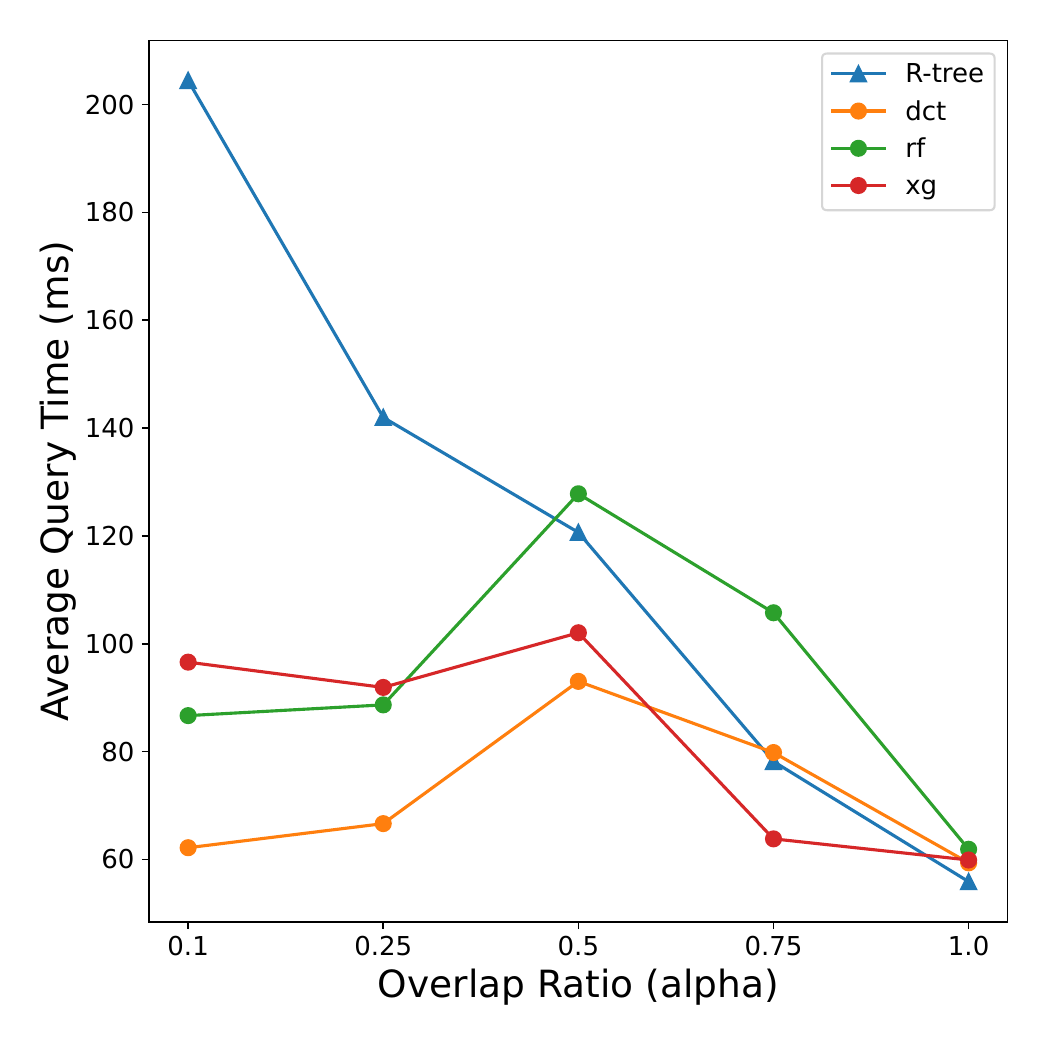}}
     %\hfill
     \hspace{1em}
     \subfloat[Query Selectivity=0.0001\label{fig:gowalla_0.0001_time}]{
      \includegraphics[width=0.25\textwidth] {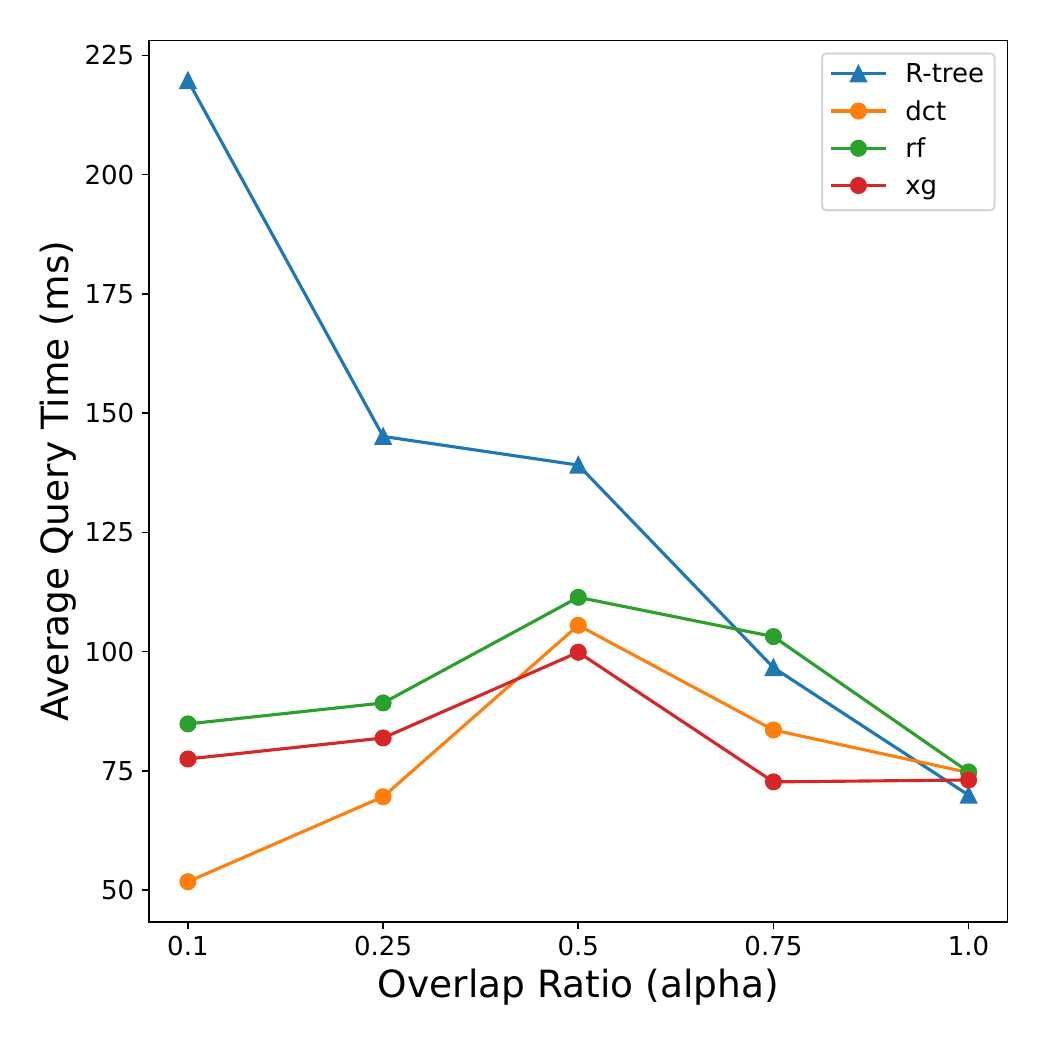}}
    %\hfill
    \hspace{1em}
     \subfloat[Query Selectivity=0.0002\label{fig:gowalla_0.0002_time}]{
      \includegraphics[width=0.25\textwidth] {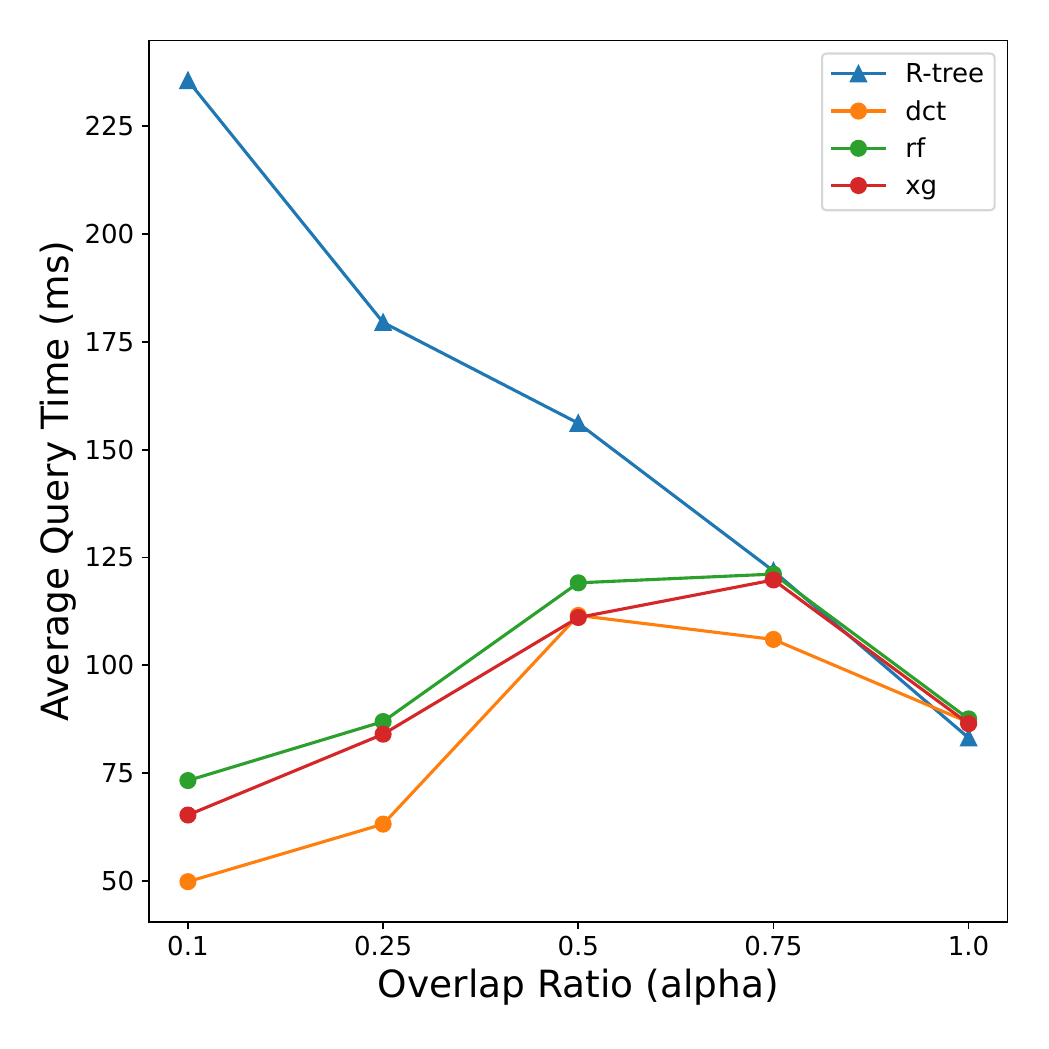}}    
        \caption{Average query processing time of DT-based classifiers for the Gowalla dataset} 
        \label{fig:gowalla_Average query time}
\end{figure*}

In Figure~\ref{fig:gowalla_Average query time}, we observe that the DT-based models maintains low query latency for processing high-overlap queries across all selectivities. Particularly, the DCT classifier enhances the query processing performance by up to 3.2X to 4.7X for queries with overlap ratio $0.10$.

\subsubsection{Effect of Loss Functions of the NN Model for the Gowalla Dataset}

\paragraph{Average Query Recall}
\begin{figure*}[ht] %[h!]
     \centering
     \captionsetup[subfloat]{labelfont=scriptsize,textfont=scriptsize}
     \subfloat[Query Selectivity=0.00005\label{fig:gowalla_NN_0.00005_percent_found}]{
      \includegraphics[width=0.25\textwidth] {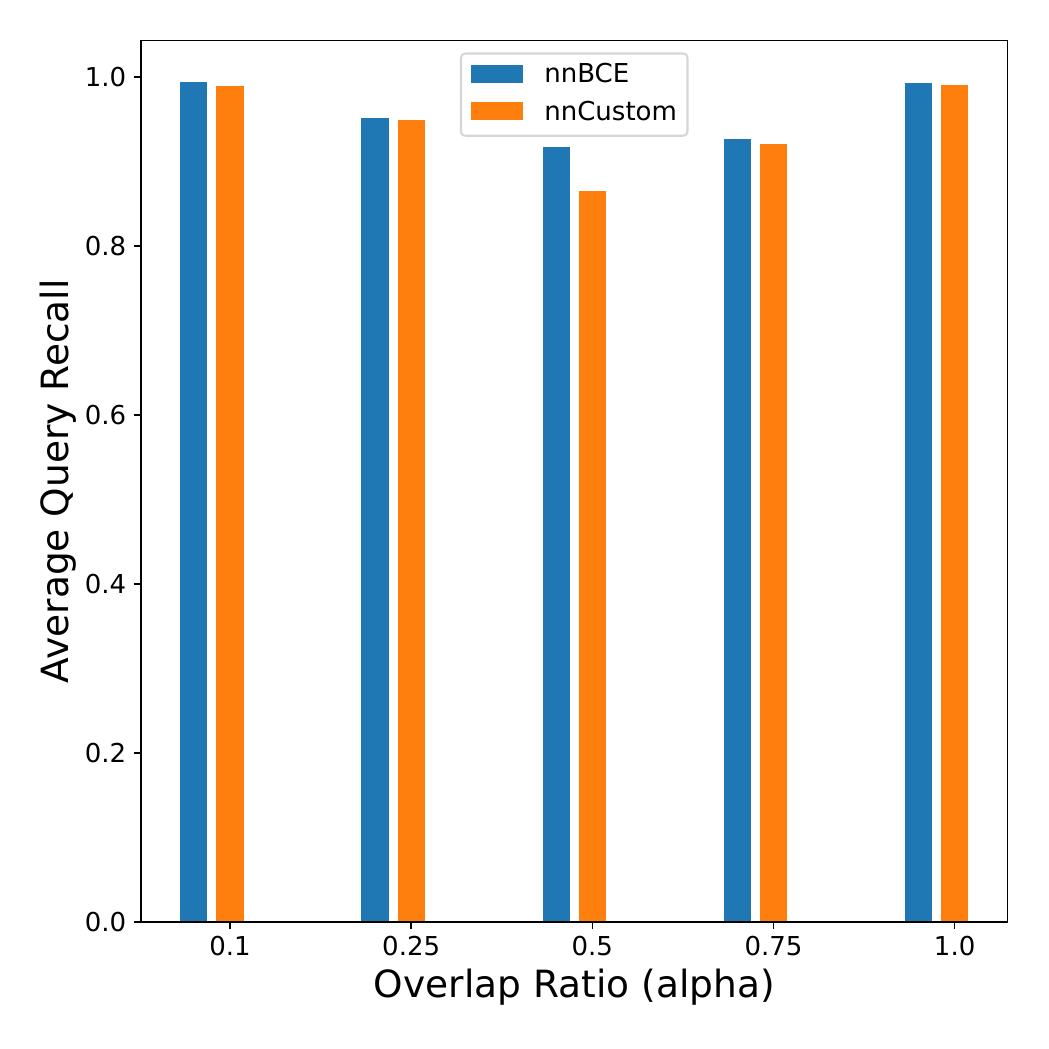}}
      %\hfill
      \hspace{1em}
      \subfloat[Query Selectivity=0.0001\label{fig:gowalla_NN_0.0001_percent_found}]{
      \includegraphics[width=0.25\textwidth] {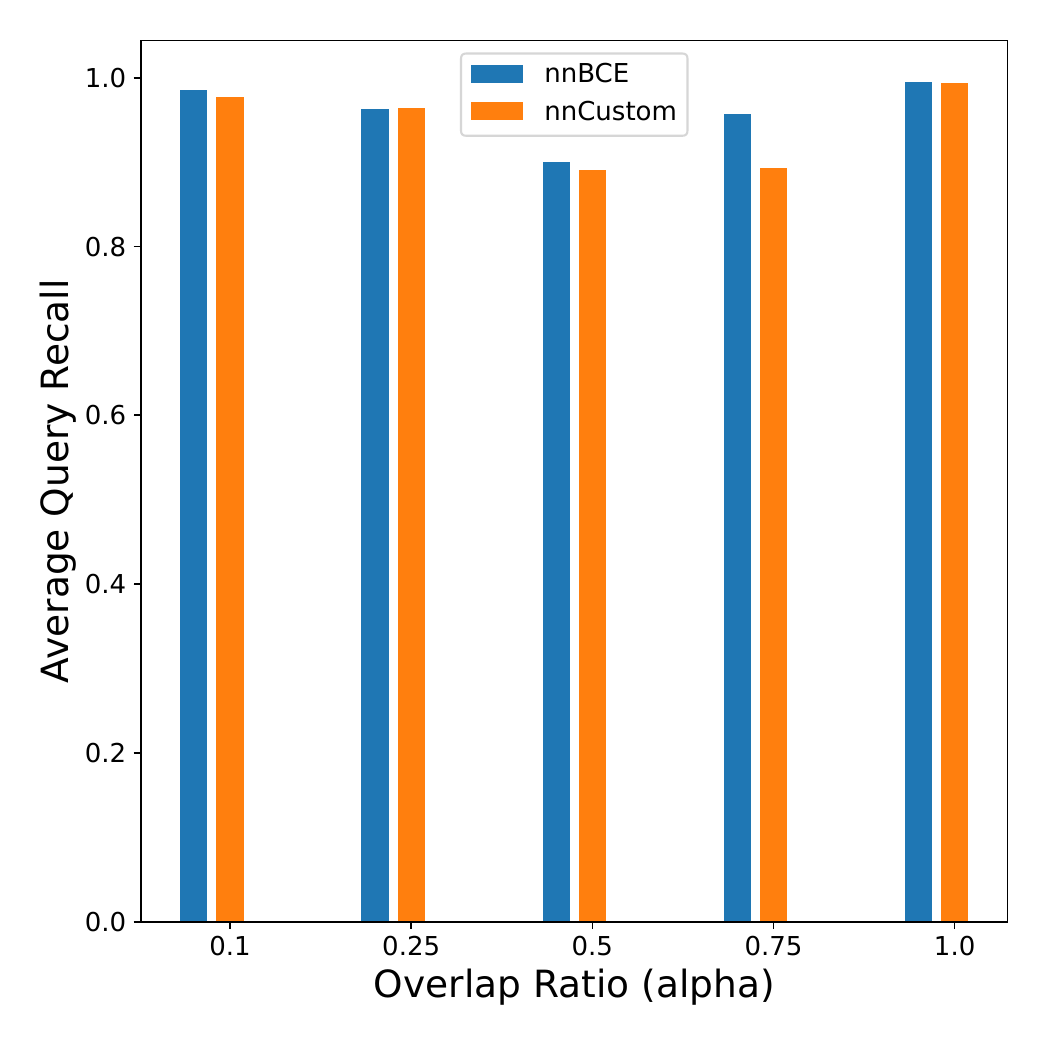}}
      %\hfill
      \hspace{1em}
      \subfloat[Query Selectivity=0.0002\label{fig:gowalla_NN_0.0002_percent_found}]{
      \includegraphics[width=0.25\textwidth] {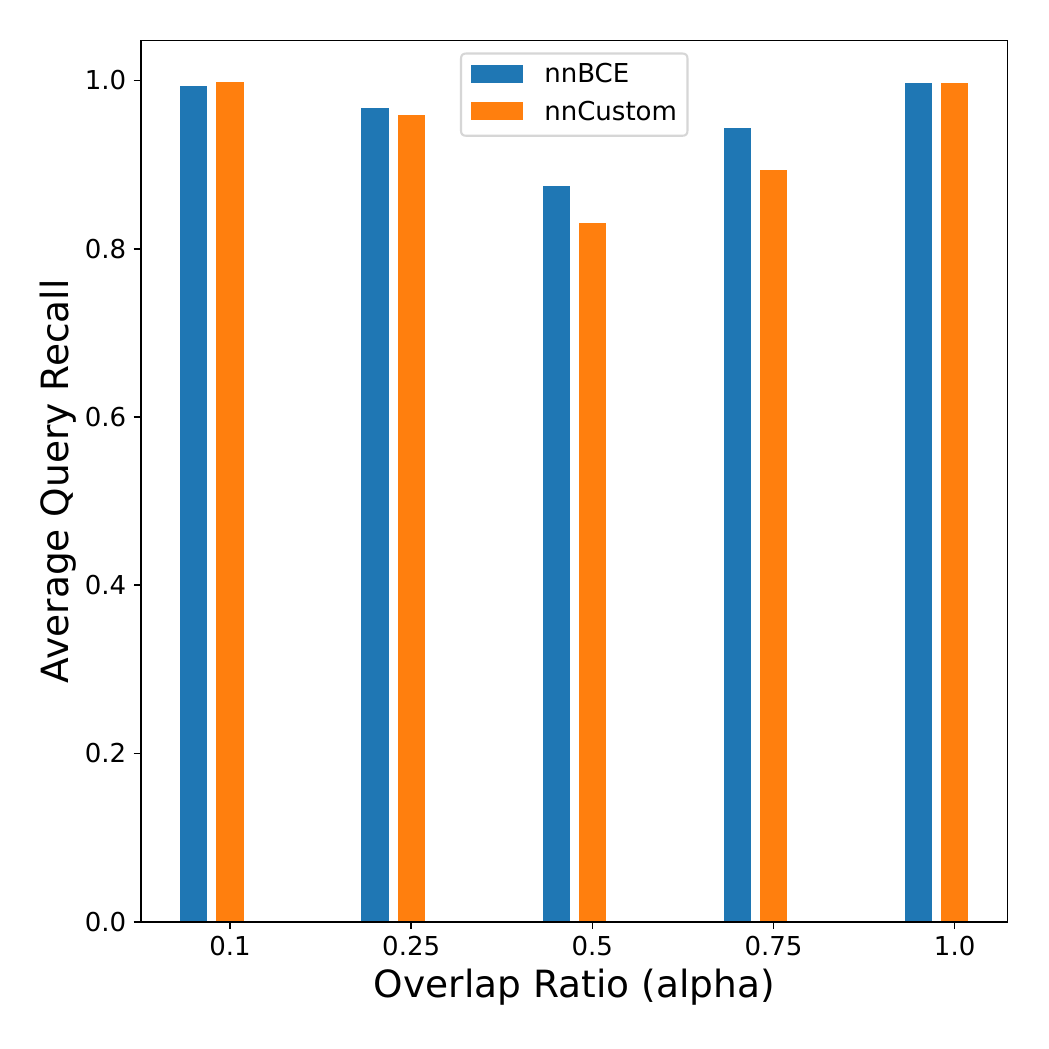}}
      \caption{Average query recall of NN based classifiers for the Gowalla dataset} 
        \label{fig:gowalla_NN_Average query recall}
\end{figure*}

In Figure~\ref{fig:gowalla_NN_Average query recall}, we observe that the average query recall of both nnBCE and the nnCustom model are similar across all values of $\alpha$. However, similar to the Tweets location dataset, the nnBCE model achieves higher recall than 
that of
the nnCustom model for some values of $\alpha$.   

\paragraph{Average Query Processing Time}
\begin{figure*}[ht] %[h!]
     \centering
     \captionsetup[subfloat]{labelfont=scriptsize,textfont=scriptsize}
     \subfloat[Query Selectivity=0.00005\label{fig:gowalla_NN_0.00005_time}]{
      \includegraphics[width=0.25\textwidth] {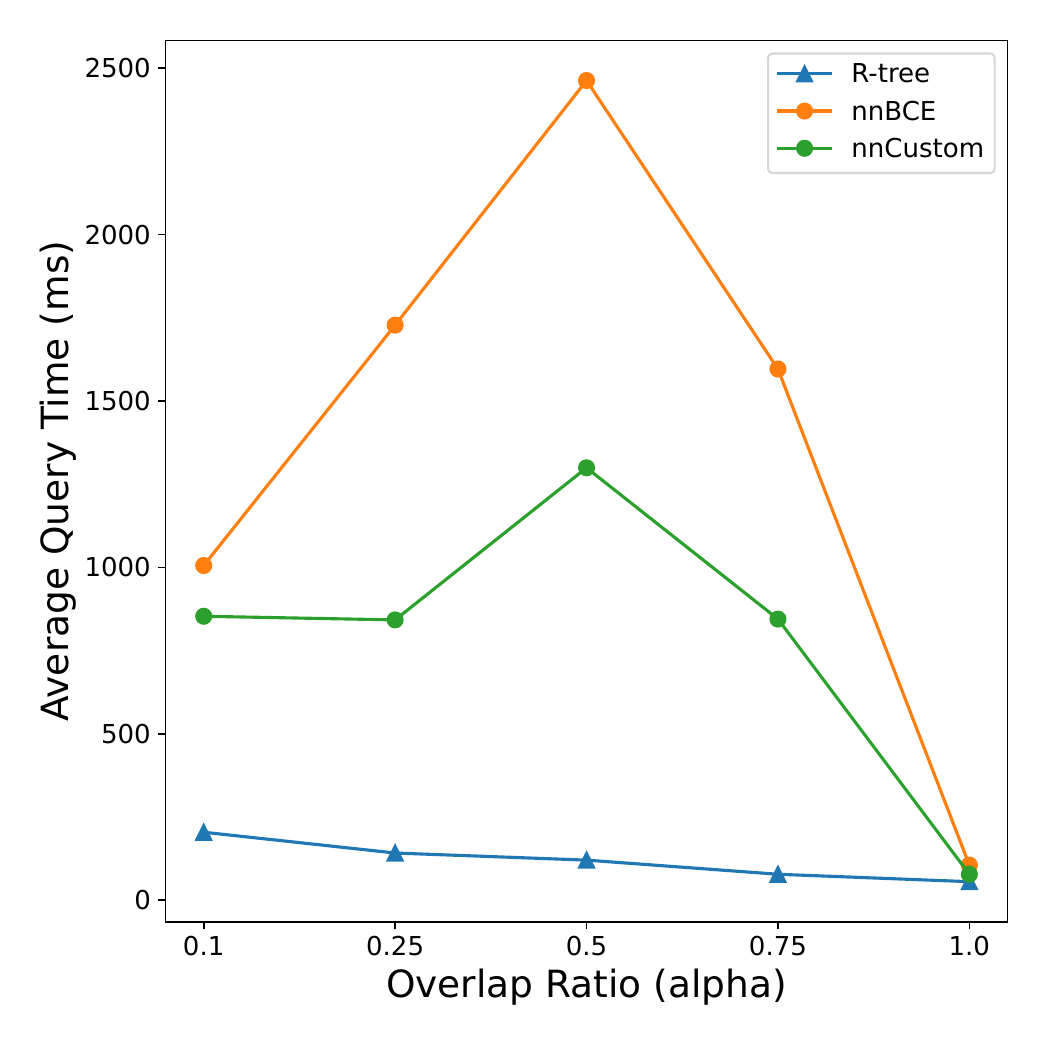}}
      %\hfill
      \hspace{1em}
       \subfloat[Query Selectivity=0.0001\label{fig:gowalla_NN_0.0001_time}]{
      \includegraphics[width=0.25\textwidth] {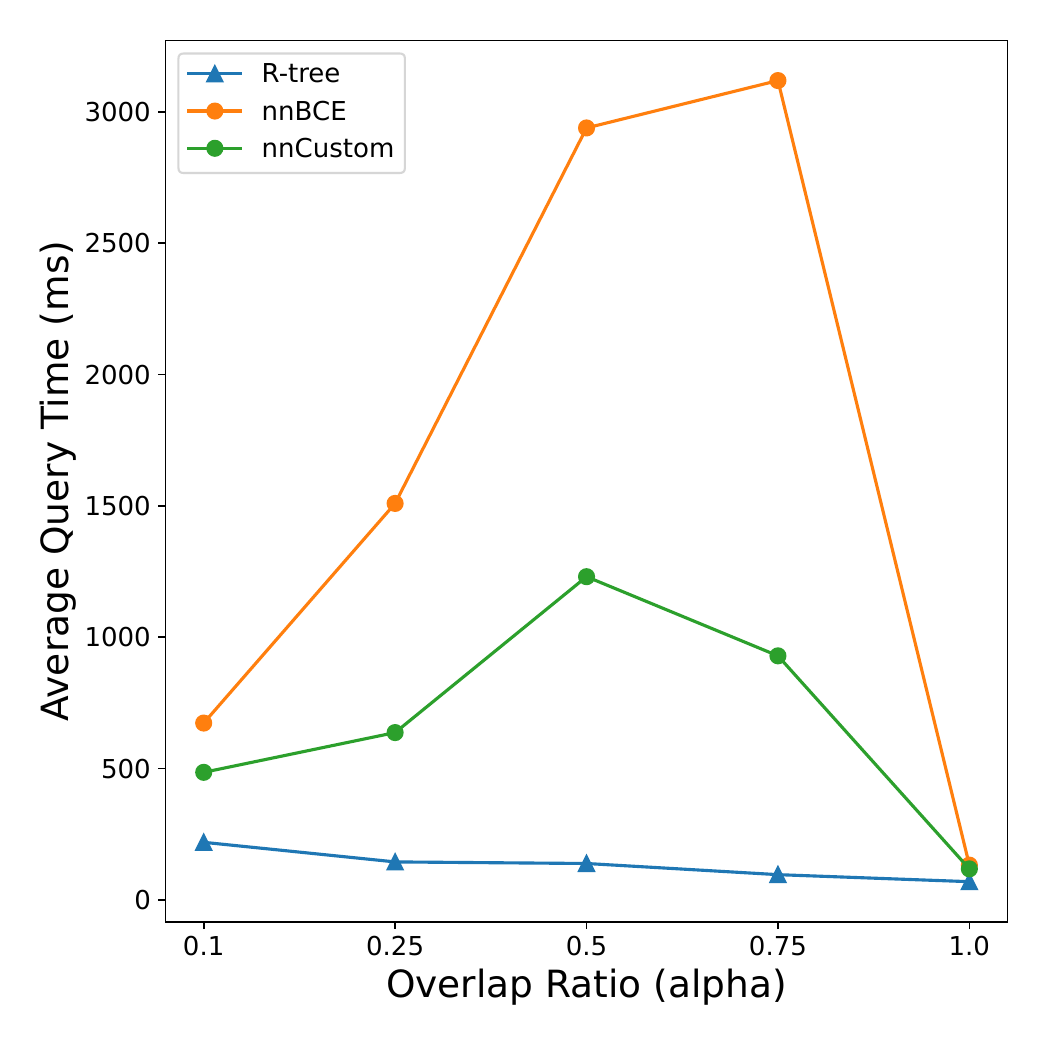}}
      %\hfill
      \hspace{1em}
       \subfloat[Query Selectivity=0.0002\label{fig:gowalla_NN_0.0002_time}]{
      \includegraphics[width=0.25\textwidth] {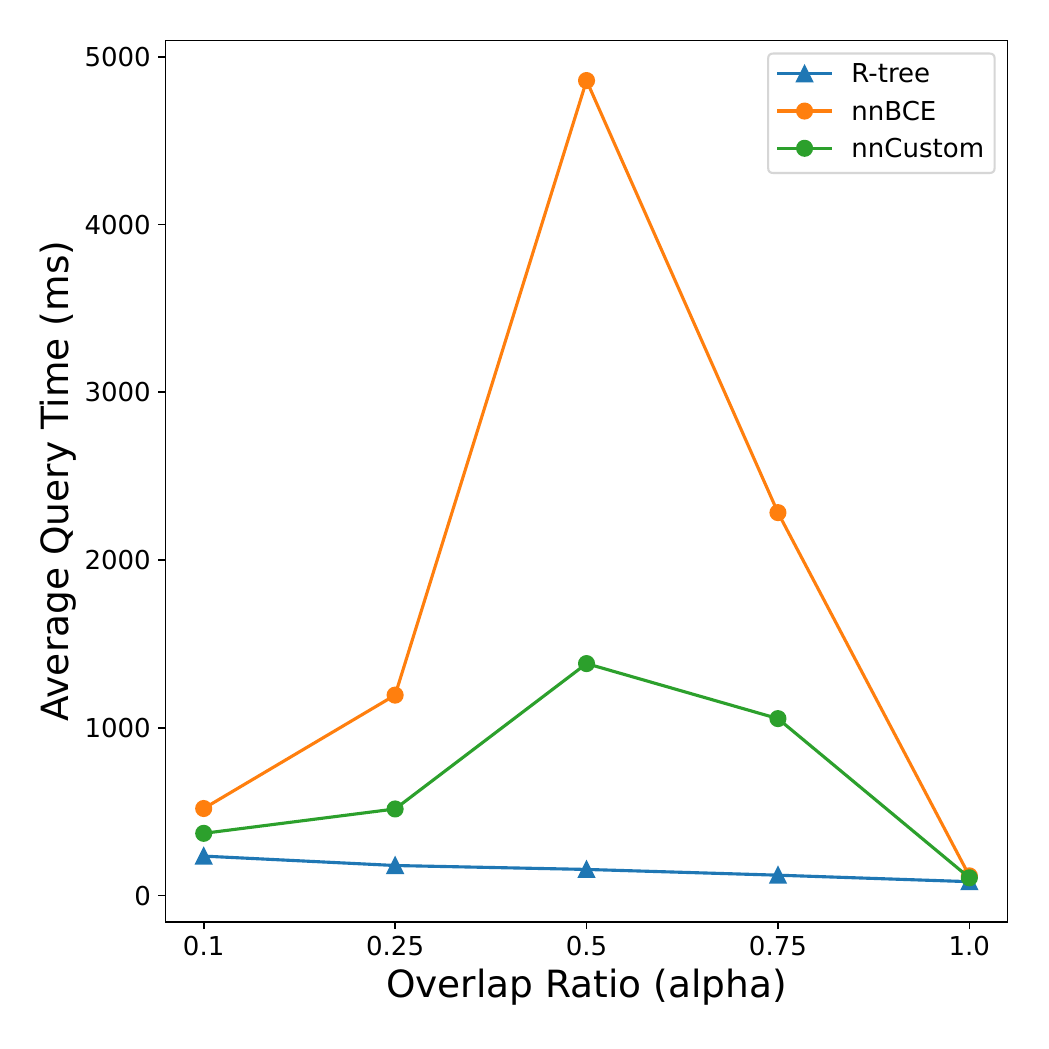}}
      \caption{Average query time of NN based classifiers for the Gowalla dataset} 
        \label{fig:gowalla_NN_Average query time}
\end{figure*}

In Figure~\ref{fig:gowalla_NN_Average query time}, the average query processing time is shown for each NN-based ML models for queries with various selectivities. Similar to the Tweets location dataset, the nnBCE models achieves higher recall with significant penalty on the query processing time. On the other hand, the nnCustom model achieves similar recall with significantly less query latency.

\subsubsection{Space Consumption of the ML Models for the Gowalla Dataset}

\begin{table}[h!]
    \caption{Average ML model size of the AI+R-tree for the gowalla dataset across all $\alpha$ values (in MBs)}
    \label{tab:modelsize_gowalla}
    \begin{tabular}{p{3em} p{3em} p{1.5em} p{1.5em} p{1.5em} c c}
    \toprule
        &  &\multicolumn{3}{c}{DT based models}& \multicolumn{2}{c}{NN based models}\\
        %&  & \multicolumn{3}{c}{NN based models}\\
        \cmidrule(lr){3-5}
        \cmidrule(lr){6-7}
        Selectivity & R-tree & DCT & RF & XG & nnBCE & nnCustom\\
    \midrule
        %dct
        %0.00005 & 695.61 & 6.20 & 5.87 & 9.26 & 9.24 & 15.97\\
        %0.0001 & 695.61 & 4.51 & 6.20 & 8.59 & 10.20 & 14.63\\
        %0.0002 & 695.61 & 3.16 & 5.18 & 7.24 & 8.59 & 13.96\\

        %rf
        %0.00005 & 695.61 & 4.10 & 4.17 & 4.30 & 4.18 & 4.15\\
        %0.0001 & 695.61 & 4.07 & 4.14 & 4.30 & 4.26 & 4.17\\
        %0.0002 & 695.61 & 4.05 & 4.10 & 4.29 & 4.27 & 4.19\\

        %xg
        %0.00005 & 695.61 & 0.075 & 0.125 & 0.125 & 0.125 & 0.125\\
        %0.0001 & 695.61 & 0.075 & 0.125 & 0.125 & 0.125 & 0.125\\
        %0.0002 & 695.61 & 0.075 & 0.125 & 0.125 & 0.125 & 0.125\\

        %NN: the model size is expected to be independent of the underlying loss function
        %0.00005 & 1106.54 & 0.13 & 0.13 & 0.13 & 0.13 & 0.13\\
        %0.00005 & 1106.54 & 0.13 & 0.13 & 0.13 & 0.13 & 0.13\\
        %0.00005 & 1106.54 & 0.13 & 0.13 & 0.13 & 0.13 & 0.13\\
        
        %Average accros all alpha values
        0.00005 & 695.61 & 9.30 & 4.18 & 0.11 & 0.13 & 0.13\\
        0.0001 & 695.61 & 8.82 & 4.18 & 0.11 & 0.13 & 0.13\\
        0.0002 & 695.61 & 7.62 & 4.18 & 0.11 & 0.13 & 0.13\\
    \bottomrule
\end{tabular}
\end{table}

In Table~\ref{tab:modelsize_gowalla}, we present the R-tree size and the average ML model overhead accross all $\alpha$ values. For the Gowalla dataset, the model overhead varies between $0.01\%$ to $1.33\%$. Based on the query workload distribution, the DCT model normally requires a finer grid to create more decision trees than the RF and XG models. As a result, the DCT model consumes more space than the other model type for the Gowalla dataset.

\subsubsection{Effect of DT-based Models for the Chicago Crimes Dataset}

\paragraph{Average Query Recall}
\begin{figure*}[ht] %[h!]
     \centering
     \captionsetup[subfloat]{labelfont=scriptsize,textfont=scriptsize}
     \subfloat[Query Selectivity=0.00005\label{fig:chicago_0.00005_percent_found}]{
      \includegraphics[width=0.25\textwidth] {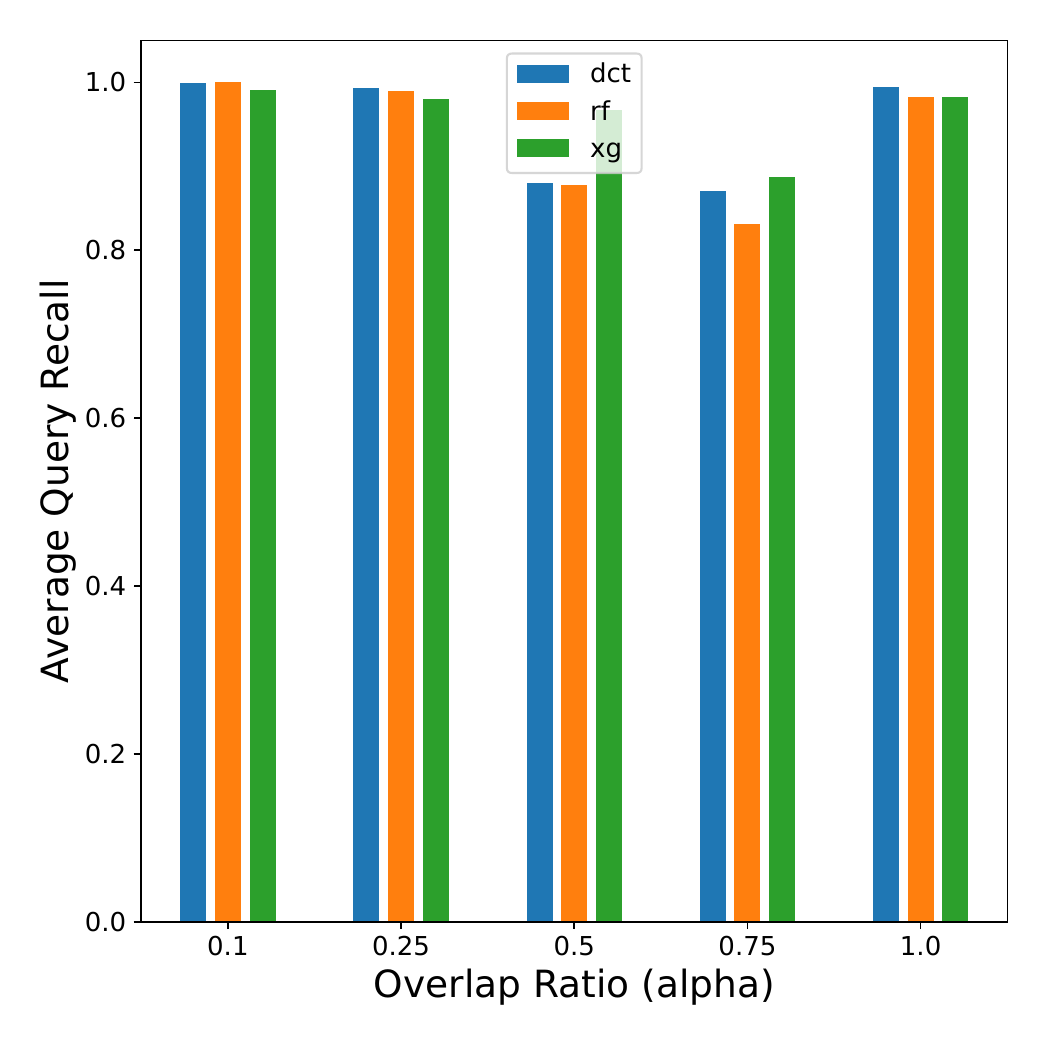}}
     %\hfill
     \hspace{1em}
     \subfloat[Query Selectivity=0.0001\label{fig:chicago_0.0001_percent_found}]{
      \includegraphics[width=0.25\textwidth] {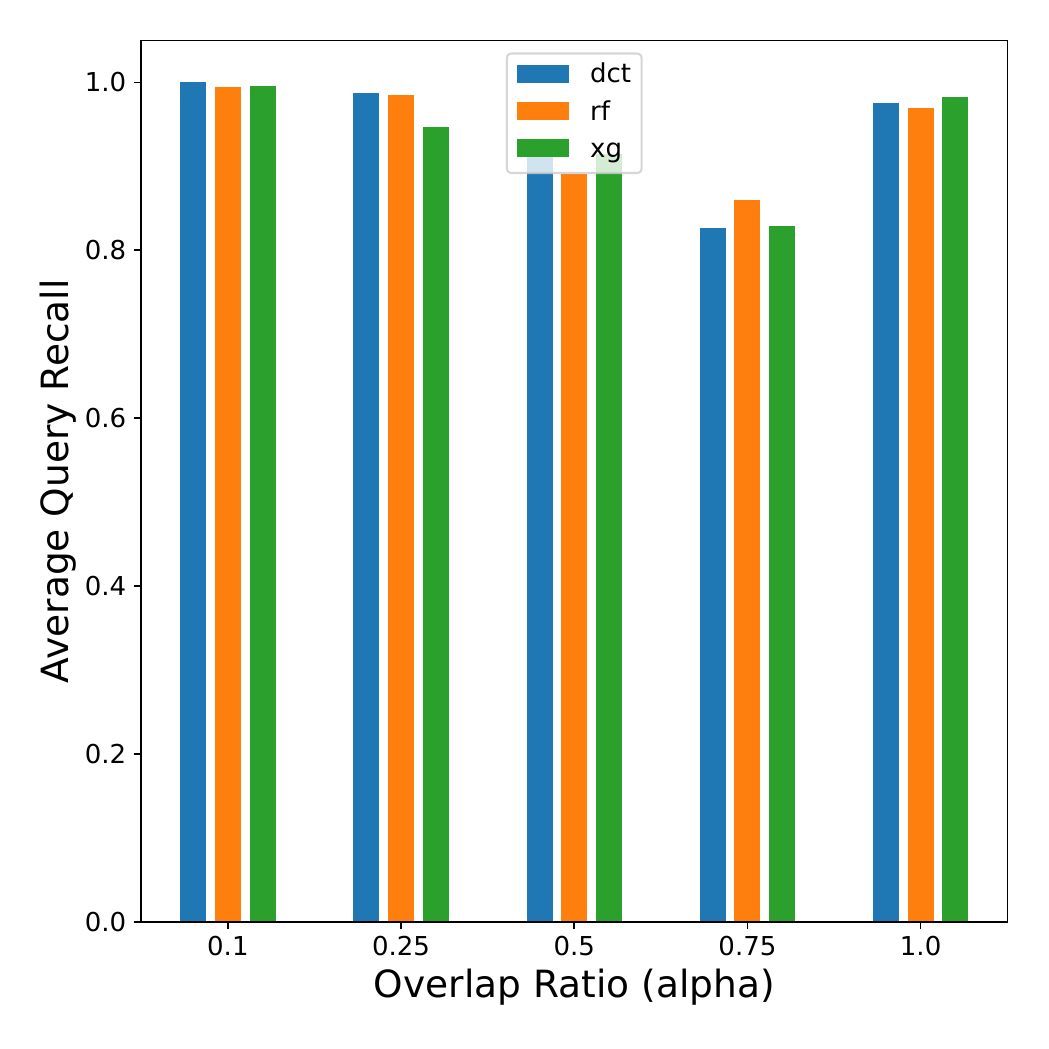}}
       %\hfill
       \hspace{1em}
     \subfloat[Query Selectivity=0.0002\label{fig:chicago_0.0002_percent_found}]{
      \includegraphics[width=0.25\textwidth] {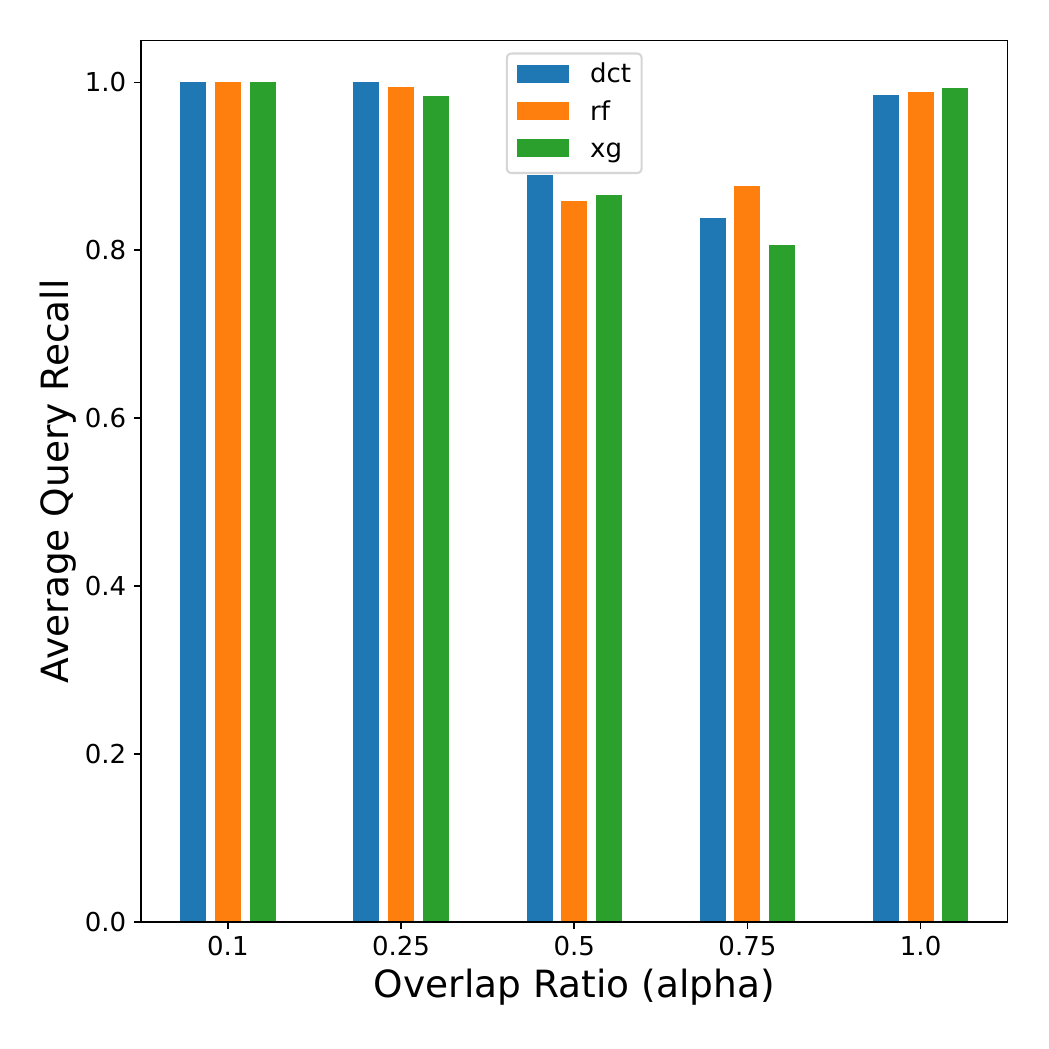}} 
        \caption{Average query recall of DT-based classifiers for the Chicago Crimes dataset. The R-tree recall is always 1 and is not shown in the figures.} 
        \label{fig:chicago_Average query recall}
\end{figure*}

In Figure~\ref{fig:chicago_Average query recall}, we observe that the DT-based models maintain average query recall over 98\% for high-overlap queries with overlap ratio in $0.10$ and $0.25$ across all selectivities. For the queries with selectivity $0.0002$, the recall reaches up to 99\%. However, similar to the Tweets location and Gowalla datasets, for queries with overlap ratio closer to the threshold, the recall deteriorates for all variants of the DT-based model.

\paragraph{Average Query Processing Time}
\begin{figure*}[ht] %[h!]
     \centering
     \captionsetup[subfloat]{labelfont=scriptsize,textfont=scriptsize}
     \subfloat[Query Selectivity=0.00005\label{fig:chicago_0.00005_time}]{
      \includegraphics[width=0.25\textwidth] {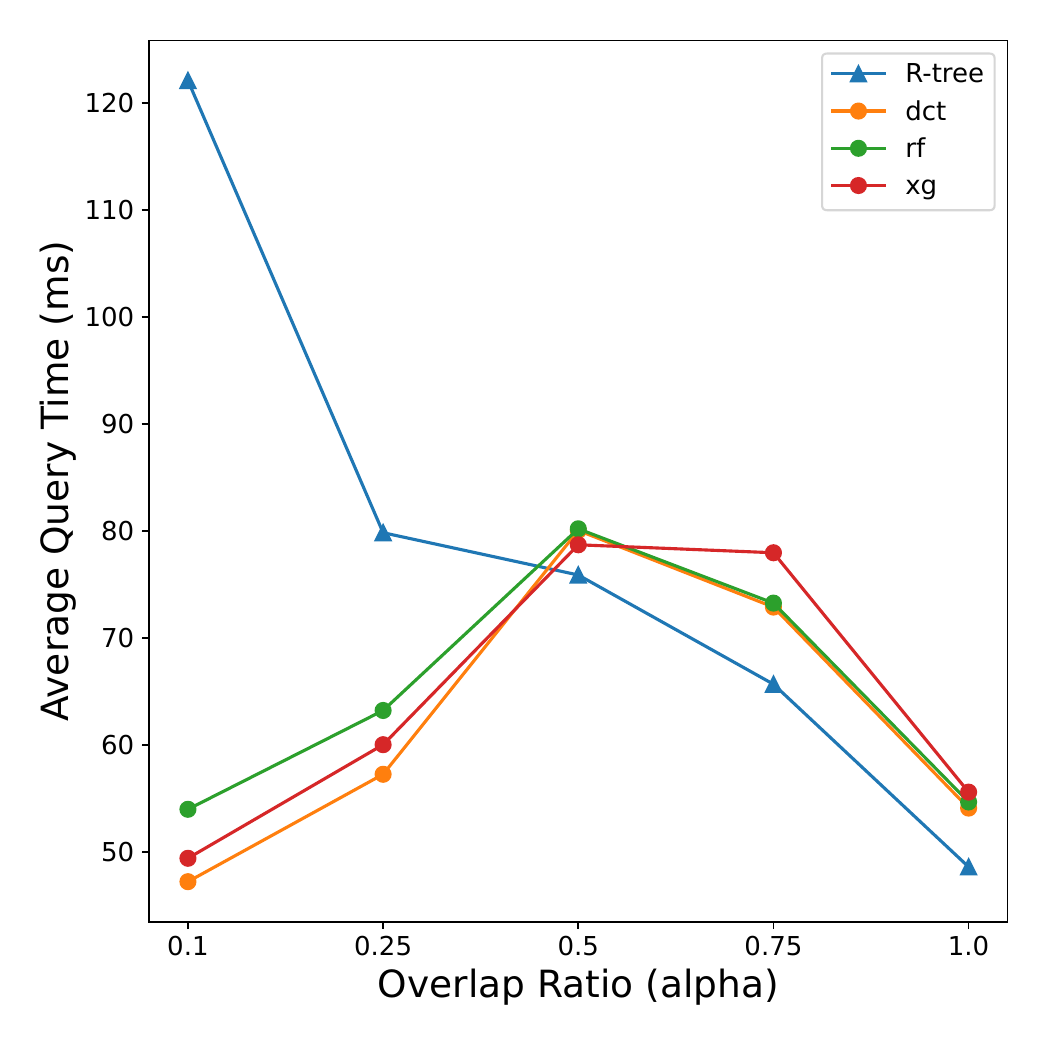}}
     %\hfill
     \hspace{1em}
     \subfloat[Query Selectivity=0.0001\label{fig:chicago_0.0001_time}]{
      \includegraphics[width=0.25\textwidth] {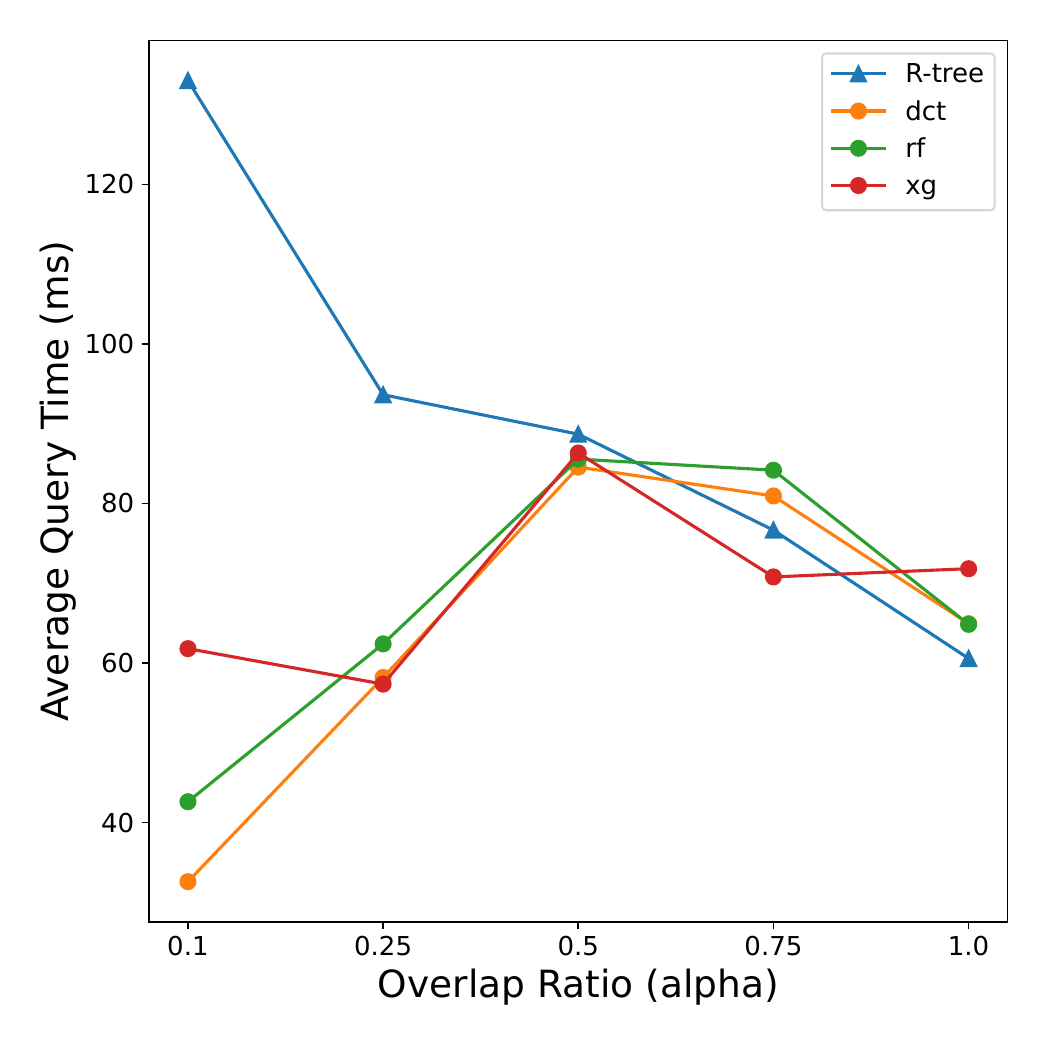}}
       %\hfill
       \hspace{1em}
     \subfloat[Query Selectivity=0.0002\label{fig:chicago_0.0002_time}]{
      \includegraphics[width=0.25\textwidth] {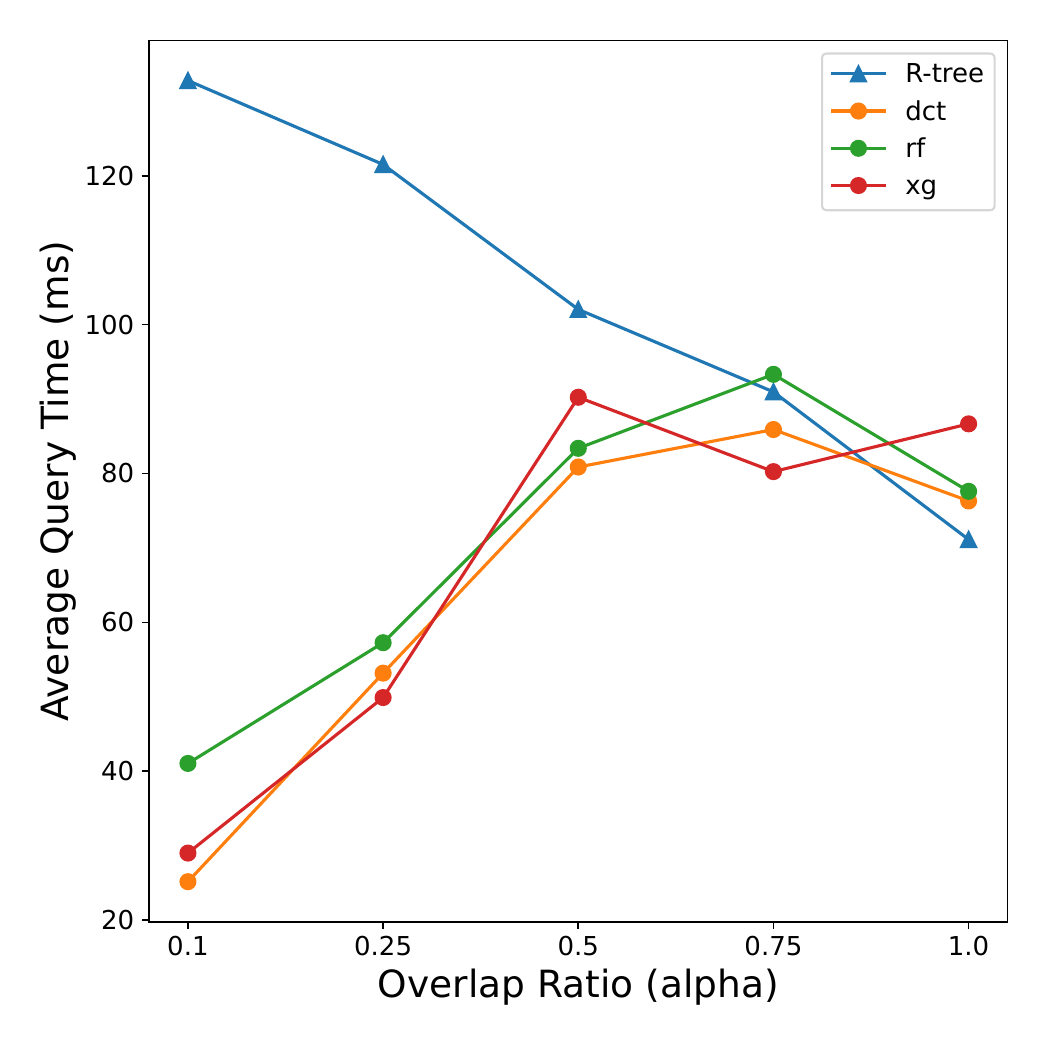}} 
        \caption{Average query time of DT-based classifiers for the Chicago Crimes dataset} 
        \label{fig:chicago_Average query time}
\end{figure*}

In Figure~\ref{fig:chicago_Average query time}, we observe that the DT-based models maintains low query latency for processing high-overlap queries across all selectivities. Particularly, the DCT classifier enhances the query processing performance by up to 2.6X to 5.2X for queries with overlap ratio $0.10$.

\subsubsection{Effect of Loss Functions of the NN Model for the Chicago Crimes Dataset}
%\textcolor{red}{Including the impact of different configuration of NN models}

\paragraph{Average Query Recall}
\begin{figure*}[ht] %[h!]
     \centering
     \captionsetup[subfloat]{labelfont=scriptsize,textfont=scriptsize}
     \subfloat[Query Selectivity=0.00005\label{fig:chicago_NN_0.00005_percent_found}]{
      \includegraphics[width=0.25\textwidth] {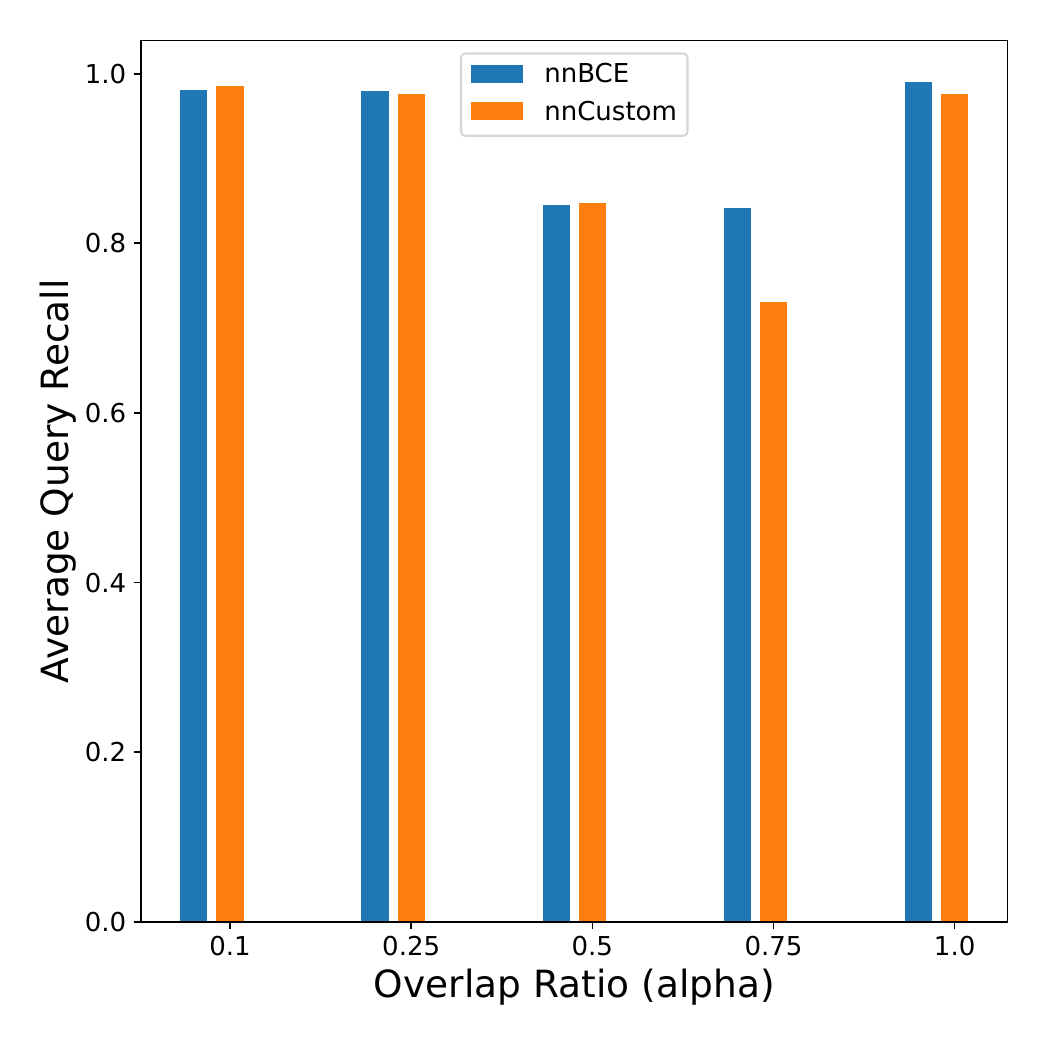}}
      %\hfill
      \hspace{1em}
      \subfloat[Query Selectivity=0.0001\label{fig:chicago_NN_0.0001_percent_found}]{
      \includegraphics[width=0.25\textwidth] {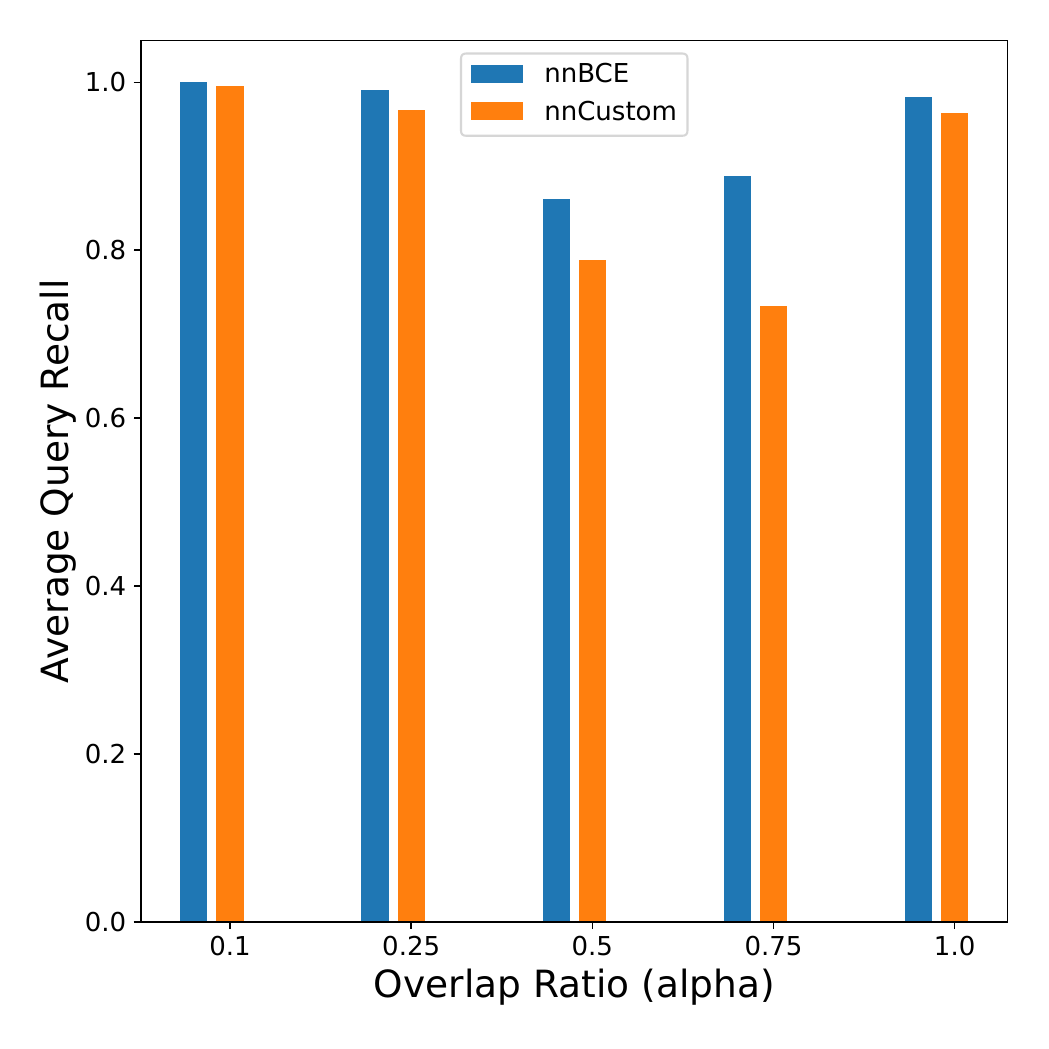}}
      %\hfill
      \hspace{1em}
      \subfloat[Query Selectivity=0.0002\label{fig:chicago_NN_0.0002_percent_found}]{
      \includegraphics[width=0.25\textwidth] {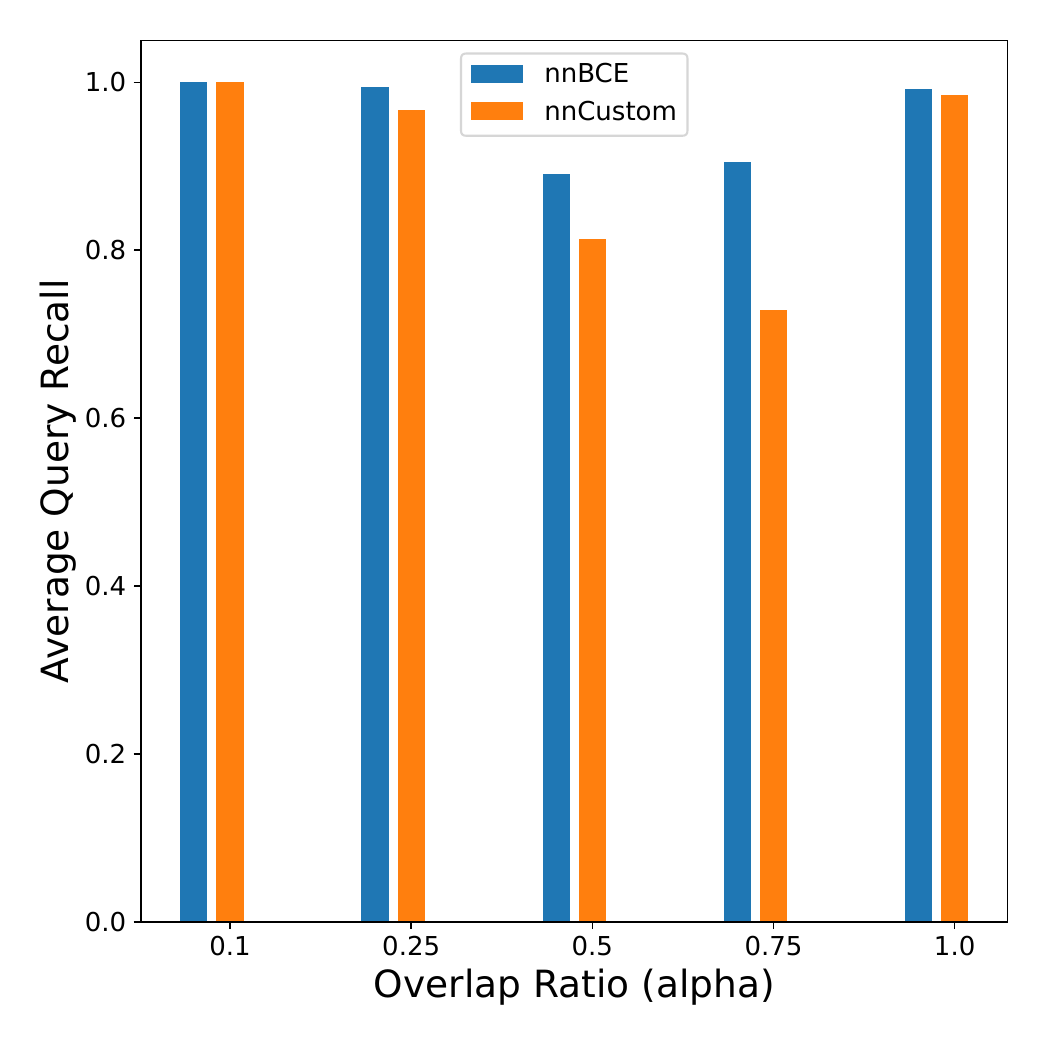}}
      \caption{Average query recall of NN based classifiers for the Chicago Crimes dataset} 
        \label{fig:chicago_NN_Average query recall}
\end{figure*}

In Figure~\ref{fig:chicago_NN_Average query recall}, we observe that the average query recall of both nnBCE and the nnCustom model are similar across all values of $\alpha$. However, similar to the Tweets location and Gowalla datasets, the nnBCE model achieves higher recall than the nnCustom model for some values of $\alpha$. 

\paragraph{Average Query Processing Time}
\begin{figure*}[ht] %[h!]
     \centering
     \captionsetup[subfloat]{labelfont=scriptsize,textfont=scriptsize}
     \subfloat[Query Selectivity=0.00005\label{fig:chicago_NN_0.00005_time}]{
      \includegraphics[width=0.25\textwidth] {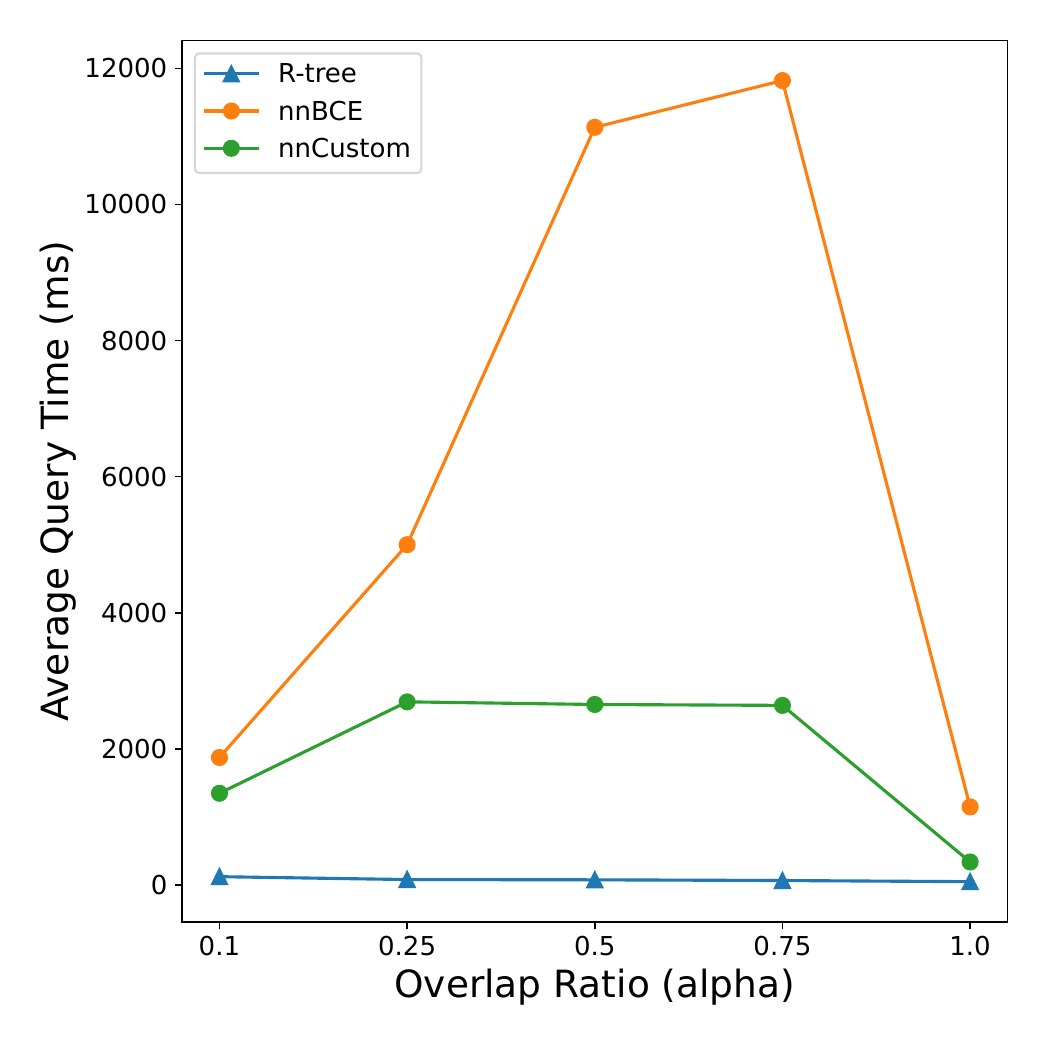}}
      %\hfill
      \hspace{1em}
       \subfloat[Query Selectivity=0.0001\label{fig:chicago_NN_0.0001_time}]{
      \includegraphics[width=0.25\textwidth] {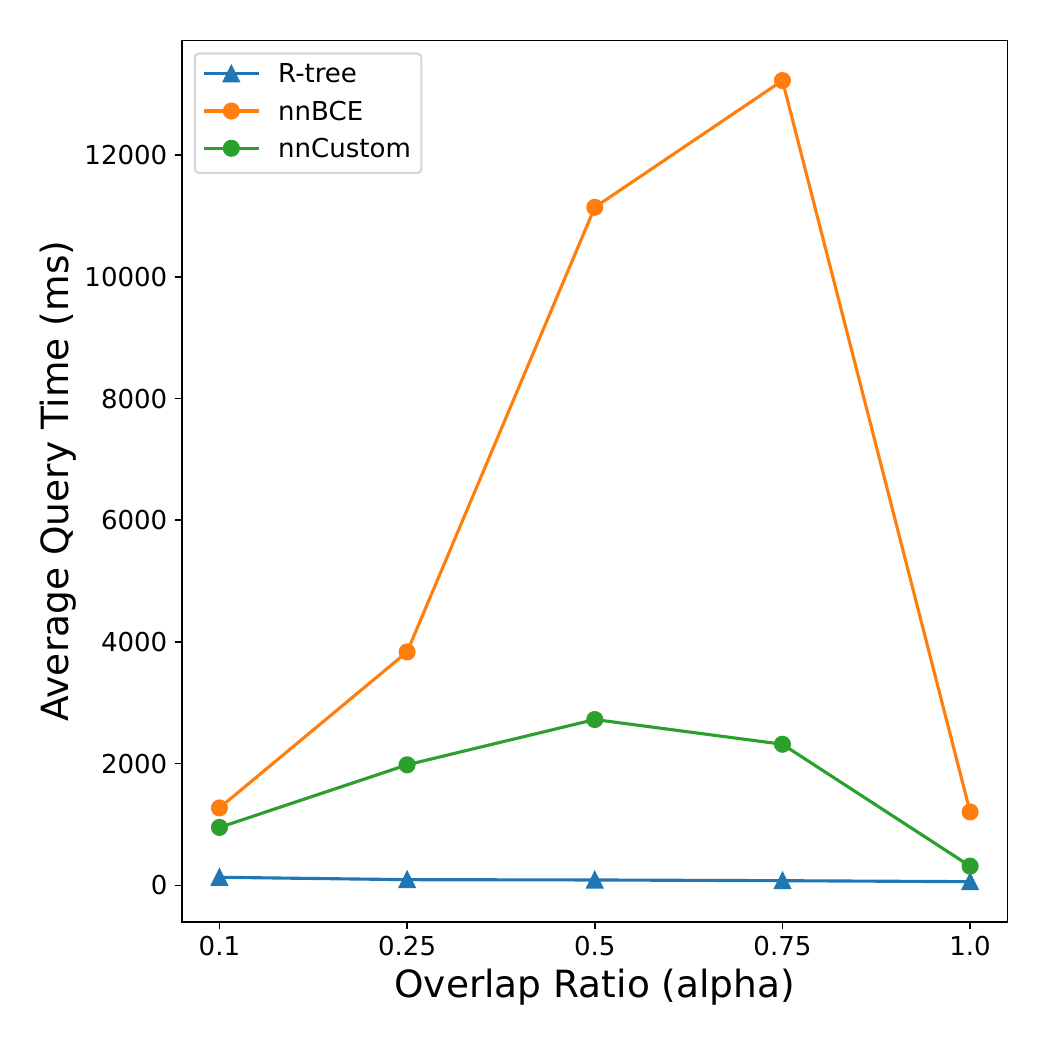}}
      %\hfill
      \hspace{1em}
       \subfloat[Query Selectivity=0.0002\label{fig:chicago_NN_0.0002_time}]{
      \includegraphics[width=0.25\textwidth] {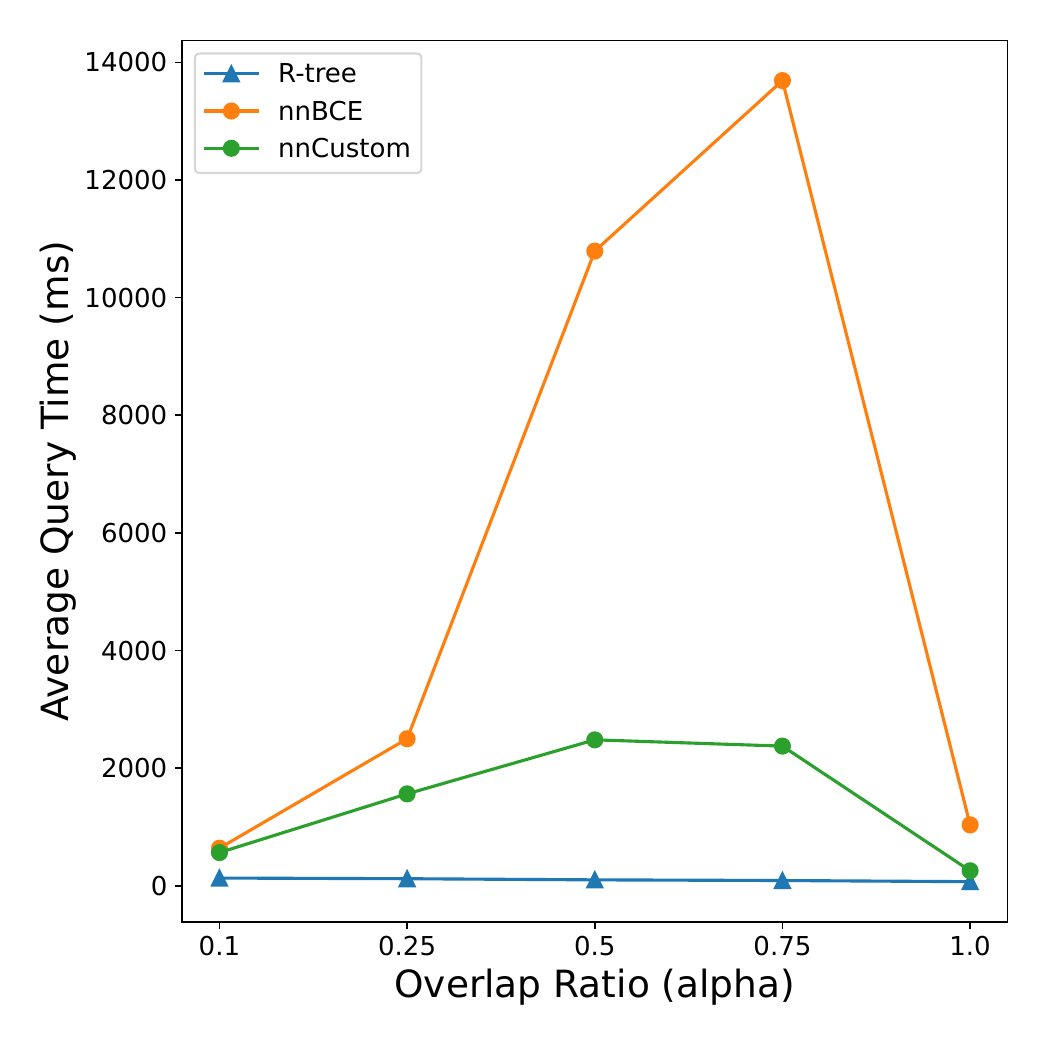}}
      \caption{Average query time of NN based classifiers for the chicago crimes dataset} 
        \label{fig:chicago_NN_Average query time}
\end{figure*}

In Figure~\ref{fig:chicago_NN_Average query time}, similar to the Tweets location and Gowalla datasets, we observe that the nnBCE models achieves higher recall with significant penalty on the query processing time. On the other hand, the nnCustom model achieves similar recall with significantly less query latency.

\subsubsection{Space Consumption of the ML Models for the Chicago Crimes Dataset}
\begin{table}[h!]
    \caption{Average ML model size of the AI+R-tree for the Chicago Crimes dataset across all $\alpha$ values (in MBs)}
    \label{tab:modelsize_chicago_crimes}
    \begin{tabular}{p{3em} p{3em} p{1.5em} p{1.5em} p{1.5em} c c}
    \toprule
        &  &\multicolumn{3}{c}{DT-based models}& \multicolumn{2}{c}{NN-based models}\\
        %&  & \multicolumn{3}{c}{NN based models}\\
        \cmidrule(lr){3-5}
        \cmidrule(lr){6-7}
        Selectivity & R-tree & DCT & RF & XG & nnBCE & nnCustom\\
    \midrule
        %dct per alpha
        %0.00005 & 482.64 & 0.83 & 0.84 & 0.85 & 1.09 & 1.09\\
        %0.0001 & 482.64 & 0.82 & 0.83 & 0.86 & 1.10 & 1.09\\
        %0.0002 & 482.64 & 0.82 & 0.83 & 0.86 & 1.10 & 1.10\\

        %rf per alpha
        %0.00005 & 482.64 & 1.49 & 1.55 & 1.66 & 1.68 & 1.66\\
        %0.0001 & 482.64 & 1.48 & 1.53 & 1.66 & 1.70 & 1.68\\
        %0.0002 & 482.64 & 1.47 & 1.51 & 1.66 & 1.72 & 1.71\\

        %xg per alpha
        %0.00005 & 482.64 & 0.072 & 0.121 & 0.121 & 0.121 & 0.121\\
        %0.0001 & 482.64 & 0.072 & 0.121 & 0.121 & 0.121 & 0.121\\
        %0.0002 & 482.64 & 0.072 & 0.121 & 0.121 & 0.121 & 0.121\\

        %NN: the model size is expected to be independent of the underlying loss function
        %0.00005 & 1106.54 & 0.13 & 0.13 & 0.13 & 0.13 & 0.13\\
        %0.00005 & 1106.54 & 0.13 & 0.13 & 0.13 & 0.13 & 0.13\\
        %0.00005 & 1106.54 & 0.13 & 0.13 & 0.13 & 0.13 & 0.13\\

        0.00005 & 482.64 & 0.94 & 1.60 & 0.11 & 0.13 & 0.13\\
        0.0001 & 482.64 & 0.94 & 1.61 & 0.11 & 0.13 & 0.13\\
        0.0002 & 482.64 & 0.94 & 1.61 & 0.11 & 0.13 & 0.13\\
    \bottomrule
\end{tabular}
\end{table}

In Table~\ref{tab:modelsize_chicago_crimes}, we present the R-tree size and the average ML model overhead across all $\alpha$ values. Notice that the model size includes both the binary and the multi-label classifiers. For the Chicago Crimes dataset, the model overhead varies between $0.02\%$ to $0.33\%$.

%\abdullah{is here for the final check and proof reading}

\section{Related Work} \label{section:related_work}
Many variants of the R-tree exist, e.g., see~\cite{guttman1984r, sellis1987r+, beckmann1990r, beckmann2009revised, samet2006foundations, manolopoulos2010r, sidlauskas2018improving, arge2008priority, kamel1993hilbert}. On the other hand, in the context of learned multi-dimensional and spatial indexes, there are several variants of ML-enhanced R-trees, e.g.,~\cite{gu2021rlr, yang2023platon, huang2023acr}. In~\cite{gu2021rlr, yang2023platon, huang2023acr}, the goal is to leverage ML techniques to build a better R-tree to replace the traditional heuristic presented in the index construction algorithm (e.g., choosing sub-tree during new data insertions).
As a result, the query processing performance is expected to be improved. Moreover, these ML-enhanced variants of R-trees do not modify the query processing algorithms 
but 
rather advocate for re-using the existing query processing techniques. Notice that regardless of the type of the R-tree, all R-tree variants attempt to reduce the amount of node overlap. However, with dynamic updates, the shape of 
%an 
a
constructed R-tree deteriorates. As a result, our design principles for the AI+R-tree can be applied to any of the R-tree variants.

A class of learned multi-dimensional indexes is designed to replace a traditional multi-dimensional index structure~\cite{al2024survey,al2020tutorial,liu2024good,li2024survey}. However, in the context of the AI+R-tree, our target is not to replace the existing index structure rather to enhance its performance using ML models. On the other hand, there are several learned multi-dimensional indexes that leverage a projection function to map the multi-dimensional points into one-dimension~\cite{al2024survey,al2020tutorial,liu2024good,li2024survey}. After that any one-dimensional learned index can be used in the projected space. However, the proposed AI+R-tree avoids using a projection function, and operate directly on the original multi-dimensional representation of the spatial data objects.

The idea of using helper ML models inside traditional multi-dimensional and spatial indexes to enhance their query processing performance 
%have 
has
been presented in~\cite{hadian2020handsoff, pandey2020case, kang2021the}. In the context of ML-enhanced multi-dimensional indexes, the focus of the above mentioned techniques is not on analyzing (i.e., high- vs. low-overlap queries) and optimizing the index for the given query workload. Notice that the AI+R-tree focuses on analyzing the query workload to identify the queries for which a traditional disk-based spatial index (in this case, the R-tree) does not perform well. Moreover, we propose to adopt a hybrid approach to leverage the benefit of both the proposed AI-tree, and the traditional R-tree.

The tradeoffs between In-place and Out-of-place strategies for supporting updates in the context of learned one-dimensional indexes have been identified in~\cite{chatterjee2024limousine}. In contrast, in this paper, we present the design tradeoffs for supporting both In-place and Out-of-place strategies in the context of the (learned multi-dimensional) AI+R-tree. On the other hand, the benefit of designing a custom loss function in the context of Space Partitioning has been presented in~\cite{fahim2023unsupervised}. In this paper, we present a case study by identifying the potential benefit of designing a custom loss function in the context of the (learned multi-dimensional) AI+R-tree. 

\section{Future Research Directions}~\label{section: future directions}

\subsection{Investigating the Benefit of the Hybrid AI+X-tree Structure}
Similar to the hybrid structure~\cite{davitkova2024learning}, the AI+R-tree can fall back to the traditional R-tree during the re-training process or in case of a significant distribution shift. Moreover, 
%the the 
the
benefit of a 2-tree architecture in the context of larger than memory indexes has been presented in~\cite{zhou2023two}. Although the motivation and application of the proposed index in~\cite{zhou2023two} is different than the AI+R-tree, the hybrid 2-tree (i.e., the AI-tree and R-tree) design of the AI+R-tree structure might be beneficial for other index structures as long as node overlaps exist, and hence multiple tree paths are explored during search. As a result, it is an interesting direction to investigate the benefit of the AI+X-tree structure, where $X$ is a traditional spatial 
%of 
or 
a multi-dimensional index structure (e.g., the SS-tree~\cite{white1996similarity}).

\subsection{Efficient Query Processing over the Mutable AI+R-tree}
We have presented the design tradeoffs for realizing a mutable AI+R-tree. As a result, analyzing the empirical tradeoffs of the mutable AI+R-tree, and developing efficient query processing techniques over the mutable structure are interesting directions for future research. Moreover, incorporating the concept of Machine Unlearning~\cite{kurmanji2024machine} in the context of the mutable AI+R-tree is an interesting future research direction. 

\subsection{Efficient Multi-label NN-based Classifier} 
An NN-based model provides an opportunity to design custom loss functions tailored to the requirement of the underlying learning task~\cite{fahim2023unsupervised}. The proposed NN-based custom loss function shows promising results over the standard loss function. However, the query processing performance of the NN-based models is inferior to the DT-based models in the context of the AI+R-tree. As a result, based on the proposed custom loss function, investigating the techniques to develop an efficient multi-label NN-based classifier in the context of the AI+R-tree is an interesting direction for future research. 

\section{Conclusion}~\label{section:conclusion}
We investigate the empirical tradeoffs in processing dynamic query workloads over the AI+R-tree. Particularly, we investigate the impact of the choice of ML models on the AI+R-tree query processing performance. We present a case study of designing a custom loss function for a neural-network-based model tailored to the requirement of the AI+R-tree query processing. Furthermore, we present the design tradeoffs for supporting inserts/updates/deletes with the vision of realizing a mutable AI+R-tree. Experiments demonstrate that the AI+R-tree can maintain high recall and low query latency while processing dynamic query workloads.
In summary, this paper takes an important step towards realizing the AI+R-tree in various practical settings, and opens several directions for future research within 
%the AI+R-tree 
ML-enhanced R-tree structures.

\section*{Acknowledgments}
The authors thank Dr. Raymond Yeh for his valuable discussion and feedback during the initial stages of this work. Jianguo Wang acknowledges the support of the National Science Foundation under Grant Number 2337806.

%AAM: added bibliography
%\balance
\bibliographystyle{IEEEtran}
\bibliography{reference}

%\newpage
\section{Biography Section}
%If you have an EPS/PDF photo (graphicx package needed), extra braces are
%needed around the contents of the optional argument to biography to prevent
%the LaTeX parser from getting confused when it sees the complicated
%$\backslash${\tt{includegraphics}} command within an optional argument. (You can create
%your own custom macro containing the $\backslash${\tt{includegraphics}} command to make things
%simpler here.)
 
%\vspace{11pt}

%\bf{If you include a photo:}\vspace{-33pt}
%\begin{IEEEbiography}[{\includegraphics[width=1in,height=1.25in,clip,keepaspectratio]{fig1}}]{Michael Shell}
%Use $\backslash${\tt{begin\{IEEEbiography\}}} and then for the 1st argument use $\backslash${\tt{includegraphics}} to declare and link the author photo.
%Use the author name as the 3rd argument followed by the biography text.
%\end{IEEEbiography}

%\vspace{11pt}

%\bf{If you will not include a photo:}\vspace{-33pt}
\begin{IEEEbiographynophoto}{Abdullah-Al-Mamun}
is a PhD candidate at the Department of Computer Science, Purdue University. His research interest is in the area of Machine Learning for Database Systems, particularly, in the field of Learned Multi- and High-dimensional Index Structures.
\end{IEEEbiographynophoto}

\begin{IEEEbiographynophoto}{Ch. Md. Rakin Haider}
%Abdullah: place holder
is a PhD candidate at the Department of Computer Science, Purdue University. His research interests are in the area of improving traditional systems using machine learning tools while maintaining certain desirable properties such as fairness and efficiency.
\end{IEEEbiographynophoto}

% \begin{IEEEbiographynophoto}{Raymond A. Yeh}
% %Abdullah: place holder
% is a tenure-track assistant professor in the Department of Computer Science at Purdue University. He completed his Ph.D. and M.S. in Electrical and Computer Engineering at the University of Illinois at Urbana-Champaign (UIUC). 
% He is interested in research relating to machine learning and computer vision.
% \end{IEEEbiographynophoto}

\begin{IEEEbiographynophoto}{Jianguo Wang}
%Abdullah: place holder
is a tenure-track assistant professor in the Department of Computer Science at Purdue University. He received his Ph.D. degree in Computer Science from UC San Diego. His research interests include disaggregated databases and vector databases. His research has won multiple prestigious awards, including the NSF CAREER Award and the SIGMOD Research Highlight Award.
\end{IEEEbiographynophoto}

\begin{IEEEbiographynophoto}{Walid G. Aref}
%Abdullah: place holder
is a professor  of Computer Science at Purdue University. His research focus is in extending the functionality of database systems in support of emerging applications, e.g., spatial, spatio-temporal and graph databases. He is also interested in query processing, indexing, and data streaming. He has served as Editor-in-Chief of the ACM Transactions of Spatial Algorithms and Systems (ACM TSAS), associate editor of the VLDB Journal and the ACM Transactions of Database Systems (ACM TODS). Walid has won several best paper awards including the 2016 VLDB ten-year best paper award. He is a Fellow of the IEEE, and a member of the ACM.

\end{IEEEbiographynophoto}

\vfill

\end{document}